\documentclass[fleqn,usenatbib]{mnras}

\usepackage{newtxtext,newtxmath}
\usepackage[table]{xcolor}
\usepackage[T1]{fontenc}
\usepackage{ae,aecompl}
\usepackage{url}
\usepackage{lscape} 
\usepackage{multirow}
\usepackage{soul}

\usepackage{graphicx}	
\usepackage{amsmath}	
\usepackage{color}
\usepackage{etoolbox}
\usepackage{journals}

\definecolor{lightgrey}{rgb}{0.85,0.9,0.95}

\title[Massive compact galaxies in MaNGA]{The puzzling origin of massive compact galaxies in MaNGA} %

\author[A. Schnorr-M\"uller et al.]{
A. Schnorr-M\"uller$^{1},$\thanks{E-mail: allan.schnorr@ufrgs.br}
M. Trevisan$^{1}$,
R. Riffel$^{1}$,
A. L. Chies-Santos$^{1}$,
\newauthor
C. Furlanetto$^{2}$,
T. V. Ricci$^{3}$,
F. S. Lohmann$^{1}$,
R. Flores-Freitas$^{1}$,
\newauthor
N. D. Mallmann${^1}$,
and  K. A. Alamo-Mart\'inez$^{4}$\\
$^1$Universidade Federal do Rio Grande do Sul -- Departamento de Astronomia -- 91501-970, Porto Alegre-RS, Brazil\\
$^2$Universidade Federal do Rio Grande do Sul -- Departamento de F\'isica -- 91501-970, Porto Alegre-RS, Brazil\\
$^3$Universidade Federal da Fronteira Sul -- Campus Cerro Largo -- 97900-000, Cerro Largo-RS, Brazil\\
$^4$Departamento de Astronom\'ia, Universidad de Guanajuato, Apartado Postal 144, 36000, Guanajuato, Guanajuato, Mexico \\
}

\date{Accepted XXX. Received YYY; in original form ZZZ}

\pubyear{2020}

\begin{document}
\label{firstpage}
\pagerange{\pageref{firstpage}--\pageref{lastpage}}
\maketitle

\begin{abstract}

We characterized the kinematics, morphology, and stellar population (SP) properties of a sample of massive compact quiescent galaxies (MCGs, $10\,\lesssim\,\log M_\star$/M$_\odot\,\lesssim\,11$ and $r_{\rm e}$\,$\sim$\,1--3\,kpc) in the MaNGA Survey, with the goal of constraining their formation, assembly history and assessing their relation with non-compact quiescent galaxies.  We compared their properties with those of a control sample of median-sized quiescent galaxies ($r_{\rm e}$\,$\sim$\,4--8\,kpc) with similar effective velocity dispersions. MCGs have elevated rotational support, as evidenced by a strong anti-correlation between the Gauss-Hermite moment $h_3$ and $V/\sigma$. In contrast, 30$\%$ of control sample galaxies (CSGs) are slow rotators, and fast-rotating CSGs generally show a weak $h_3$--$V/\sigma$ anti-correlation. MCGs and CSGs have similar ages, but MCGs are more metal-rich and $\alpha$-enhanced. Both MCGs and CSGs have shallow negative metallicity gradients and flat [$\alpha$/Fe] gradients. On average, MCGs and CSGs have flat age gradients, but CSGs have a significantly larger dispersion of gradient values. The kinematics and SP properties of MCGs suggest that they experienced highly-dissipative gas-rich events, such as mergers, followed by an intense, short, and centrally concentrated burst of star formation, between 4 to 10\,Gyr ago ($z\sim0.4-2$), and had a quiet accretion history since then. This sequence of events might be analogous to, although less extreme than, the compaction events which formed compact quiescent galaxies at $z$\,$\sim$\,2. The small sizes of MCGs, and the high efficiency and short duration of their last star formation episode suggest that they are descendants of compact post-starburst galaxies. 

\end{abstract}




\section{Introduction}
\label{sec:intro}

 Large Extragalactic Surveys have shown that the local galaxy population is strongly bimodal in color \citep{strateva01,baldry04,baldry06}. In the color-magnitude diagram, local galaxies are mainly divided into two groups, the blue cloud, populated by galaxies where the star formation rate and the stellar mass are tightly correlated \citep{brinchmann04,noeske07}, and the red sequence, populated by quiescent galaxies whose star formation has been quenched. Compared to star-forming galaxies of the same mass, quiescent galaxies have larger bulge-to-total mass ratios \citep{bamford09,bluck14}, higher mass surface densities within 1\,kpc \citep{cheung12}, smaller sizes \citep{shen03} and more concentrated light profiles \citep{Kauffmann03,baldry06,driver06}. 
 
 Observations of the high-redshift universe (z\,$\gtrsim$\,1) revealed that the color bimodality of the galaxy population already existed at early epochs: the red sequence is in place by $z$\,$\sim$\,3 and quiescent galaxies become dominant in the high mass end of the galaxy mass function by $z$\,$\sim$\,2 \citep{muzzin13,ilbert13}. At these epochs, quiescent galaxies were already smaller and denser than their star-forming counterparts \citep{szomoru12,newman12,bell12,vanderwel14,whitaker17}. High redshift quiescent galaxies, however, are structurally unlike the local quiescent galaxy population. They are extremely compact \citep{daddi05,trujillo06,buitrago08,vandokkum08,damjanov09}, having half-light radii smaller that their local counterparts by a factor of $\approx$\,2.5\,--\,3 \citep{mclure13,vandesande13,vanderwel14}. In addition, these objects have significant rotational support \citep{belli17,bezanson18,newman18b}, and the majority have prominent stellar disks \citep{bundy10,bruce12,chang13,huertas-company16,davari17,hill19}. This implies that the quiescent galaxy population experienced a significant structural evolution in the last 10 billions years.
 
The size evolution of quiescent galaxies, in particular, has been the subject of many studies. Theoretical studies and cosmological simulations suggest that the growth of massive quiescent galaxies is driven by repeated minor mergers which build an envelope around a compact quiescent galaxy \citep{naab09,hopkins10,oser10,hilz12,hilz13,furlong15,wellons16}. The increase of the fraction of stellar mass stored in the haloes of massive early-type galaxies with decreasing redshift \citep{vandokkum10,buitrago17,hill17a}, the identification of multiple stellar components in the light profiles of massive elliptical galaxies \citep{huang13a,huang13b,oh17,huang18},  the flattening of age and metallicity gradients at large radii \citep{oyarzun19,zibetti20} and the presence of metal-poor and $\alpha$-enhanced stars in their outskirts \citep{greene13} have been interpreted as evidence in favor of the accretion of small satellites being the driving process behind the size evolution.

As star-forming galaxies also evolve in size, although moderately \citep{vanderwel14}, it is possible that the size evolution of the quiescent population might be, at least partially, due to a progenitor bias effect: the mean size of quiescent galaxies increases as larger star-forming galaxies become quiescent at later epochs \citep{vandokkum01,carollo13}. If progenitor bias is an important factor in the size evolution of the quiescent galaxy population, then, at fixed stellar mass, larger quiescent galaxies should be younger. A dependence of age on size has indeed been observed for galaxies with $\log M_\star$/M$_\odot$\,$\lesssim$\,11.0, but at higher masses size and age are not related \citep{mcdermid15,fagioli16}. Thus, at the high mass end, dry mergers appear to be the main driver of evolution \citep{faisst17}, while at lower masses the structural evolution of quiescent galaxies is more complex, with both progenitor bias and dry mergers playing role \citep{Damjanov19}. 

A similar conclusion can be reached from the evolution of the number density of compact quiescent galaxies. It has been found that their number density is approximately constant since $z$\,$\sim$\,1.5 \citep{carollo13,damjanov15,Charbonnier.etal:2017}, as expected if the size evolution is driven by progenitor bias. However, the number density of the most extreme compact galaxies (those with $r_{\rm e}$\,$<$\,1.5\,$\times$\,($M_\star$/10$^{11}$\,M\,$_\odot$)$^{0.75}$ and $\log M_\star$/M$_\odot$\,$>$\,10.5) decreases by a factor of $\sim$\,20 in this redshift interval \citep{Charbonnier.etal:2017} and compact galaxies continue to form at intermediate and low redshifts \citep{ferremateu12,damjanov13}, so the size growth of individual objects must also play a significant role.

Despite significant progress in the characterization of the size evolution of quiescent galaxies, some aspects of the structural evolution of these objects still are poorly constrained. For example, while the most massive ($\log M_\star$/M$_\odot$\,$\gtrsim$\,11.0) compact quiescent galaxies are thought to be the progenitors of slow-rotating giant ellipticals \citep{bezanson09,patel13,newman18b}, the z\,$\sim$\,0 descendants of compact quiescent galaxies with masses in the range $\log M_\star$/M$_\odot$\,$\sim$\,10.5\,--\,10.8 \citep{dominguezsanchez16} are yet to be identified. It has been suggested that, besides becoming the cores of lower mass ellipticals, lower mass compact galaxies might become compact bulges in early-type spirals and S0s \citep{graham15,delaRosa16,gao20}, as rejuvenation events regrow a large disk \citep{mancini19}. However, as of this moment, only tentative evidence links these populations. While old, compact and massive bulges in the center of local galaxies have been identified, it is not known if the stellar population properties and structure of these bulges are consistent with an origin as a high redshift compact galaxy. This is because little is known about the stellar population properties of high redshift compact quiescent galaxies, as the faintness of the stellar continuum and their small angular size make spatially resolved observation of these objects extremely challenging. 

An alternative to the study of high-$z$ compact quiescent galaxies emerged with the discovery of a population of local analogues, the so called relic galaxies: ultra-compact and massive ($r_{\rm e}$\,$\sim$\,1--2\,kpc, $\log M_\star$/M$_\odot$\,$\sim$\,11.0) quiescent galaxies which quenched at $z$\,$\sim$\,2 and have had a quiet accretion history since then, thus retaining their structure and stellar population properties since formation \citep{Trujillo.etal:14}. So far, the study of compact quiescent galaxies at z\,$\sim$\,0 have been focused on the search for these elusive relic galaxies, and as a result, only old massive ultra-compact galaxies have been studied in detail \citep{FerreMateu.etal:2017,Yildirim.etal:2017,scognamiglio20,spiniello21}, yet the population of massive ultra-compact galaxies is composed of a mix of old and younger objects \citep{Buitrago.etal:2018}. How these younger galaxies formed is not clear, but a connection to compact post-starburst galaxies has been suggested \citep{zahid16}.

Our goal is to build up on the aforementioned works by extending the study of z\,$\sim$\,0 compact galaxies to lower masses ($10\,\lesssim\,\log M_\star$/M$_\odot\,\lesssim\,11$) 
and slightly larger sizes ($r_{\rm e}$\,$\sim$\,1--3\,kpc). By a combined analysis of the kinematics, stellar populations, structural properties and environment, we aim to constrain the formation and accretion history of compact galaxies, assessing if they can be considered analogues of intermediate and high-redshift quiescent galaxies and probing their relation to z\,$\sim$\,0 non-compact quiescent galaxies. We also intend to investigate how relic galaxies are linked to the general compact population: are they just the high-mass and small-size tail of the compact population, or are they another class of objects, despite the similarities?

This paper is structured as follows: in Section\,\ref{sec:data} we describe the sample selection and methodology and in Section\,\ref{sec:results} we present our results. In Section\,\ref{sec:discussion} we discuss the formation and assembly history of compact galaxies and their relation to local and higher redshift quiescent galaxies. In Section\,\ref{sec:conclusions} we summarize our results and present our conclusions. In this paper we adopt a standard simplified $\Lambda$CDM cosmology  with  $\Omega_{\rm M}$\,=\,0.3, $\Omega_\Lambda$\,=\,0.7 and $H_0$\,=\,70\,km\,s$^{-1}$\,Mpc$^{-1}$.  

\section{Data and Sample Selection}
\label{sec:data}

\subsection{The MaNGA survey}

In this work we make use of data from the Mapping Nearby Galaxies at Apache Point Observatory (MaNGA), an integral-field spectroscopic survey which observed $\sim$\,10,000 galaxies from 2014 to 2020 \citep{smee13,bundy15,drory15}. The MaNGA sample was designed to have a flat distribution in log\,M$_\star$ and a roughly uniform radial spatial coverage. The MaNGA sample is divided into three samples: the primary sample, the secondary sample and the color-enhanced sample. Galaxies in the primary sample are selected so that 80\% of the galaxies can be covered by the integral field unit (IFU) to a major axis radius of 1.5 effective radii, while galaxies in secondary sample are selected so that 80\% can be covered to a major axis radius of 2.5 effective radii. The  Color-Enhanced sample was designed to add galaxies in under-represented regions of the NUV$ - i$ versus $M_i$ color–magnitude plane \citep{bundy15,wake17}. We note that the effective radii adopted by MaNGA were extracted from the NASA-Sloan Atlas, and they were measured based on elliptical Petrosian photometry. These do not necessarily agree with the effective radii we adopt in this work (which were extracted from the \citealt{simard11} catalogue). The data analysed in this work was released as part of data release 15 from the Sloan Digital Sky Survey (SDSS), which includes datacubes for a total of 4621 galaxies \citep{aguado19}. The datacubes have a spaxel size of 0.5\arcsec/spaxel, and the wavelength coverage of the observations is 3600--10300\,\r{A} at R\,$\sim$\,2700.

\subsection{Sample Selection}

\subsubsection{Compact galaxies}

We selected massive compact quiescent galaxies (MCGs) based on their effective radii ($r_{\rm e}$) at fixed effective stellar velocity dispersion ($\sigma_{\rm e}$, the velocity dispersion measured within an aperture of radii $r_{\rm e}$). We considered a galaxy as quiescent if its specific star formation rate (sSFR) is below 10$^{-11}$\,Gyr$^{-1}$. Both sSFR and stellar masses were extracted from the GALEX-SDSS-WISE Legacy Catalog \citep{salim18}. We adopted as $r_{\rm e}$ the semi-major axis of the half-light ellipse obtained from S\'ersic$+$Exponential fits to the 2D surface brightness profiles of SDSS-DR7 $r$-band images performed by \citet{simard11}. To obtain $\sigma_{\rm e}$, we converted the stellar velocity dispersions ($\sigma_{\rm ap}$) provided by the SDSS collaboration \citep{Abazajian.etal:2009}, which are measured through an aperture with a fixed diameter of 3\arcsec, using $\sigma_{\rm ap}$\,=\,$\sigma_{\rm e}$[$r_{\rm ap}$/$r_{\rm e}$]$^{-0.066}$ \citep{Cappellari.etal:2006}. The interval 150\,$<$\,$\sigma_{\rm e}$\,$<$\,350\,km\,s$^{-1}$ was divided into eight bins of 25\,km\,s$^{-1}$ width and for each bin we computed the $r_{\rm e}$ corresponding to the 10th percentile ($r_{10{\rm th}}$) of the quiescent galaxy size distribution. Compact galaxies were defined as those satisfying $r_{\rm e}$\,$<$\,$r_{10{\rm th}}$ for a given $\sigma_{\rm e}$ bin. There are 15024 galaxies in SDSS DR7 satisfying these criteria, of which 70 have integral field spectroscopy data publicly available as part of MaNGA-DR15. 

\subsubsection{Control sample} \label{sec:control}

Considering that we aim to explore differences in the stellar population properties of MCGs and median-sized quiescent galaxies, we chose to build a control sample of galaxies with velocity dispersions comparable to those of MCGs, as several studies showed that velocity dispersion is a better predictor of the past star formation history (SFH) than either stellar or dynamical mass \citep{thomas05,graves09a,graves09b,vanderwel09,shankar09,mcdermid15}. The control sample galaxies (CSGs) were selected from a sample of quiescent galaxies with $r_{\rm e}$ within the 40th and 60th percentiles of the $r_{\rm e}$ the distribution.

We defined the control sample by applying the Propensity Score Matching (PSM) technique (\citealp{PSM:1983}; see also \citealp{deSouza.etal:2016} for more details on the PSM method). 
We have used the  {\sc MatchIt} package \citep{MatchIt:2011}, written in 
R\footnote{\url{https://cran.r-project.org/}} \citep{R:2015}.  This
technique allows us to select from the sample of average sized quiescent galaxies in MaNGA
a control sample of 70 galaxies in which the distribution of the effective velocity dispersion is as close as possible to that of the MCGs.  
We have adopted the Mahalanobis distance approach \citep{Mahalanobis36} and the nearest-neighbour method to perform the matching.
We note that the PSM technique is not necessary when only one variable
is used; other simpler matching methods would work as well as PSM.
However, PSM allows us to test how the results change when taking other galaxy properties into account for the matching procedure and test the robustness of our results using other control samples. 

\begin{figure}
\centering
    \includegraphics[width=0.95\hsize]{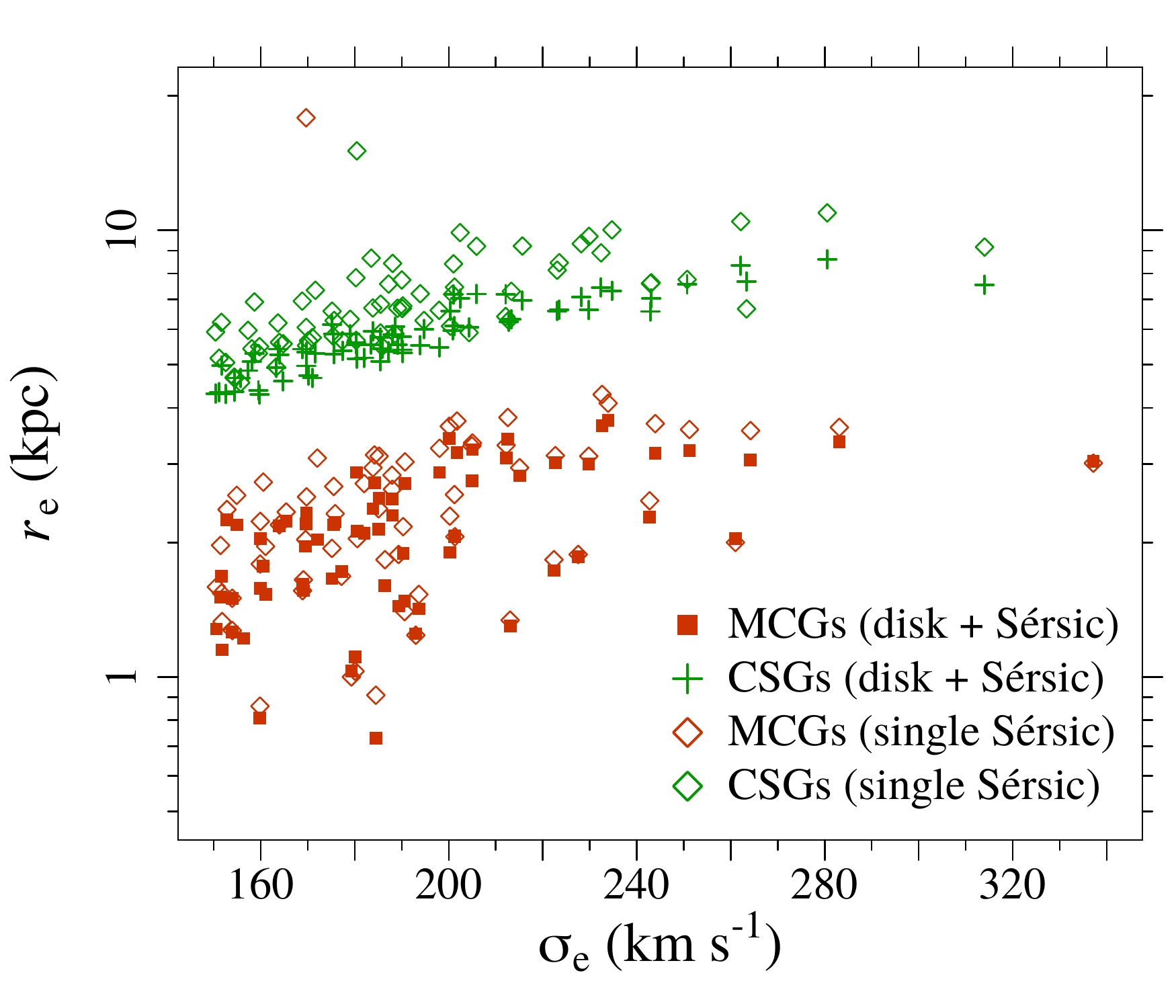}
    \caption{Effective radius as a function of velocity dispersion for compact (\emph{red symbols}) and control (\emph{green symbols}) galaxies. The \emph{squares} and \emph{crosses} indicate the radii obtained from a disk ($n = 1$) and bulge ($n =\,$free) decomposition. Radii measured by fitting a single Sérsic profile are shown as \emph{open diamonds}.}
    \label{fig:re_sersic}
\end{figure}

In Fig.\,\ref{fig:re_sersic} we show the effective radius as a function of the velocity dispersion, comparing the effective radii measured by fitting a single Sérsic profile (extracted from the \citealt{simard11} catalogue) with effective radii obtained from a disk ($n = 1$) and bulge ($n =\,$free) decomposition. The differences between the effective radii obtained with these two methods are smaller in MCGs than in CSGs. Two objects are outliers, having effective radii larger than 10\,kpc. This is due to bad fits, as their effective radii extracted from the MaNGA PyMorph photometric catalog are consistent with the trends followed by their respective samples for both single Sérsic and disk+bulge fits.

\begin{figure}
\centering
    \includegraphics[width=0.95\hsize]{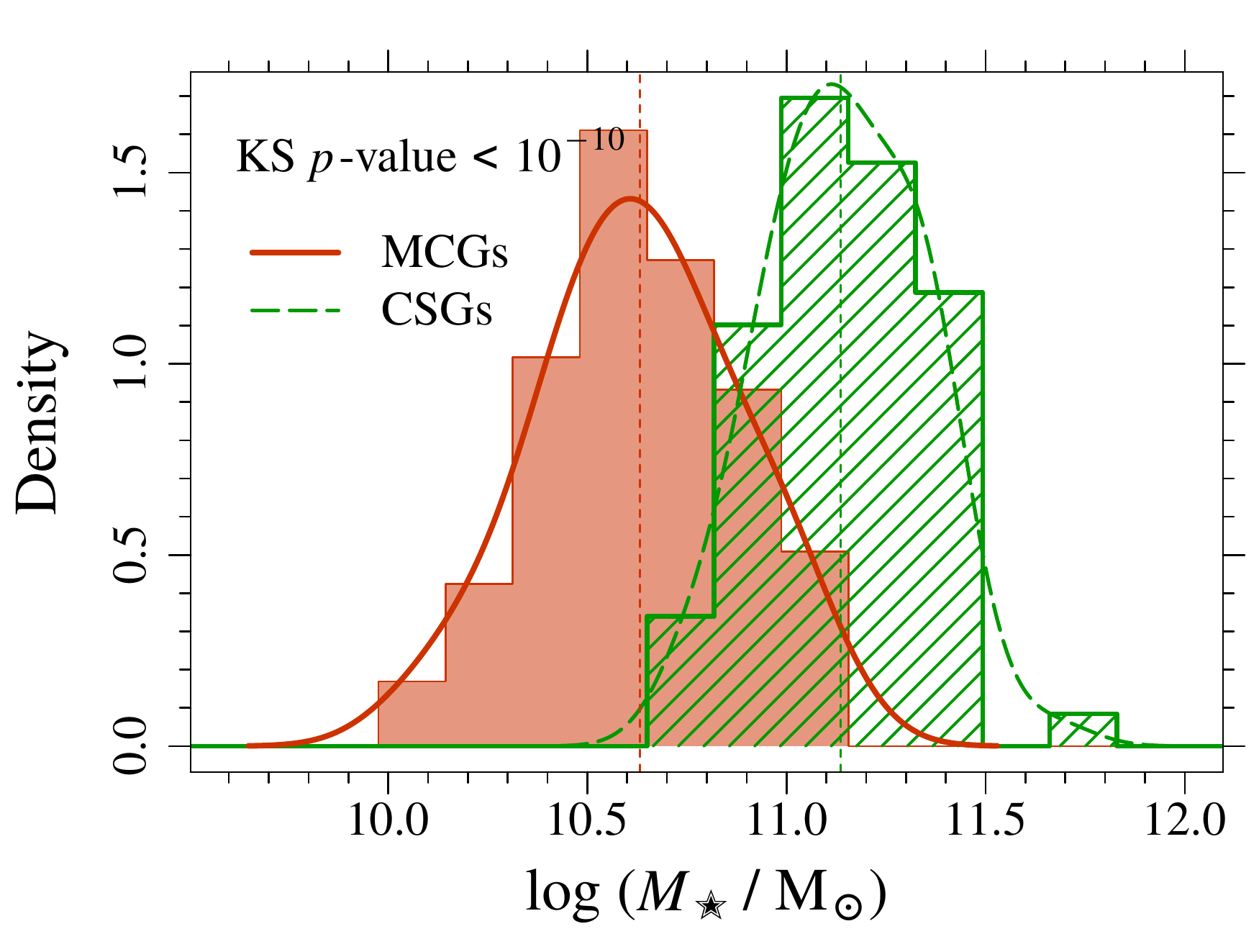}
    \caption{Distributions of stellar masses of MCGs (\emph{solid red histogram}) and CSGs (\emph{hashed green histogram}). The $p$-value of a KS test comparing the two distribution is indicated in the plot. The \emph{vertical dashed lines} indicate the median values. The curves are obtained by smoothing the positions of the data points (not the histograms) using a Gaussian kernel with the standard deviation equal to half of the standard deviation of the data points.}
    \label{fig:mass}
\end{figure}

In Fig.\,\ref{fig:mass} we show the mass distributions of MCGs and CSGs. We note that the significant difference between the mass distribution of the two samples is expected, as stellar mass is correlated with both size and central velocity dispersion \citep{shen03,bernardi11}. The implications of the difference in mass distribution of both samples are discussed in Section\,\ref{sec:results:M-CSGs}.   

\begin{figure}
\centering
    \includegraphics[width=0.95\hsize]{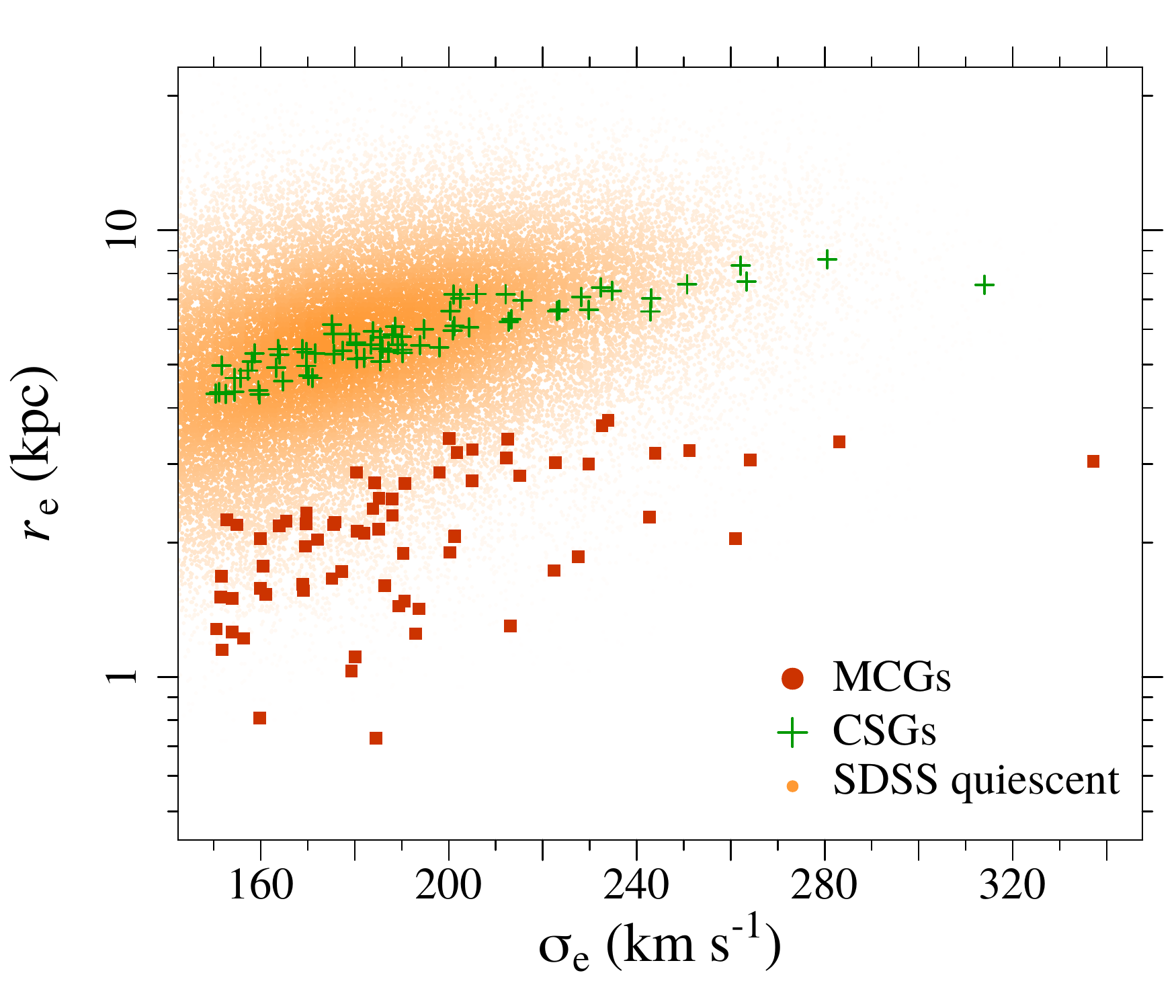}
    \caption{Distribution of the compact (\emph{red squares}) and control (\emph{green crosses}) samples on the $\sigma_{\rm e}$ vs. $r_{\rm e}$ plane. Quiescent SDSS galaxies are indicated as \emph{orange symbols} (1 out of each 5 SDSS galaxies are drawn). }
    \label{fig:sig_re}
\end{figure}

\begin{figure*}
\centering
    \includegraphics[width=\hsize]{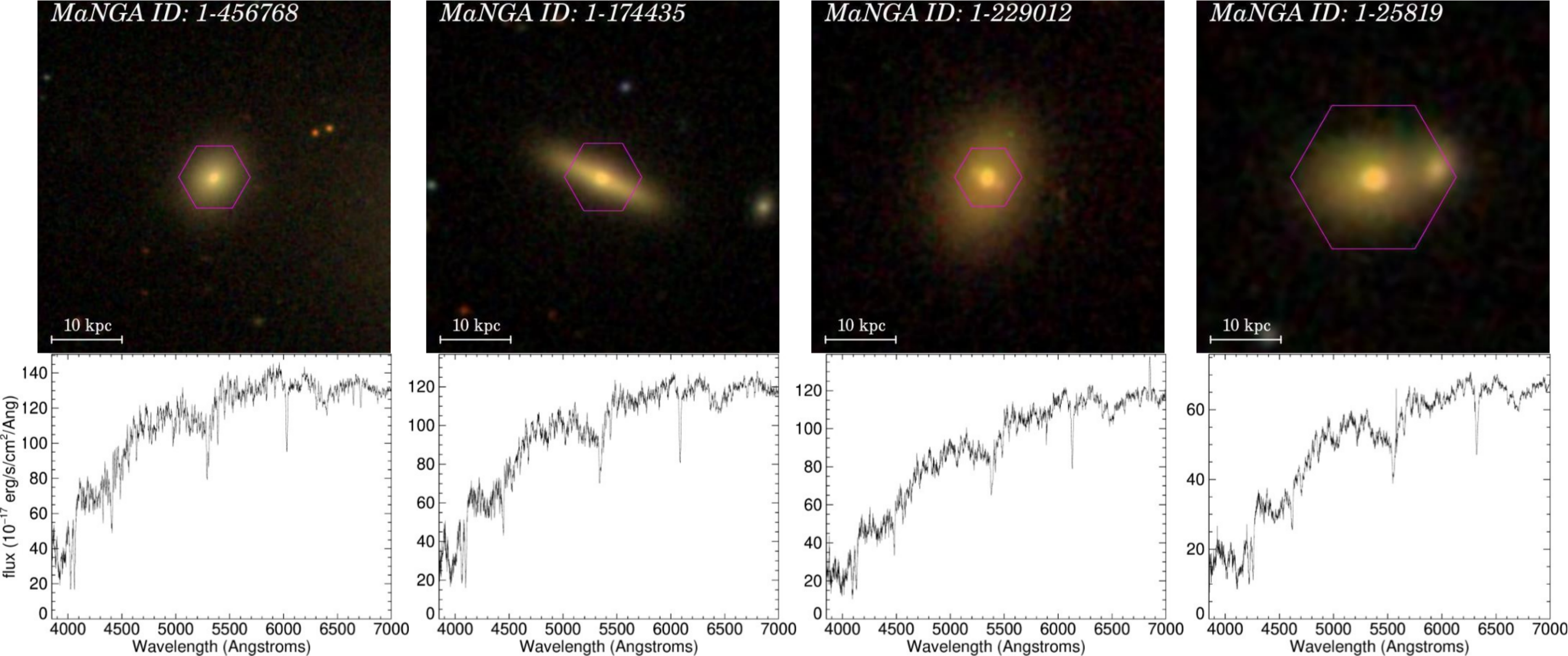}
    \caption{SDSS images and spectra of 4 MCGs: MaNGA ID\,=\,1-456768 ($r_{\rm e}$\,=\,1.6\,kpc, $\sigma_{\rm e}$\,=\,169\,km\,s$^{-1}$), 1-174435 ($r_{\rm e}$\,=\,3.0\,kpc, $\sigma_{\rm e}$\,=\,223\,km\,s$^{-1}$), 1-229012 ($r_{\rm e}$\,=\,3.1\,kpc, $\sigma_{\rm e}$\,=\,264\,km\,s$^{-1}$) and 1-25819 ($r_{\rm e}$\,=\,3.0\,kpc, $\sigma_{\rm e}$\,=\,337\,km\,s$^{-1}$). All images are 50 x 50 kpc and the MaNGA field of view is shown in magenta.}
    \label{fig:compact}
\end{figure*}

\begin{figure*}
\centering
    \includegraphics[width=\hsize]{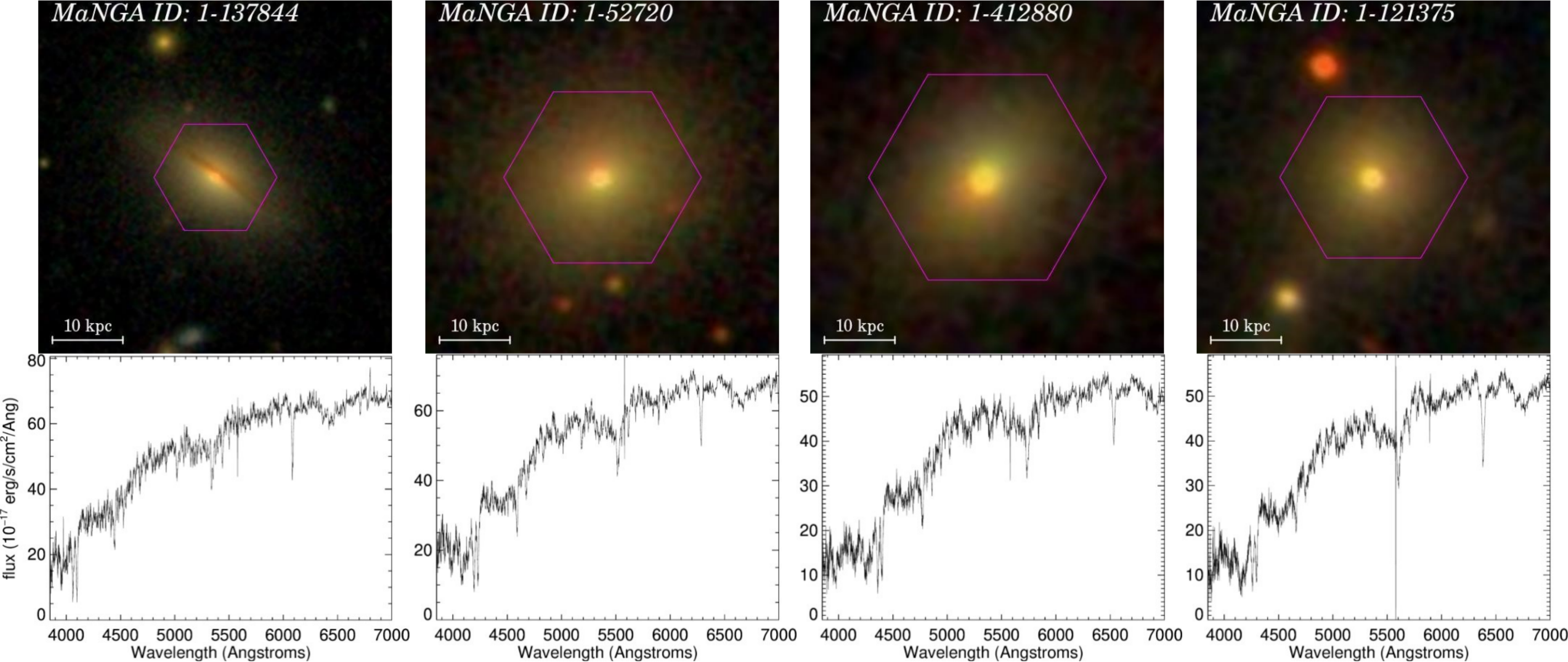}
    \caption{SDSS images and spectra of 4  CSGs: 1-137844 ($r_{\rm e}$\,=\,5.3\,kpc, $\sigma_{\rm e}$\,=\,164\,km\,s$^{-1}$), 1-52720 ($r_{\rm e}$\,=\,7.1\,kpc, $\sigma_{\rm e}$\,=\,228\,km\,s$^{-1}$), 1-412880 ($r_{\rm e}$\,=\,7.7\,kpc, $\sigma_{\rm e}$\,=\,263\,km\,s$^{-1}$) and 1-121375 ($r_{\rm e}$\,=\,7.6\,kpc, $\sigma_{\rm e}$\,=\,314\,km\,s$^{-1}$). All images are 50 x 50 kpc and the MaNGA field of view is shown in magenta.}
    \label{fig:control}
\end{figure*}

 In Fig.\,\ref{fig:sig_re} we show the final compact and control samples in the $r_{\rm e}$ vs. $\sigma_{\rm e}$ plane, in Fig.\,\ref{fig:mass} we show their stellar mass distribution and in Figs.\,\ref{fig:compact} and \ref{fig:control} we show a SDSS color image and spectra of four compact and control sample galaxies respectively. 
 
 Properties of compact and control galaxies, as well as the parameters derived in this study, will be provided in machine-readable form as supplementary data available online. See Table\,\ref{tab:general} in appendix\,\ref{app:ap_table}  for an example of the data format.

\section{Methodogy}

\subsection{Stellar population properties}

The galaxy ages and metallicities were derived through stellar population synthesis (SPS) analysis using the \textsc{STARLIGHT} code \citep{CidFernandes.etal:2005} with \citet[][V15]{Vazdekis.etal:2015} models. We used V15 models computed  with the \cite{Kroupa01} IMF, and stellar evolution tracks from BaSTI 
(Bag of Stellar Tracks and Isochrones, \citealp{Pietrinferni+04,Pietrinferni+06}).
We ran \textsc{STARLIGHT} considering 15 bins of ages, ranging from 30\,Myr up to 13.5\,Gyr,
and 6 metallicity bins between [M/H]$\,=\,-1.3$ and $+0.4$. 
Before running the code, the observed spectra were corrected
for foreground extinction and de-redshifted, and the SSP models degraded to match the wavelength-dependent mean resolution of the MaNGA data cubes. 
The fit was performed in the wavelength region from 3900 to
5800\,\AA\ to avoid IMF-sensitive features\footnote{ We find that the MCG IMF-sensitive spectral features tend to be stronger than those of the CSGs, suggesting differences in the IMF (see \citealt{martin-navarro15} for a study of the IMF in a nearby massive compact galaxy). Since we do not take variations in the IMF into account, we avoid spectral reagions containing IMF-sensitive features in our SPS analysis. We will investigate the IMF of MCGs in a forthcoming paper. } and normalized at 4020\,\AA.
We adopted the \citet{Cardelli.etal:1989}
extinction law, assuming $R_{\rm V} = 3.1$. 

To estimate the [$\alpha$/Fe] ratios of the MCGs and CSGs, 
we adopted the approach described in \citet{LaBarbera.etal:2013} and \citet{Vazdekis.etal:2015},  which is based on the spectral indices
Mg$b$ and Fe3\footnote{Fe3 = (Fe4383 + Fe5270 + Fe5335)/3 \citep{Kuntschner:2000}.}. 
We measured the line strengths with an adapted Python version of the 
code \textsc{PACCE} \citep{Riffel.BorgesVale:2011, Riffel.etal:2019},
and applied corrections for the broadening
due to the internal velocity dispersion of the galaxy following the
prescriptions of \citet{delaRosa.etal:2007}.
 
\begin{figure*}
\centering
\begin{tabular}{cc}
    \includegraphics[width=0.45\hsize]{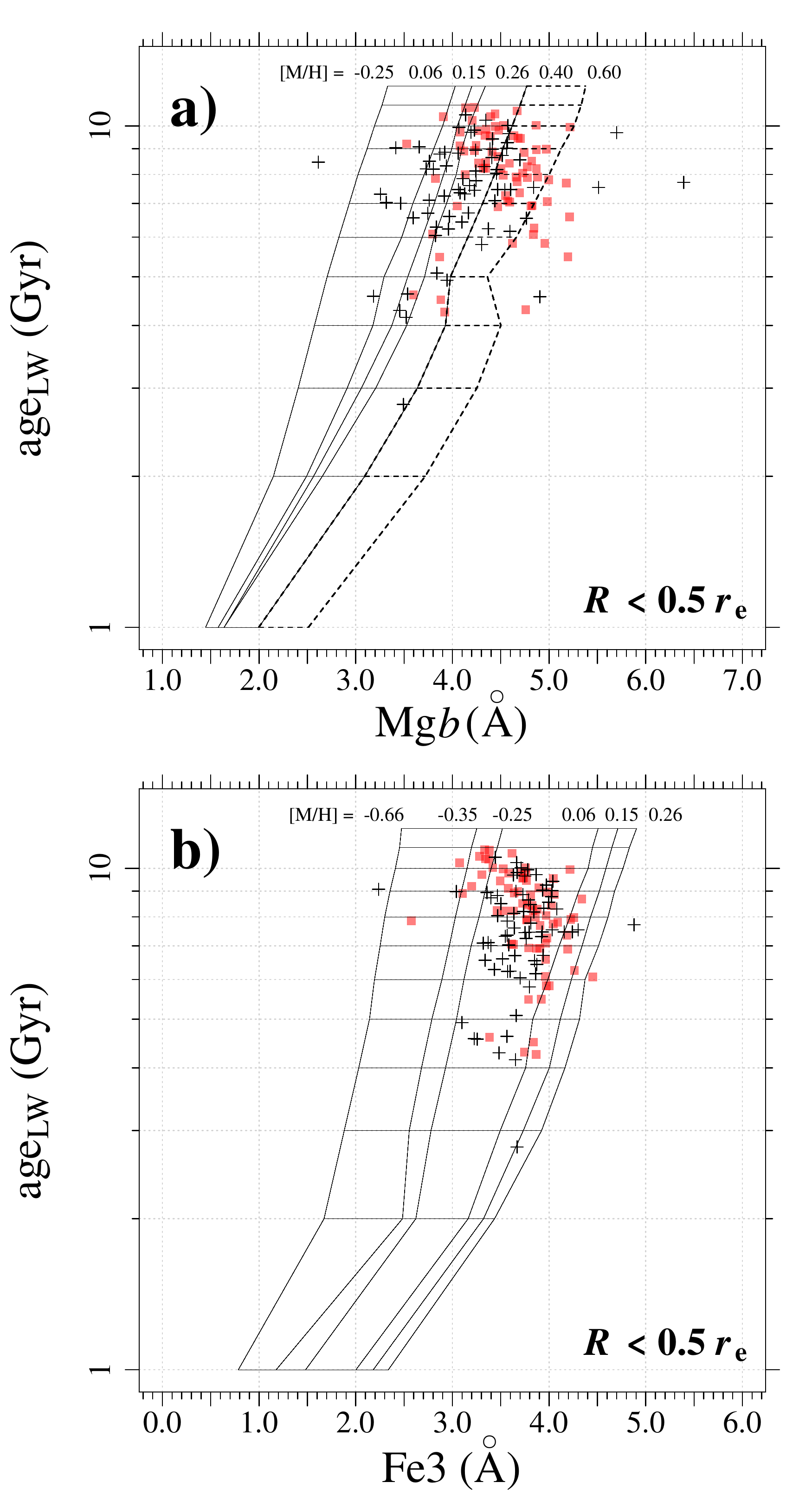} &
    \includegraphics[width=0.45\hsize]{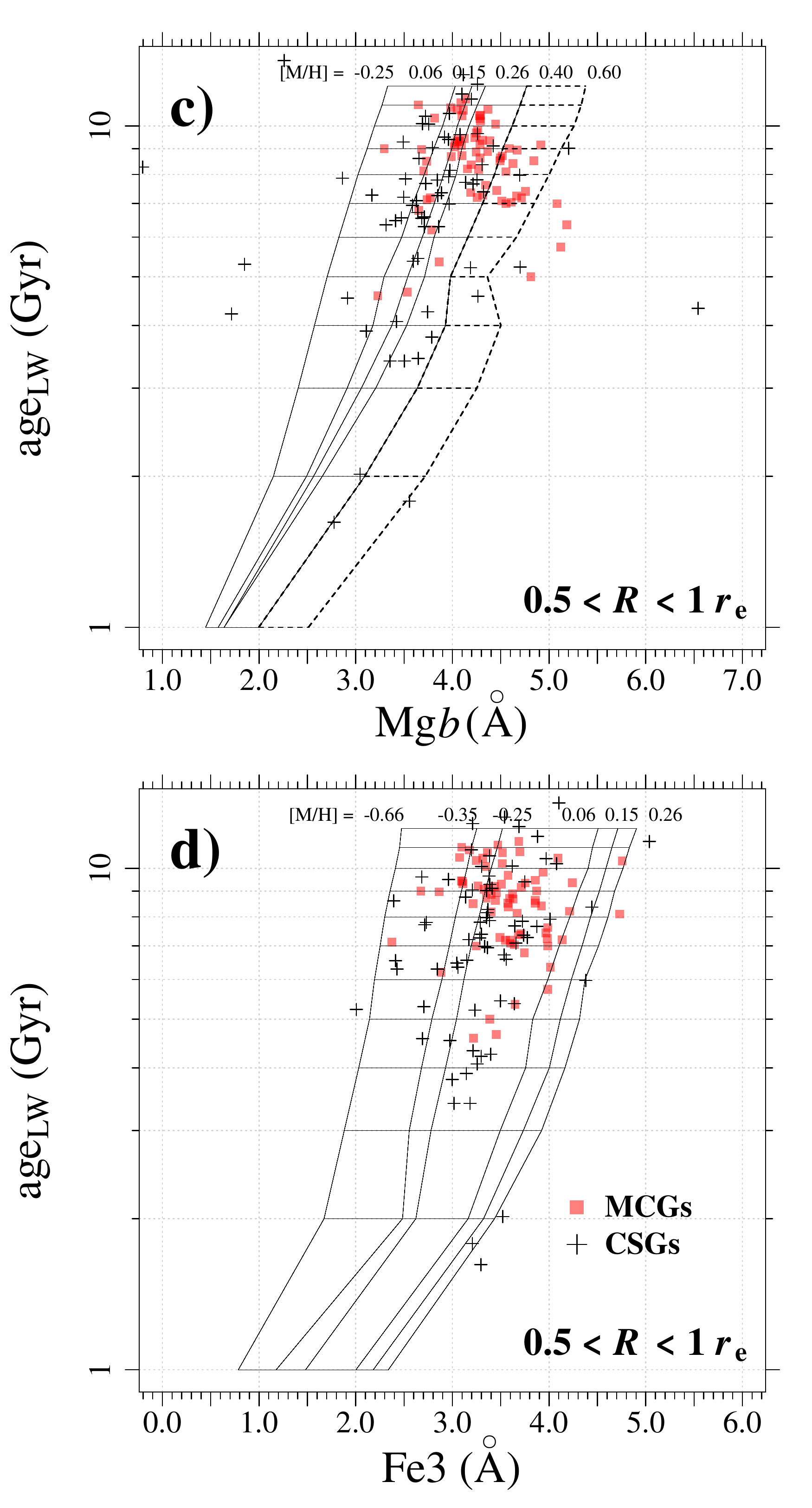} \\
\end{tabular}
\caption{The luminosity-weighted ages as a function of the spectral indices Mg$b$ (\emph{top panels}) and Fe3 (\emph{bottom panels}). The \emph{left} and \emph{right panels} show the ages and indices measured in the spectra integrated from $0$ to $0.5\,r_{\rm e}$ and from $0.5$ to $1\,r_{\rm e}$, respectively. The \citet{Vazdekis.etal:2015} model grids are indicated by the solid black lines in both panels. In the upper panels, the dashed part of the grids corresponds to our linear extrapolation of the model Mg$b$ to [Z/H]$\,=0.6$.}
\label{fig:mgb_fe3}
\end{figure*}

The procedure to determine the proxy of [$\alpha$/Fe] is illustrated in Fig.~\ref{fig:mgb_fe3}, where we show the galaxy luminosity-weighted ages (derived using {\sc STARLIGHT} with V15 models) as a function of Mg$b$ and ${\rm Fe}3$, as well as the predictions from the V15 models with different metallicities.
For each galaxy, we estimate two independent metallicities, $Z_{{\rm Mg}b}$ and $Z_{{\rm Fe}3}$, by fixing the galaxy age and interpolating the model grid. 
As discussed by \cite{LaBarbera.etal:2013}, estimating $Z_{{\rm Mg}b}$ of an $\alpha$-enhanced population may require extrapolation of the models to higher metallicities. 
This is illustrated in the upper panels of Fig.~\ref{fig:mgb_fe3}, where we show our linear extrapolation of the model Mg$b$ to metallicity $[{\rm Z}/{\rm H}] = +0.6$.
The proxy of [$\alpha$/Fe] is then defined as the difference between these two metallicities, $[Z_{{\rm Mg}b} / Z_{{\rm Fe}3}] = Z_{{\rm Mg}b} - Z_{{\rm Fe}3}$, and we use the following relation to obtain $[\alpha/{\rm Fe}]$ \citep{Trevisan.etal:2017}:
\begin{equation}
     [\alpha/{\rm Fe}] = -0.07 + 0.51\ [Z_{{\rm Mg}b} / Z_{{\rm Fe}3}]
\end{equation}

We derived the ages, metallicities and [$\alpha$/Fe] in the inner and outer parts of each galaxy by integrating the spectra within $0.5\, r_{\rm e}$ from the galaxy centre, and from $0.5\, r_{\rm e}$ to $1.0\, r_{\rm e}$. We use the differences between the inner and outer parts as a proxy for the stellar population gradients, i.e., $\Delta SP = SP_{\rm outer} - SP_{\rm inner}$, where $SP$ is the $\log($age), metallicity or [$\alpha$/Fe].

\subsection{Stellar kinematics}

\begin{figure}
\centering
    \includegraphics[width=0.95\hsize]{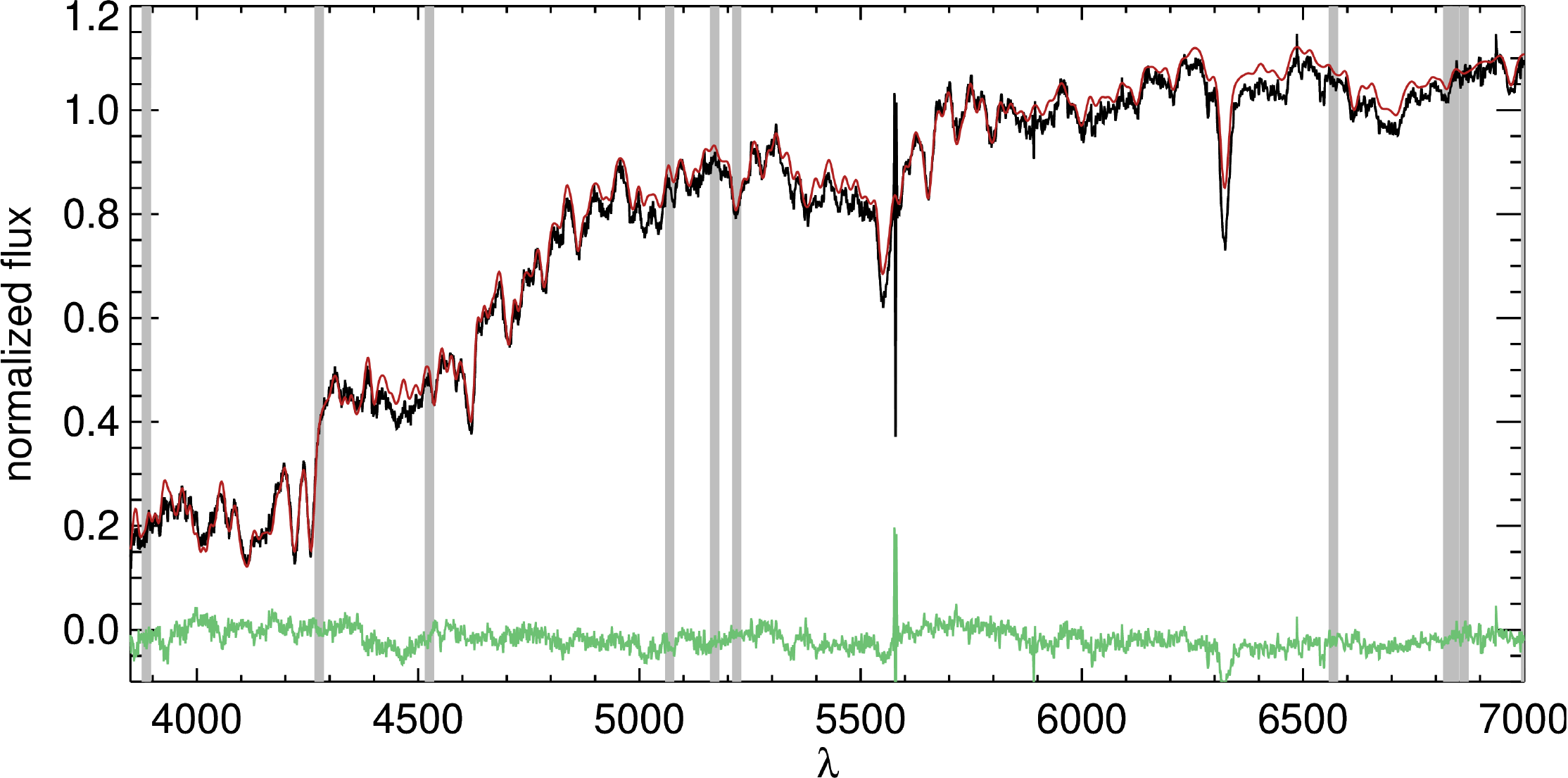} \\
    \includegraphics[width=0.95\hsize]{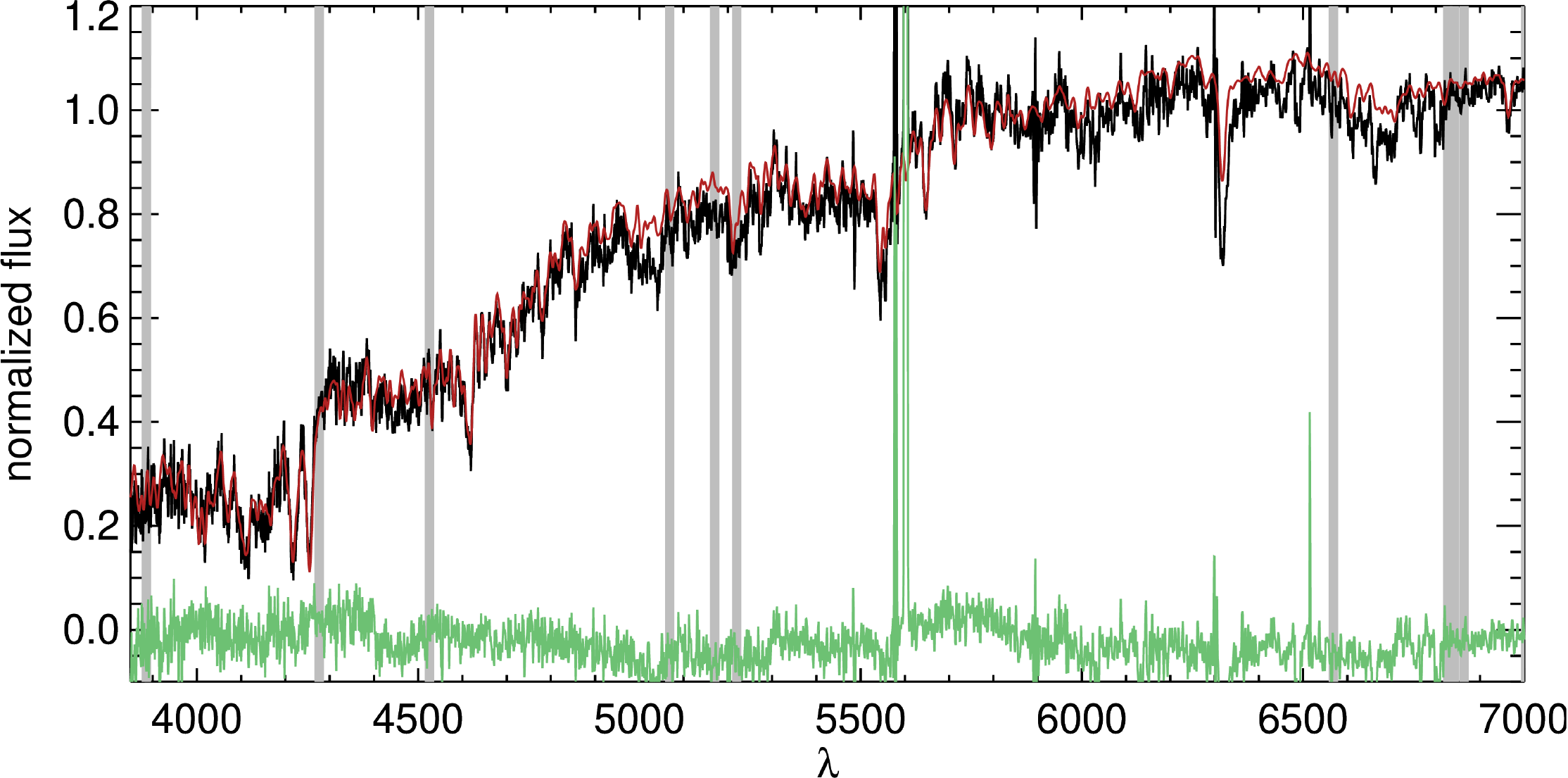} \\
\caption{Example of best fit models obtained with \textsc{pPFX} for one of the galaxies in our sample, MaNGA ID\,=\,1-25819. In the top panel we show a fit to the spectrum from the central spaxel, in the bottom panel we show a fit to a spectrum from a spaxel at $1.0\, r_{\rm e}$ from the nucleus. The observed spectra is shown in black, the best-fit model is shown in red and in green we show the residuals. Regions excluded from the fit are in grey.}
\label{fig:ppxf_best}
\end{figure}

To derive the stellar kinematics, we employed the penalized pixel fitting code (\textsc{pPXF}; \citealt{Cappellari.Emsellem:2004}). \textsc{pPXF} fits the observed stellar continuum with a convolution of a set of template spectra with a Gauss-Hermite series up to order $n$\,=\,6. In this work we fit up to n=4. The first and second moments correspond to the mean velocity and the velocity dispersion, the third moment $h_3$ is a measure of the skewness of the line of sight velocity distribution (LOSVD) and the fourth moment $h_4$ is a measure of the  kurtosis of the distribution. The spectral interval 3850\,--\,7000\r{A} was fitted using the \citet{Vazdekis.etal:2015} models as template spectra. An additive Legendre polynomial of degree 4 was used to correct the template continuum shape during the fit. To avoid contamination by emission, spectral regions of 20\,\r{A} width centered in H$\alpha$, H$\beta$, H$\delta$, H$\gamma$, [O\,II], [O\,III], [N\,II] and [S\,II] were masked. Only spaxels with signal-to-noise ratio (S/N) in the restframe wavelength interval 5600\,--\,5700\r{A} above 5 were fitted with \textsc{pPXF}. Example of best fit models are shown in Fig.\,\ref{fig:ppxf_best}.  

\subsection{Angular Momentum}

We quantify the angular momentum using the dimensionless parameter $\lambda_R$, which is a proxy for the baryon projected specific angular momentum. $\lambda_R$ is defined as \citep{emsellem07}: 
\begin{equation}
\lambda_{R}\,\equiv\,\frac{\langle R|V| \rangle}{\langle R\sqrt{V^2+\sigma^2} \rangle}
\end{equation}
Where V is the stellar velocity, $\sigma$ is the stellar velocity dispersion and R is the distance to the center. 
Throughout this work we use $\lambda_{\rm e}$, which is measured inside the half-light ellipse. 

We classify galaxies as fast rotators or slow rotators based on $\lambda_{\rm e}$ and ellipticity ($\varepsilon$). Slow rotators are defined as those systems that satisfy the relation \citep{cappellari16}:
\begin{equation}
\lambda_{\rm e}\,<\,0.08\,+\,\varepsilon/4\,\,\textrm{with}\,\,\varepsilon\,<\,0.4
\end{equation}

The ellipticities adopted in this work were extracted from the MaNGA PyMorph photometric catalog \citep{fischer19}, which provides photometric parameters from S\'ersic\,$+$\,Exponential fits to the 2D surface brightness profiles of the MaNGA DR2 galaxy sample obtained with the fitting algorithm \textsc{PyMorph} \citep{vikram10}. We note that the results presented in this paper are not particularly sensitive to our choice of ellipticity, as neither $\lambda_e$ nor the number of slow rotators vary significantly if we adopt ellipticities measured from the outer isophotes in the MaNGA datacubes.

\subsection{V/$\sigma$ and h$_3$}

Deviations of the line-of-sight velocity distribution from a Gaussian shape are measured fitting the velocity profile P($V$) with a Gaussian plus third and fourth-order Gauss-Hermite polynomials \citep{gerhard93,vanderMarel93}. The third and fourth-order amplitudes $h_3$ and $h_4$ are related to the skewness and the kurtosis of the velocity profile respectively. For $h_3$\,=\,0 and $h_4$\,=\,0, the velocity profile is a Gaussian.

In fast rotators, it has been observed that usually $h_3$ and the stellar velocity $V$ have opposite signs, and $h_3$ is anti-correlated to the ratio $V/\sigma$ \citep{krajnovic11}. To quantify the strength of the anti-correlation between $h_3$ and $V/\sigma$, we computed the Kendall correlation coefficient ($\tau$) for each object. To exclude  measurements of $h_3$ that are likely spurious, we computed $\tau$ considering only spaxels with S/N\,$>$\,10. We also computed $\tau$ with  S/N\,$>$\,20 and the results were nearly unchanged. Taking the spectral resolution of the data into account, we also exclude spaxels where the velocity dispersion is lower than 80\,km/s.

\subsection{Morphology}

For a morphological classification of the galaxies in our sample, we use the MaNGA Morphology Deep Learning catalog \citep{dominguez-sanchez18}. The classifications in this catalog are obtained with Deep Learning algorithms that use Convolutional Neural Networks trained with colour images from the Galaxy Zoo 2 \citep{willett13} and the visual classication catalogue of \citet{nair10} to provide T-types. For a better separation between pure ellipticals from S0s, the probability of a galaxy being an S0 ($P_{\rm S0}$) is also provided. We classify galaxies into three groups: S0s (T-type\,$\leq$\,0 and $P_{\rm S0}$\,$>$\,50\%), ellipticals (T-type\,$\leq$\,0 and $P_{\rm S0}$\,$\leq$\,50\%) and spirals (T-type\,$>$\,0). Bulge-to-total light ratios (B/T) were extracted from the MaNGA PyMorph photometric catalog \citep{fischer19}.

\subsection{Environment}

To investigate the environment of the MCGs, we used an updated version of the catalogue of groups and clusters by \citet{Yang.etal:2007}. The catalogue contains $473\,482$ groups drawn from a sample of $601\,751$ galaxies, mostly from the SDSS-DR7 \citep{Abazajian.etal:2009}, and provides the halo masses $M_{200,{\rm m}}$ (i.e., the mass within a sphere that is $\Delta$\,=\,200 times denser than the \emph{mean} density of the Universe), which are based on abundance matching with the group luminosities. We converted the $M_{200,{\rm m}}$ masses given in the \citeauthor{Yang.etal:2007} to $M_{\rm halo}$\,=\,$M_{100}$ (i.e., the mass within a sphere that is $\Delta$\,=\,100 times denser than the \emph{critical} density of the Universe) by assuming the \citet{Navarro.etal:1996} profile and the concentration-mass relation given by \cite{Dutton&Maccio14}. The conversion from quantities relative to the mean density to those relative to the critical density is described in appendix~A in \citet*{Trevisan.etal:2017}.

In order to quantify the influence of the local environment, we use the tidal strength parameter $Q_{\rm group}$, extracted from the Galaxy Environment for MaNGA Value Added Catalog (Argudo-Fern\'andez et al. in prep.). $Q_{\rm group}$ is an estimation of the total gravitational interaction strength produced on a given galaxy by the members of its group with respect to its internal binding forces. The tidal strength on a primary galaxy $P$ created by its neighbours in a group of $n$ galaxies is \citep{argudo-fernandez15}:
\begin{equation}
Q_{\rm group}\,\equiv\,\log\left[\sum_{i=1}^{n-1} \frac{M_i}{M_P} \left( \frac{D_P}{d_i}\right)^3 \right],
\end{equation}
\noindent
where $M_P$ and $M_i$ are the stellar masses of the primary galaxy and of the $i$th neighbour respectively; $D_P$ is the diameter of the primary galaxy and $d_i$ is the projected distance of the $i$th  neighbour to the primary galaxy. $D_P$ is defined as $D_P = 2\alpha R_{90}$, where $R_{90}$ is the radius containing 90$\%$ of the Petrosian flux and $\alpha$ is a correction factor. This correction factor is applied in order to recover the projected major axis of a galaxy at the 25 mag/arcsec$^2$ isophotal level  \citep{argudo-Fernandez13}. We note that $D_P$ does not depend on model fits. A galaxy is considered isolated if $Q_{\rm group}$\,$\le$\,$-2$, i.e., the tidal strength is at most 1\% of the internal binding force \citep{athanassoula84,verley07}.

\section{Results}\label{sec:results} 

\subsection{Stellar Kinematics}

\begin{figure*}
\centering
    \includegraphics[width=0.95\hsize]{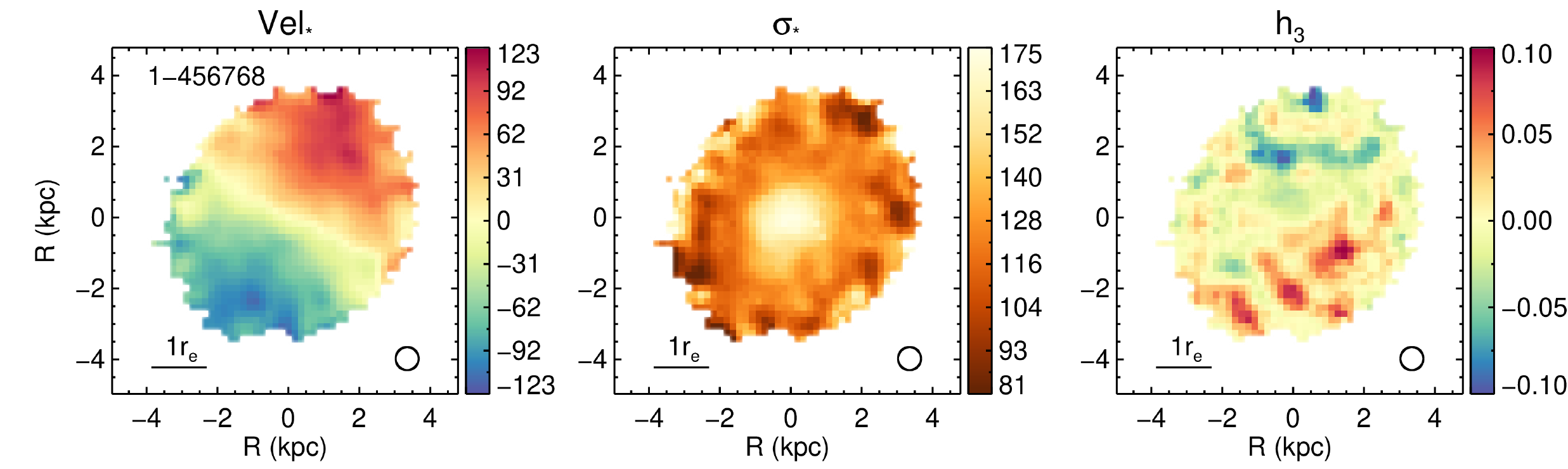}
    \includegraphics[width=0.95\hsize]{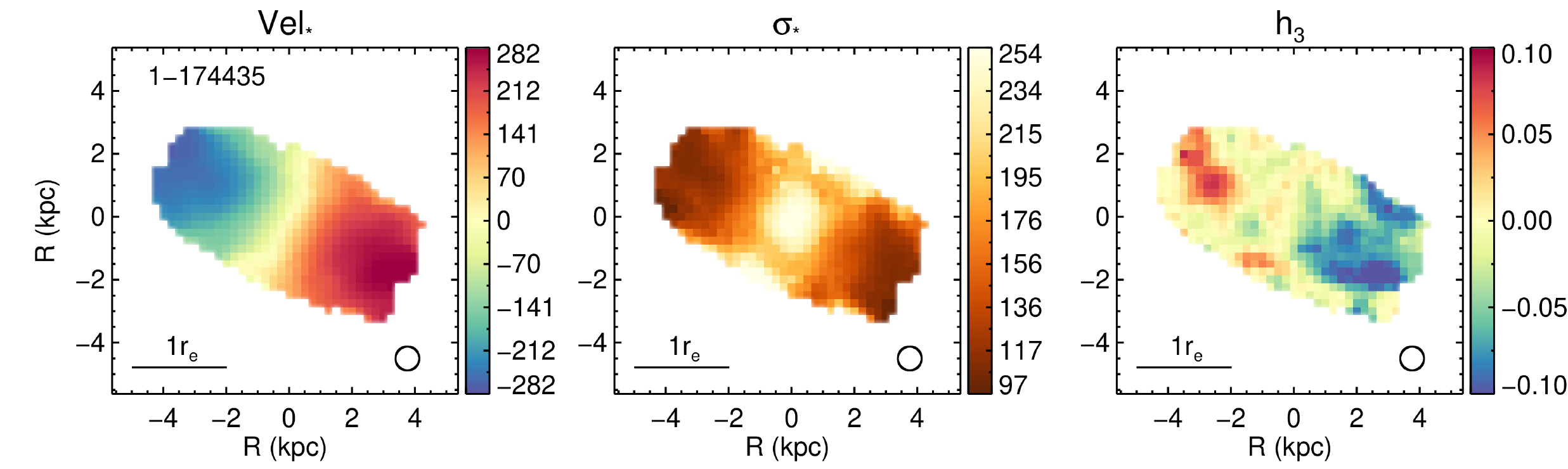}
    \includegraphics[width=0.95\hsize]{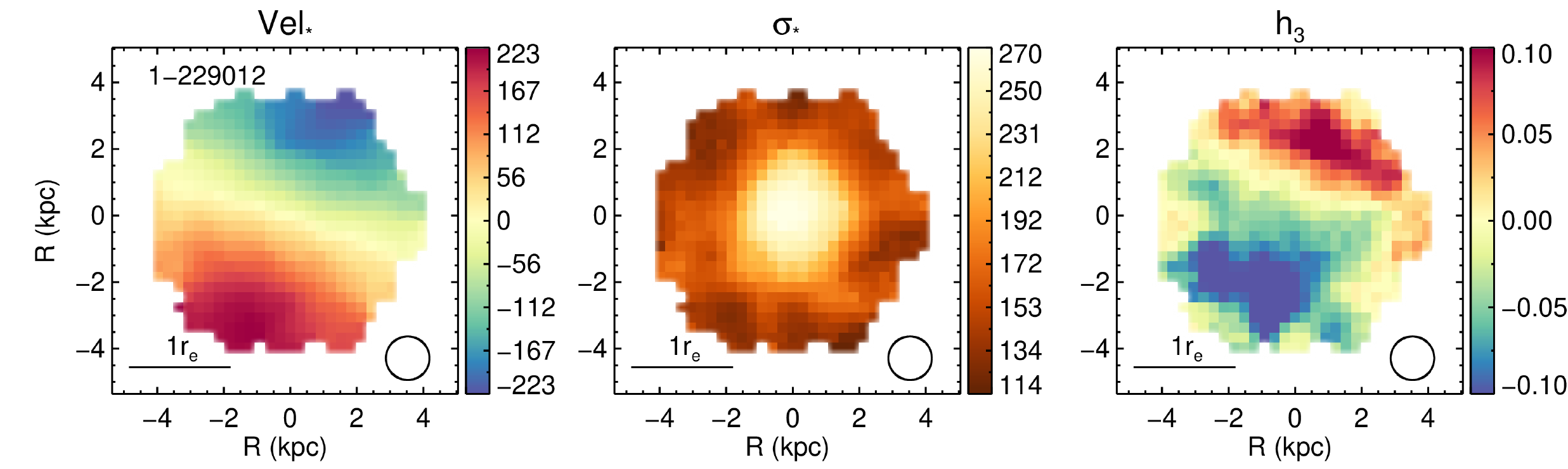}
    \includegraphics[width=0.95\hsize]{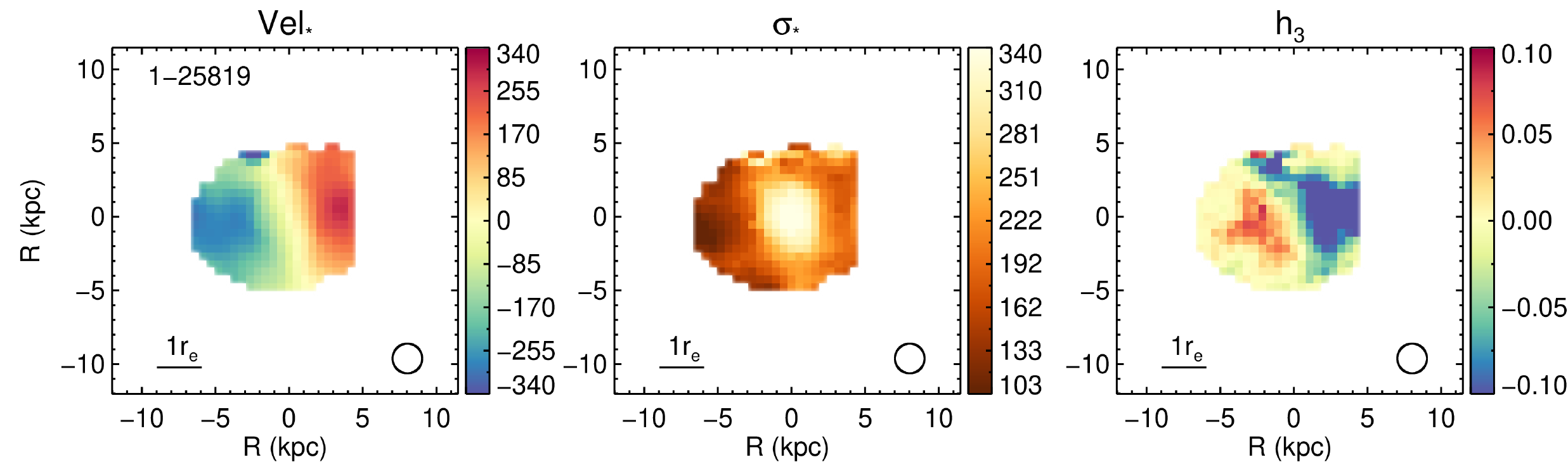}
    \caption{Stellar velocity (km\,s$^{-1}$), velocity dispersion (km\,s$^{-1}$) and Gauss-Hermite moment $h_3$ maps of four representative MCGs (same galaxies displayed in Fig.\,\ref{fig:compact}). From top to bottom: MaNGA ID\,=\,1-456768 ($r_{\rm e}$\,=\,1.6\,kpc, $\sigma_{\rm e}$\,=\,169\,km\,s$^{-1}$), 1-174435 ($r_{\rm e}$\,=\,3.0\,kpc, $\sigma_{\rm e}$\,=\,223\,km\,s$^{-1}$), 1-229012 ($r_{\rm e}$\,=\,3.1\,kpc, $\sigma_{\rm e}$\,=\,264\,km\,s$^{-1}$) and 1-25819 ($r_{\rm e}$\,=\,3.0\,kpc, $\sigma_{\rm e}$\,=\,337\,km\,s$^{-1}$). The black circle shows the angular resolution of the observations.}
    \label{fig:kinematics_compact}
\end{figure*}

\begin{figure*}
\centering
    \includegraphics[width=0.95\hsize]{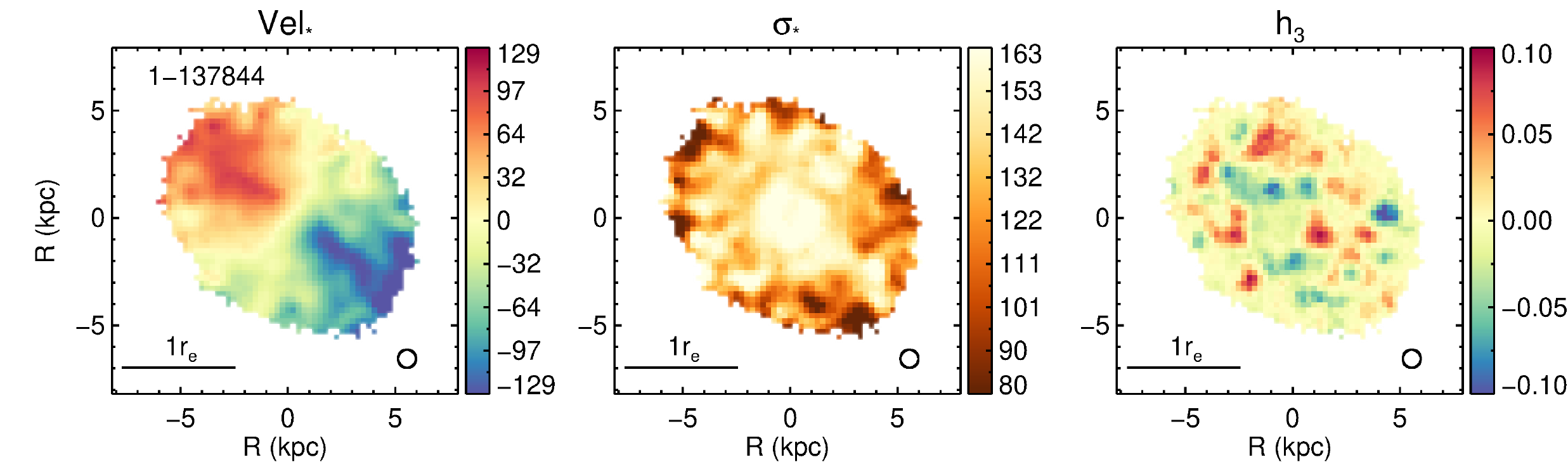}
    \includegraphics[width=0.95\hsize]{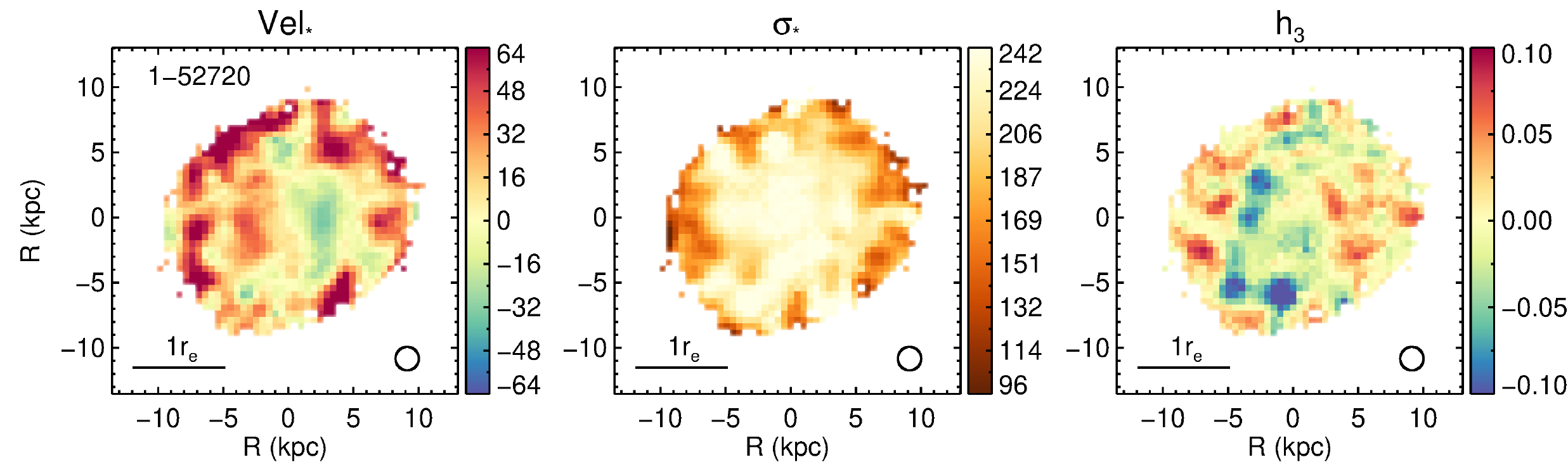}
    \includegraphics[width=0.95\hsize]{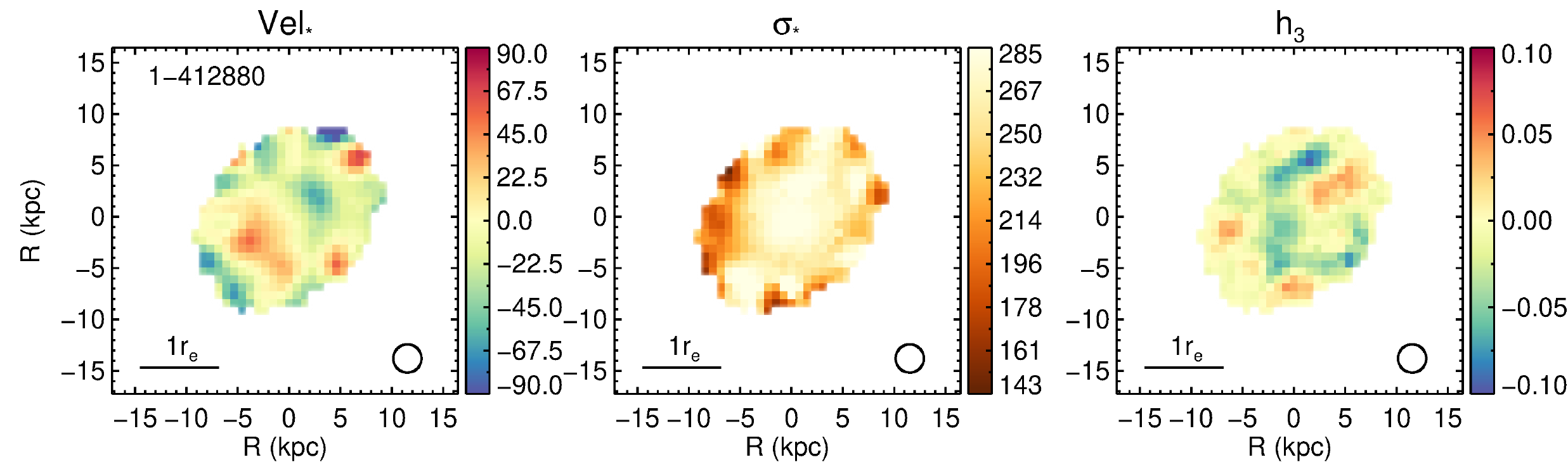}
    \includegraphics[width=0.95\hsize]{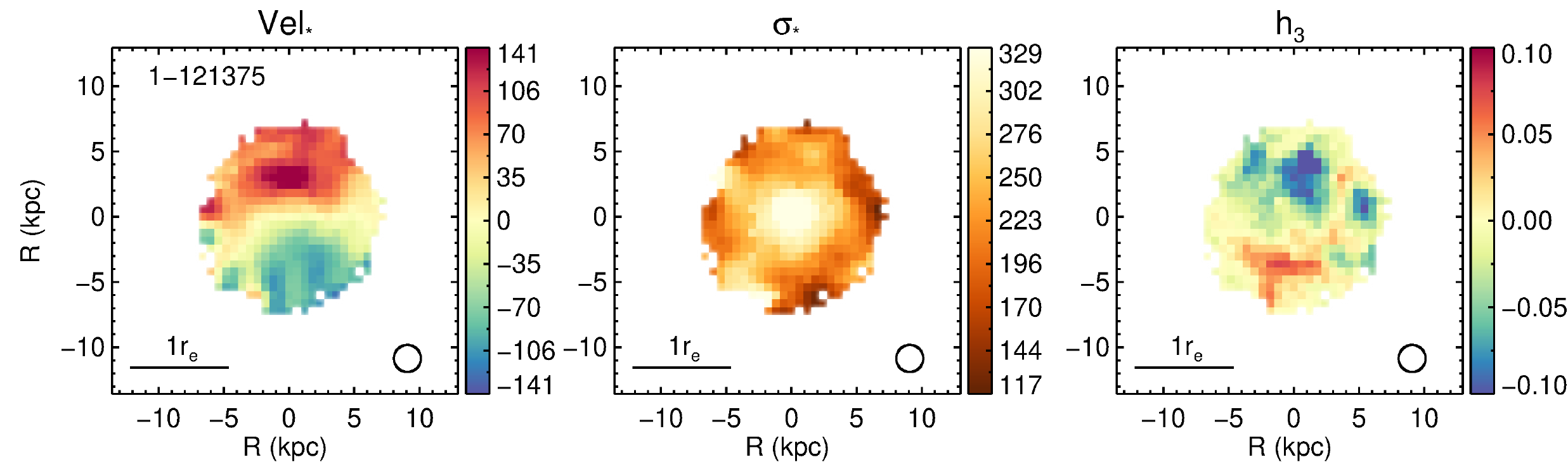}
    \caption{Stellar velocity (km\,s$^{-1}$), velocity dispersion (km\,s$^{-1}$) and Gauss-Hermite moment $h_3$ maps of four representative CSGs (same galaxies displayed in Fig.\,\ref{fig:control}). From top to bottom: MaNGA ID\,=\,1-137844 ($r_{\rm e}$\,=\,5.3\,kpc, $\sigma_{\rm e}$\,=\,164\,km\,s$^{-1}$), 1-52720 ($r_{\rm e}$\,=\,7.1\,kpc, $\sigma_{\rm e}$\,=\,228\,km\,s$^{-1}$), 1-412880 ($r_{\rm e}$\,=\,7.7\,kpc, $\sigma_{\rm e}$\,=\,263\,km\,s$^{-1}$) and 1-121375 ($r_{\rm e}$\,=\,7.6\,kpc, $\sigma_{\rm e}$\,=\,314\,km\,s$^{-1}$). The black circle shows the angular resolution of the observations.}
    \label{fig:kinematics_control}
\end{figure*}

In Fig.\,\ref{fig:kinematics_compact} and \ref{fig:kinematics_control}, we show the stellar velocity, velocity dispersion and $h_3$ maps of four MCGs and CSGs that are representative of typical objects in the samples. Maps for all MCGs are shown in appendix\,\ref{app:ap_maps} . 

MCGs display velocity maps dominated by regular motions. The velocity dispersion reaches a peak in the center and steeply drops to 100\,--\,150\,km\,s$^{-1}$ at $\sim$\,1\,$r_{\rm e}$. The $h_3$ and velocity are generally anti-correlated. These are characteristics of bulge\,$+$\,disk systems. CSGs, in contrast, present a variety of velocity maps. Some objects display ordered rotation, others are dominated by non-rotational motions. Compared to MCGs, the velocity dispersion profiles of CSGs are less steep and CSGs usually do not show a clear correlation between $h_3$ and velocity.

\subsubsection{Fast and slow rotator fractions}

With the advent of integral field spectroscopic surveys, it became apparent that, based on their kinematical maps, early-type galaxies were divided into two classes: fast rotators, with kinematical maps consistent with disks observed at various inclinations; and slow rotators, with kinematical maps inconsistent with simple disks \citep{cappellari11}. Based on the modelling of the stellar dynamics, it was found in previous studies that slow rotators are triaxial systems, while fast rotators are axisymmetric and rotationally flattened \citep{cappellari07,cappellari16}. Following \citet{emsellem07,emsellem11}, we divide the objects in our sample into fast and slow rotators based on the $\lambda_{\rm e}$ parameter, a proxy for the baryon projected specific angular momentum. 

\begin{figure}
\centering
    \includegraphics[width=0.95\hsize]{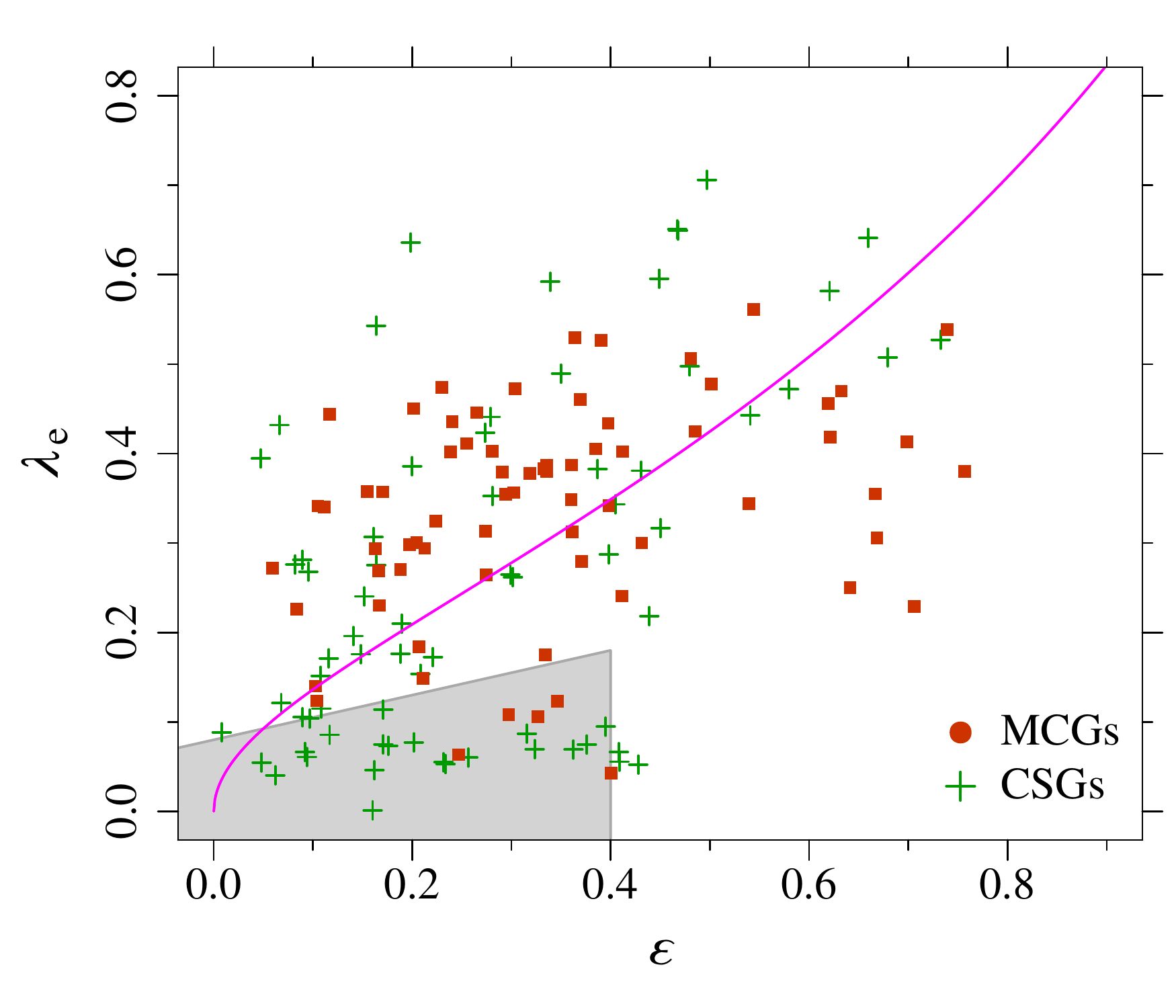}
    \caption{$\lambda_{\rm e}$\,versus\,ellipticity for the compact and control samples. MCGs are displayed as \emph{green crosses} and CSGs are displayed as \emph{red circles}. The \emph{shaded grey area} indicate the region of the $\lambda_{\rm e}-\varepsilon$ plane occupied by slow rotators. The \emph{magenta line} represents the expected $\lambda_{\rm e}$ for a galaxy with intrinsic ellipticity $\varepsilon$ (i.e., inclination of zero degrees) with constant anisotropic factor $\delta$\,=\,0.7\,$\times$\,$\varepsilon$ \citep{cappellari07}. }
    \label{fig:lambda_compact}
\end{figure}

In Fig.\,\ref{fig:lambda_compact},  we plot $\lambda_{\rm e}$ versus $\varepsilon$. Slow rotators are located inside the shaded grey area \citep{cappellari16}. MCGs are overwhelmingly fast rotators: 66 (94.3\%) are classified as fast rotators, while 50 (71.4\%) CSGs are fast rotators. 

\subsubsection{High Order Kinematical Moments}

\begin{figure}
\centering
    \includegraphics[width=0.95\hsize]{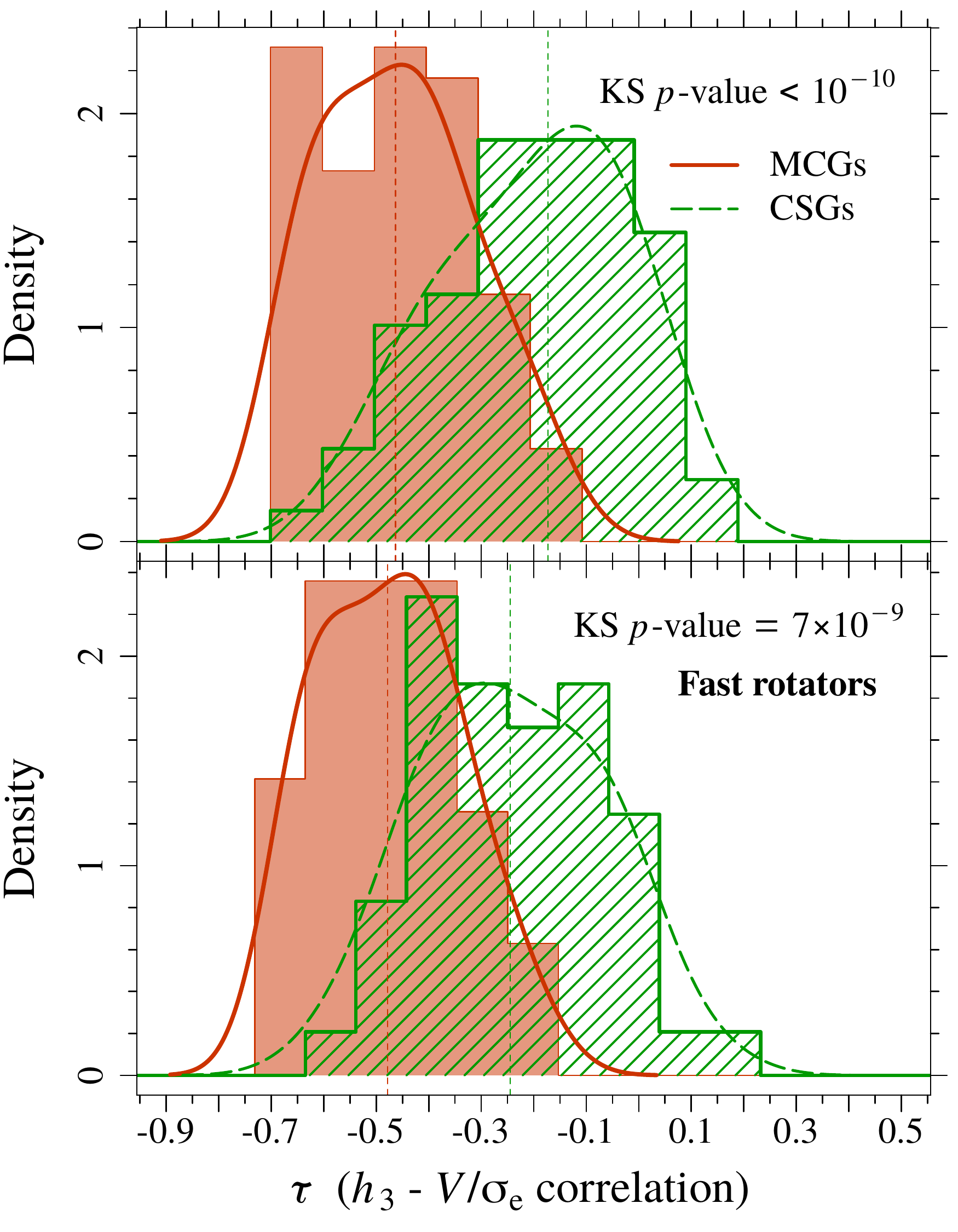}
    \caption{Distribution of the Kendall $\tau$ correlation coefficient measuring the relation between the Gauss-Hermite moment h$_3$ and the ratio $V/\sigma$ for all MCGs and CSGs (\emph{upper panel}) and only for fast rotators (\emph{lower panel}). The distributions of the compact and control samples are shown in \emph{red} and \emph{green}, respectively.
    The \emph{vertical dashed lines} indicate the median values. The curves are obtained by smoothing the positions of the data points using a Gaussian kernel with the standard deviation equal to half of the standard deviation of the data points.}
    \label{fig:tau}
\end{figure}

Cosmological hydrodynamical zoom-in simulations have shown that the formation history of early-type galaxies cannot be constrained from the $\lambda_{\rm e}$ parameter alone, as fast and slow rotators have multiple formation channels \citep{naab14}. These simulations showed, however, that when $\lambda_{\rm e}$ is analysed in combination with high order stellar kinematical moments, different merger scenarios can be distinguished. 

In fast rotators, usually $h_3$ and the ratio $V/\sigma$ are anti-correlated, which has been interpreted as a signature of stars orbiting in a disc structure \citep{vanderMarel93,bender94,Krajnovic08,krajnovic11,vandesande17a}. Simulations showed that this anti-correlation arises due to an increased relative fraction of prograde $z$-tube orbits, which creates a steep leading wing in the LOSVD \citep{hoffman09,rottgers14} and is associated with late in situ star formation due to gas-rich mergers or gas accretion \citep{naab14}. In contrast, fast-rotators which do not have anti-correlated $h_3$ and $V/\sigma$ were formed in gas-poor major mergers which lead to a spin-up of the remnant \citep{naab14}. Additionally, fast rotators with stellar disks but weak or no $h_3$-$V/\sigma$ anti-correlation were identified in observations. Recent interactions and the presence of bars have been suggested as explanations for this \citep{vandesande17a}. 

Considering that theoretical works and simulations point to box orbits being dominant in the center of galaxies while $z$-tubes dominate in outer regions, one might wonder if it is expected that a strong $h_3$--$V/\sigma$ anti-correlation is present in the inner radii probed by our observations (typically $\sim 1 r_{\rm e}$). Hydrodynamical simulations performed by \citet{rottgers14} show that, in remnants of gas rich mergers, $z$-tube orbits already account for $\sim30-40\%$ of the orbits at $\sim 0.3 r_{\rm e}$, becoming dominant at $\sim 0.5 r_{\rm e}$ (see their Fig.\,7). Thus, the spatial coverage of the MaNGA observations is not a limiting factor, and detecting a $h_3$--$V/\sigma$ anti-correlation in the MaNGA data is possible.

To quantify the strength of the anti-correlation between $h_3$ and $V/\sigma$, we computed the Kendall correlation coefficient ($\tau$) for each object. The $\tau$ distribution is shown in Fig.\,\ref{fig:tau}a. 69\% of MCGs show a strong anti-correlation between $V/\sigma$ and $h_3$ ($\tau$\,$<$\,$-0.4$), while 70\% of CSGs either show no correlation or a weak anti-correlation ($-0.3$\,$<$\,$\tau$\,$<0.3$). Considering only fast rotators (Fig.\,\ref{fig:tau}b), 73\% of fast-rotating MCGs have $\tau$\,$<$\,$-0.4$, in contrast to only 20\% of fast-rotating CSGs.

Analyses of the spatial variation of the Gauss-Hermite moment $h_4$ has also been employed to investigate evolutionary scenarios of early-type galaxies. For example, high $h_4$ values coupled with enhanced [$\alpha$/Fe] and low [Fe/H] at large radii ($R\,\gtrsim$\,1\,$r_{\rm e}$) have been interpreted as higher radial anisotropy derived from accretion of small satellites \citep{greene19}. The data analysed in this study do not reach these large radii, thus an analysis of the $h_4$ maps is beyond the scope of this paper.

\subsection{Morphology}

\begin{figure}
\centering
\begin{tabular}{c}
    \includegraphics[width=0.95\hsize]{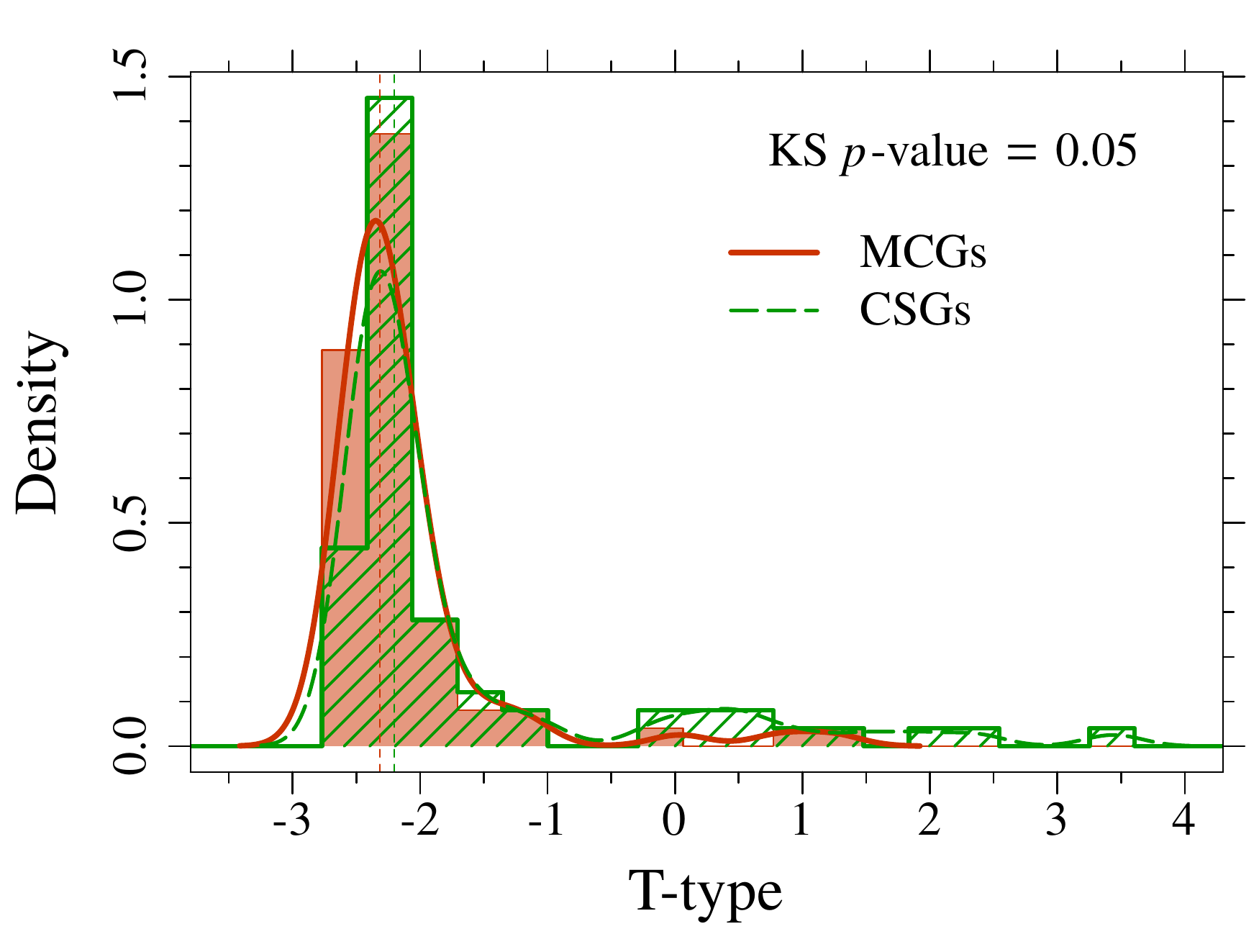} \\
    \includegraphics[width=0.95\hsize]{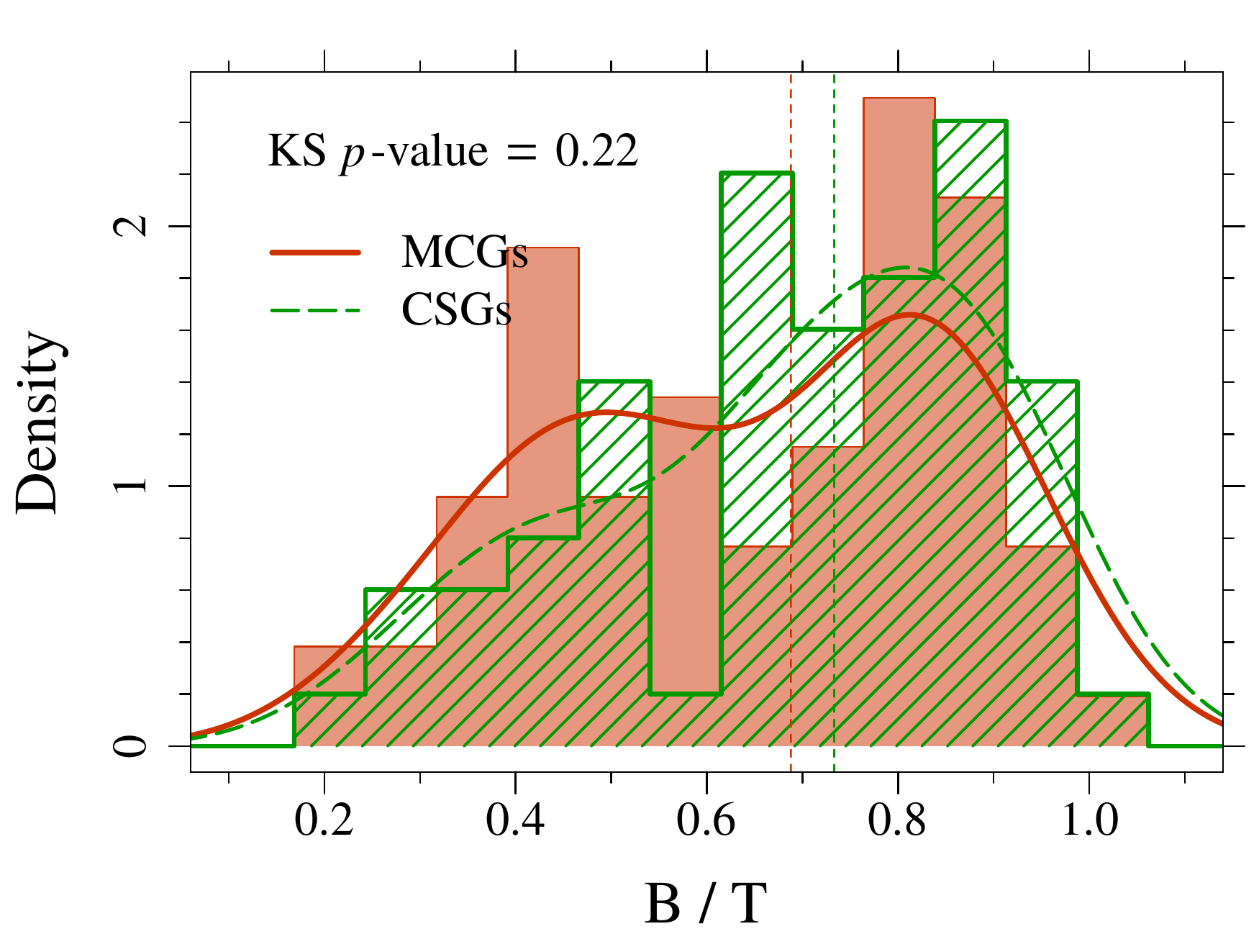} \\
    \includegraphics[width=0.95\hsize]{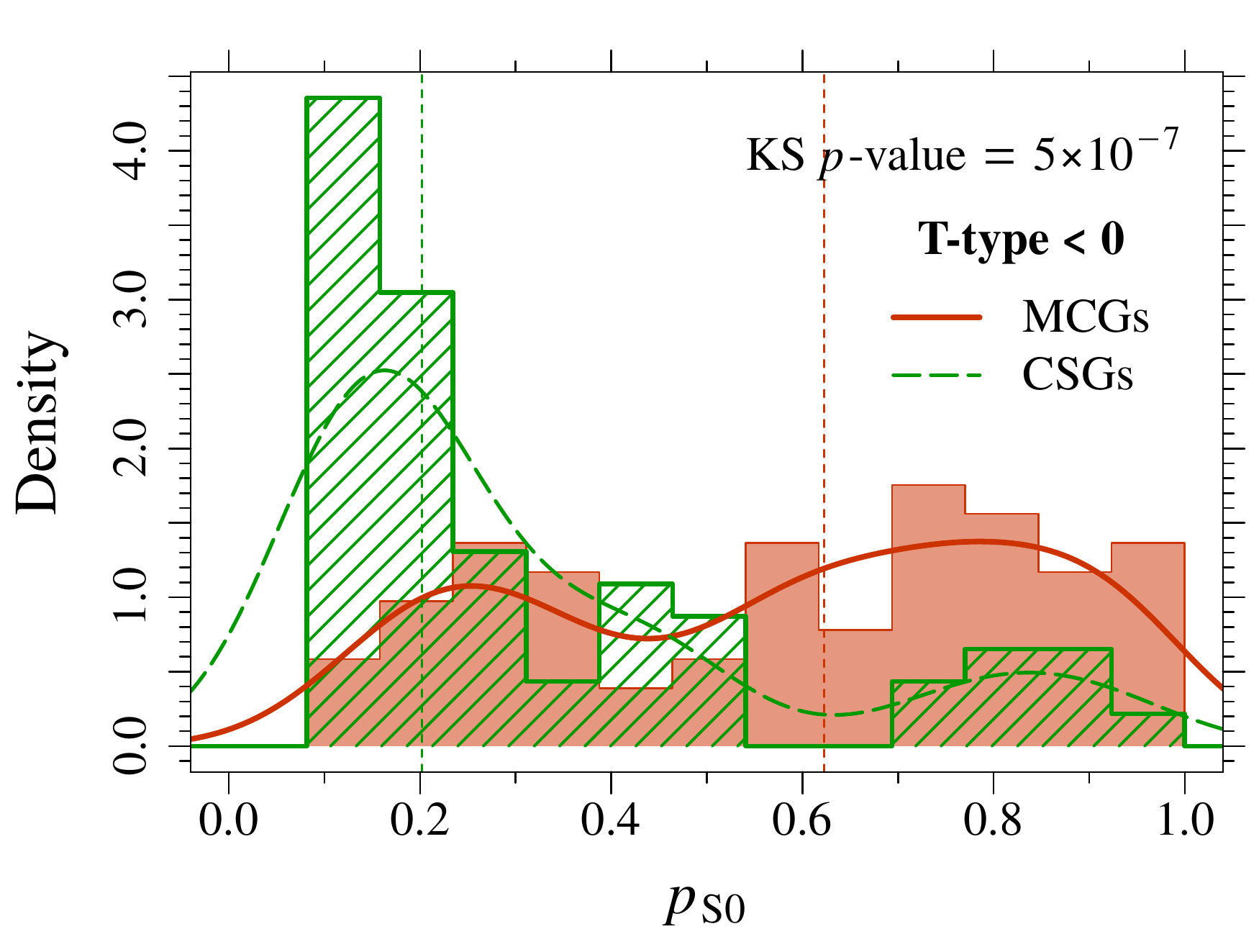} \\
\end{tabular}
    \caption{Distribution of T-types from \citet[][\emph{upper panel}]{dominguez-sanchez18}, bulge to total light ratio determined by \citet[][\emph{middle panel}]{fischer19}, and probability of having S0 morphology (\emph{lower panel}). In the lower panel, only galaxies with T-type < 0 are shown. 
     In all panels, the \emph{vertical dashed lines} indicate the median values. The curves are obtained by smoothing the positions of the data points using a Gaussian kernel with the standard deviation equal to half (for $B/T$) and one third (for T-types and $p_{\rm S0}$) of the standard deviation of the data points.
    }
    \label{fig:bt}
\end{figure}

 In Fig.\,\ref{fig:bt} we show the distributions of T-Types, bulge-to-total light ratios (B/T), and the probability of having S0 morphology ($p_{{\rm S}0}$). In terms of Hubble type, MCGs and CSGs are quite distinct. 34\% of MCGs are classified as ellipticals, 62\% as S0s and 4\% as spiral galaxies. On the other hand, 72\% of CSGs are elliptical galaxies, 14\% are S0s and 14\% are spiral galaxies. Despite differences in Hubble type, the B/T distributions of the samples are similar.
 
 
\subsection{Stellar Populations}

\begin{figure}
\centering
    \includegraphics[width=0.95\hsize]{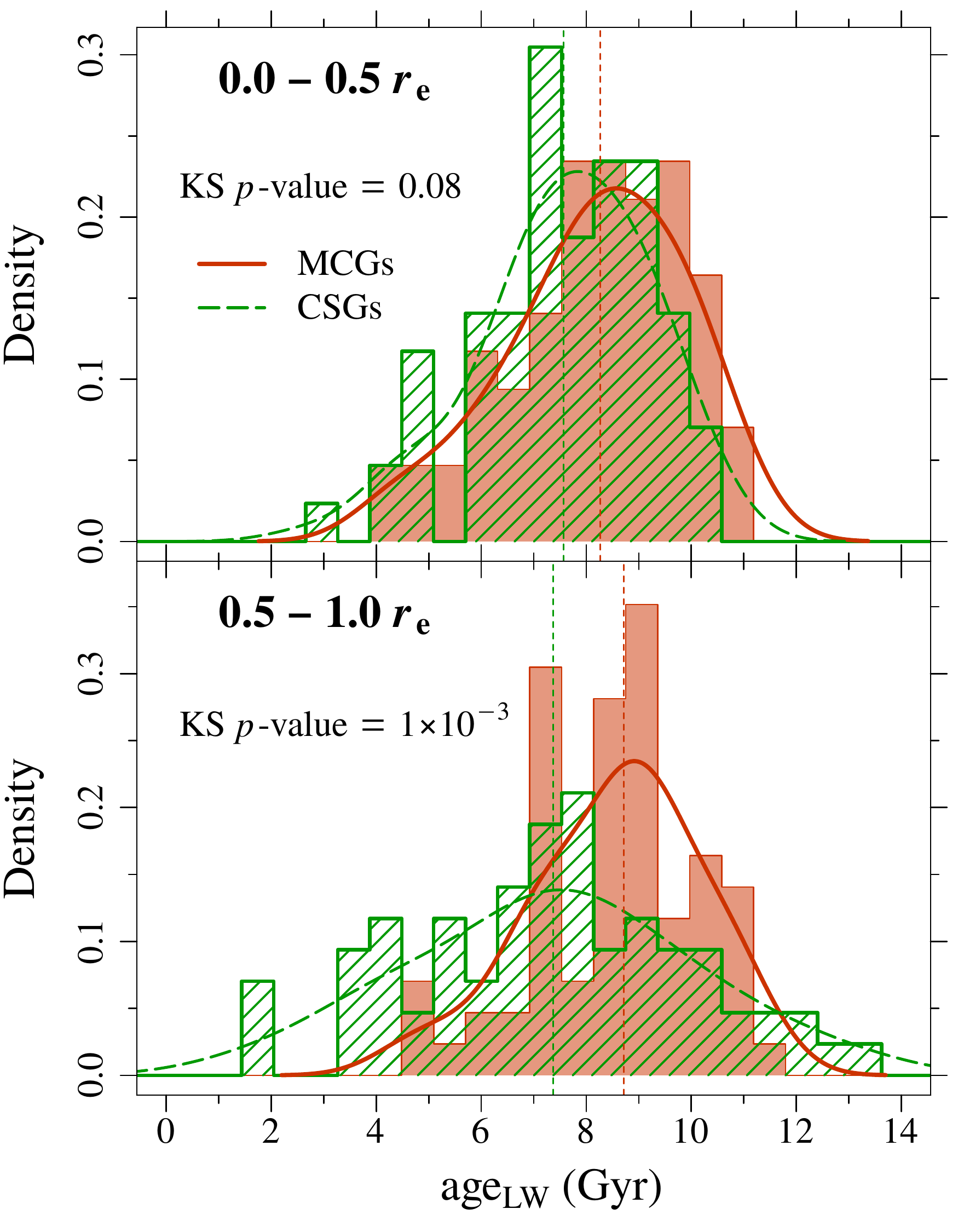}
    \includegraphics[width=0.95\hsize]{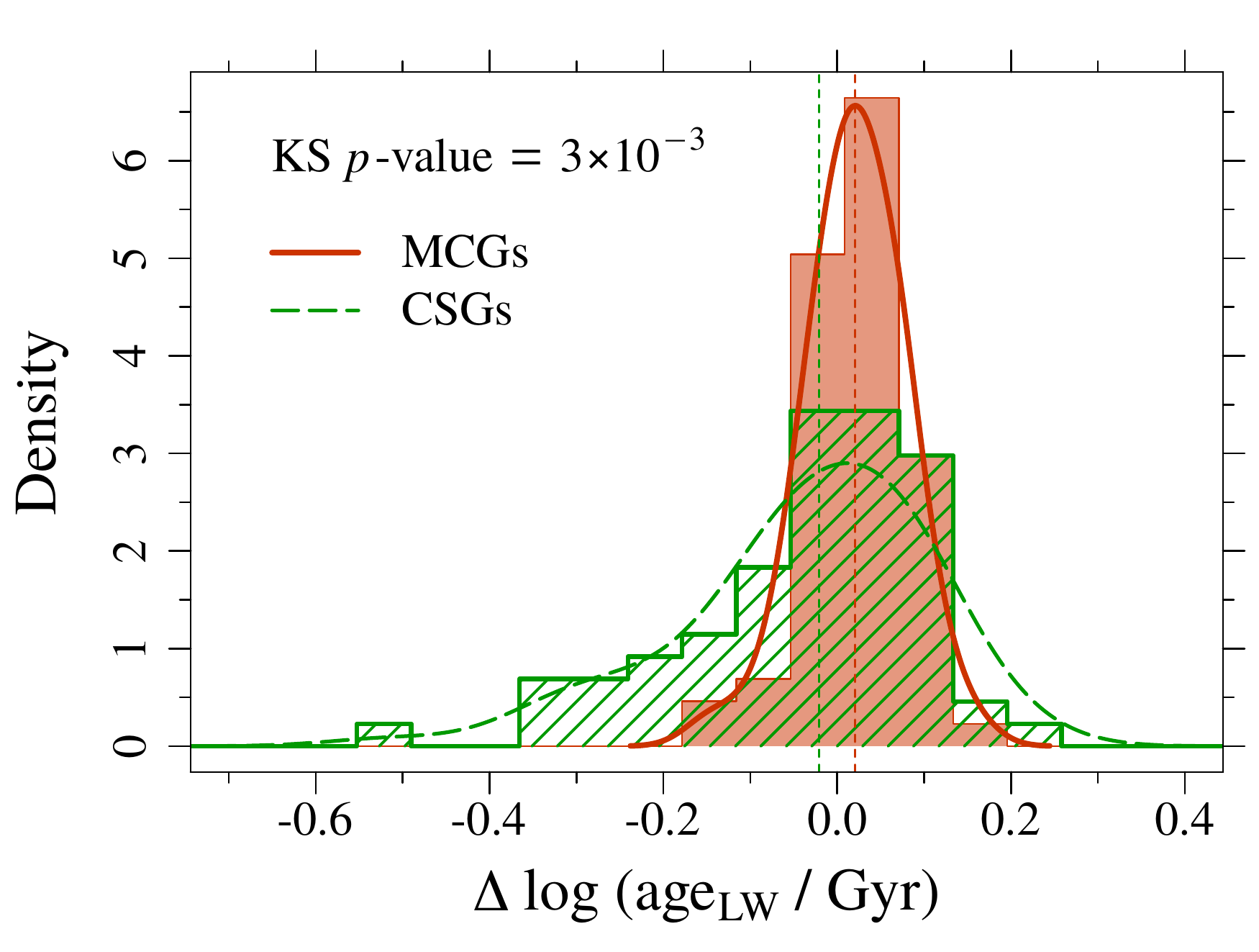}
    \caption{Distribution of the luminosity-weighted stellar age inferred from the spectra integrated from $0$ to $0.5\,r_{\rm e}$ (\emph{upper panel}) and from $0.5$ to $1\,r_{\rm e}$ (\emph{middle panel}). The \emph{lower panel} shows the distributions of the luminosity-weighted age gradients of the compact and control samples. In all panels, the \emph{vertical dashed lines} indicate the median values. The curves are obtained by smoothing the positions of the data points (not the histograms) using a Gaussian kernel with the standard deviation equal to half of the standard deviation of the data points.}
    \label{fig:age}
\end{figure}

In Fig.\,\ref{fig:age}, we show the luminosity-weighted age distribution of MCGs and CSGs, measured from spectra integrated inside an aperture of radius $0.5\,r_{\rm e}$ and within $0.5-1.0\,r_{\rm e}$. The distributions of ages of the stellar populations in the inner $0.5\,r_{\rm e}$ of MCGs and CSGs are consistent, but within $0.5-1.0\,r_{\rm e}$ the age distribution of CSGs extends to lower values compared to MCGs. The $\Delta\,\log($age) distributions of MCGs and CSGs are flat on average, while the distribution of CSG $\Delta\,\log($age) have a significantly larger tail towards negative values, as shown in the bottom panel of Fig.\,\ref{fig:age}.

\begin{figure}
\centering
    \includegraphics[width=0.95\hsize]{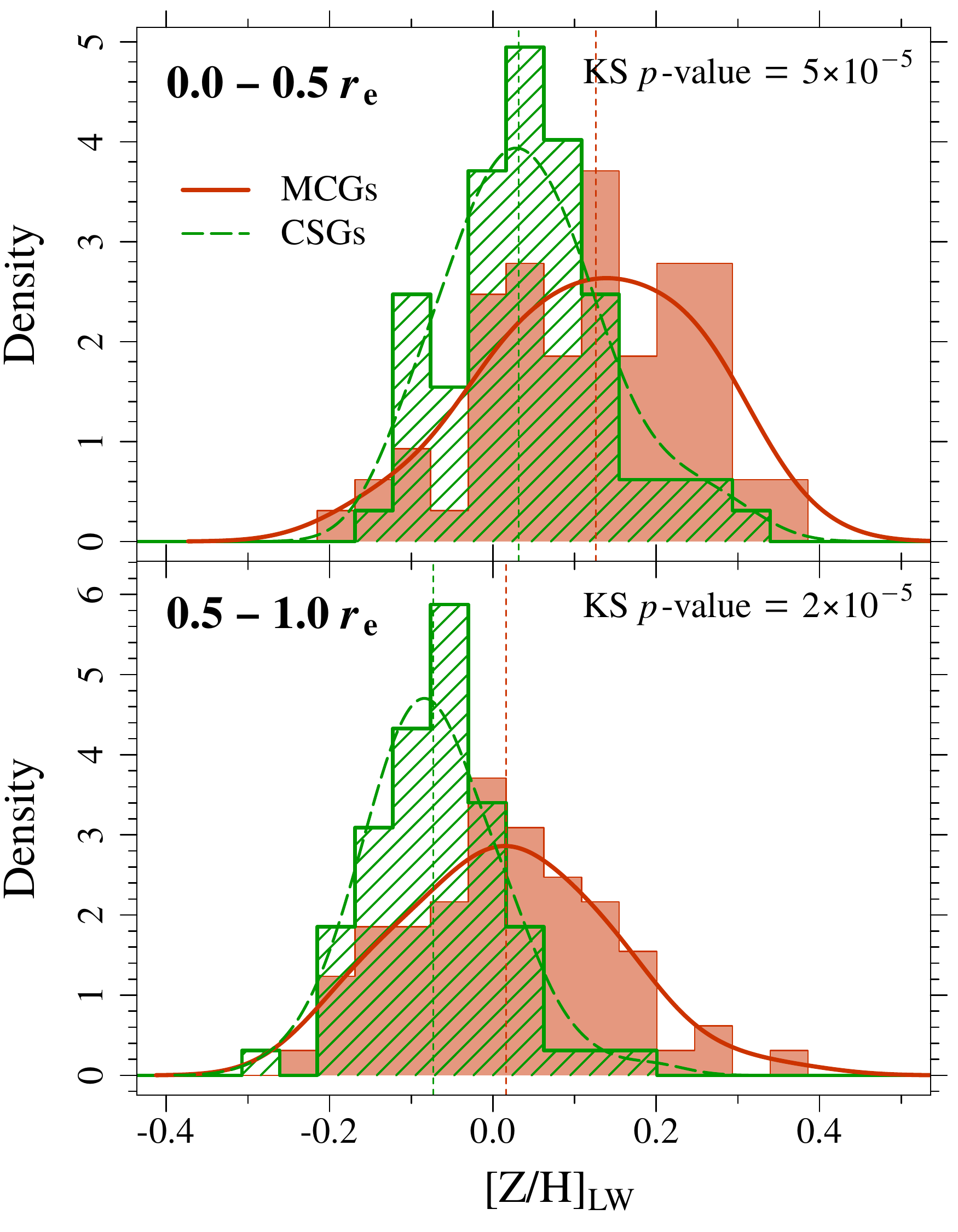}
    \includegraphics[width=0.95\hsize]{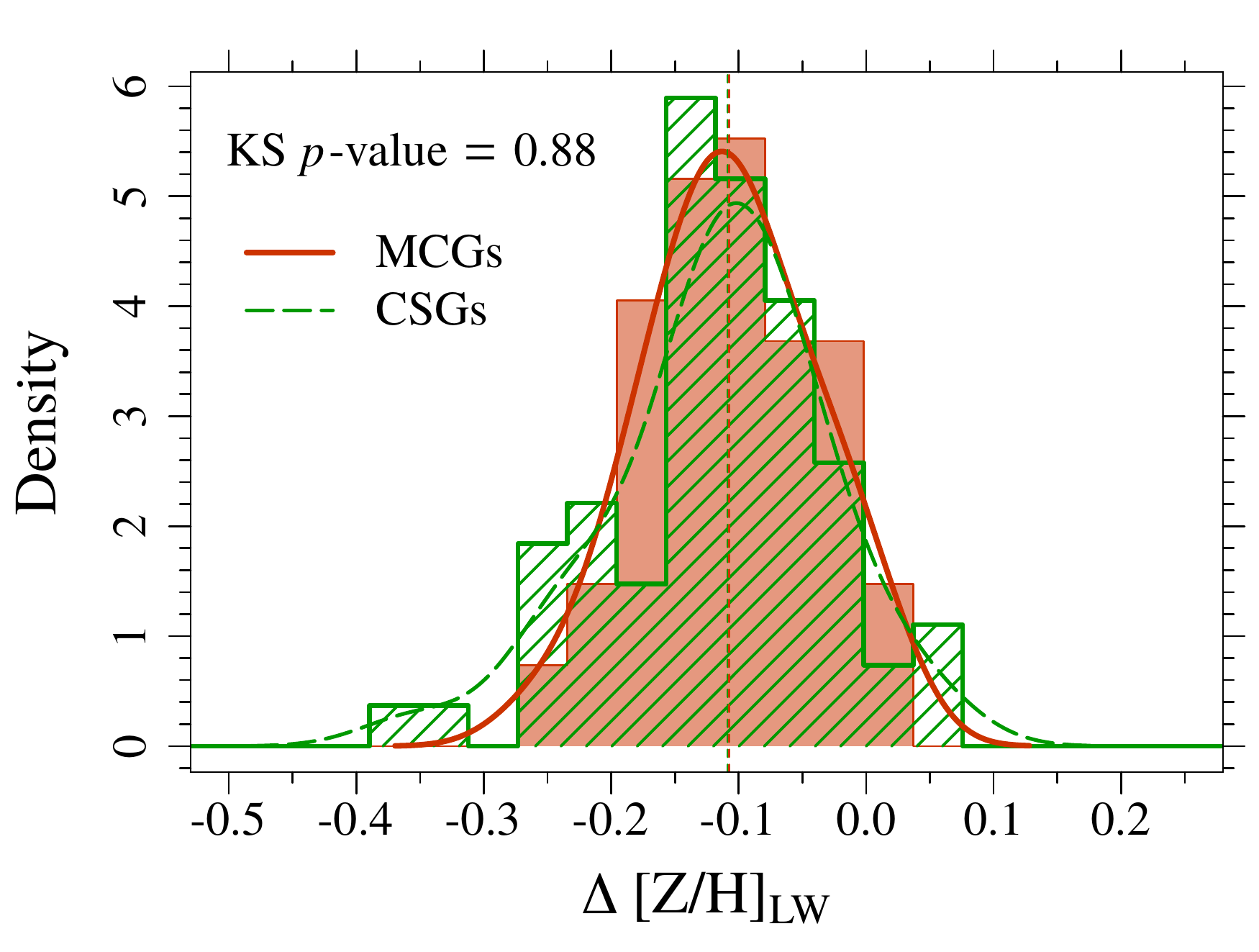}
    \caption{Distribution of the luminosity-weighted metallicity inferred from the spectra integrated from $0$ to $0.5\,r_{\rm e}$ (\emph{upper panel}) and from $0.5$ to $1\,r_{\rm e}$ (\emph{middle panel}). 
    The distributions of the luminosity-weighted metallicity gradients of the compact and control samples are shown in the \emph{lower panel}. In all panels, the \emph{vertical dashed lines} indicate the median values. The curves are obtained by smoothing the positions of the data points using a Gaussian kernel with the standard deviation equal to half of the standard deviation of the data points.}
    \label{fig:metal}
\end{figure}

MCGs tend to be more metal-rich than CSGs, as shown in Fig.\,\ref{fig:metal}.
The differences between the metallicity of the stellar populations in the inner $0.5\,r_{\rm e}$ and within $0.5-1.0\,r_{\rm e}$ shown in the bottom panel of Fig.\,\ref{fig:metal} indicate that MCGs and  CSGs have slightly negative metallicity gradients. There is no statistically-significant difference between the metallicity gradient distributions of the two samples.

\begin{figure}
\centering
    \includegraphics[width=0.95\hsize]{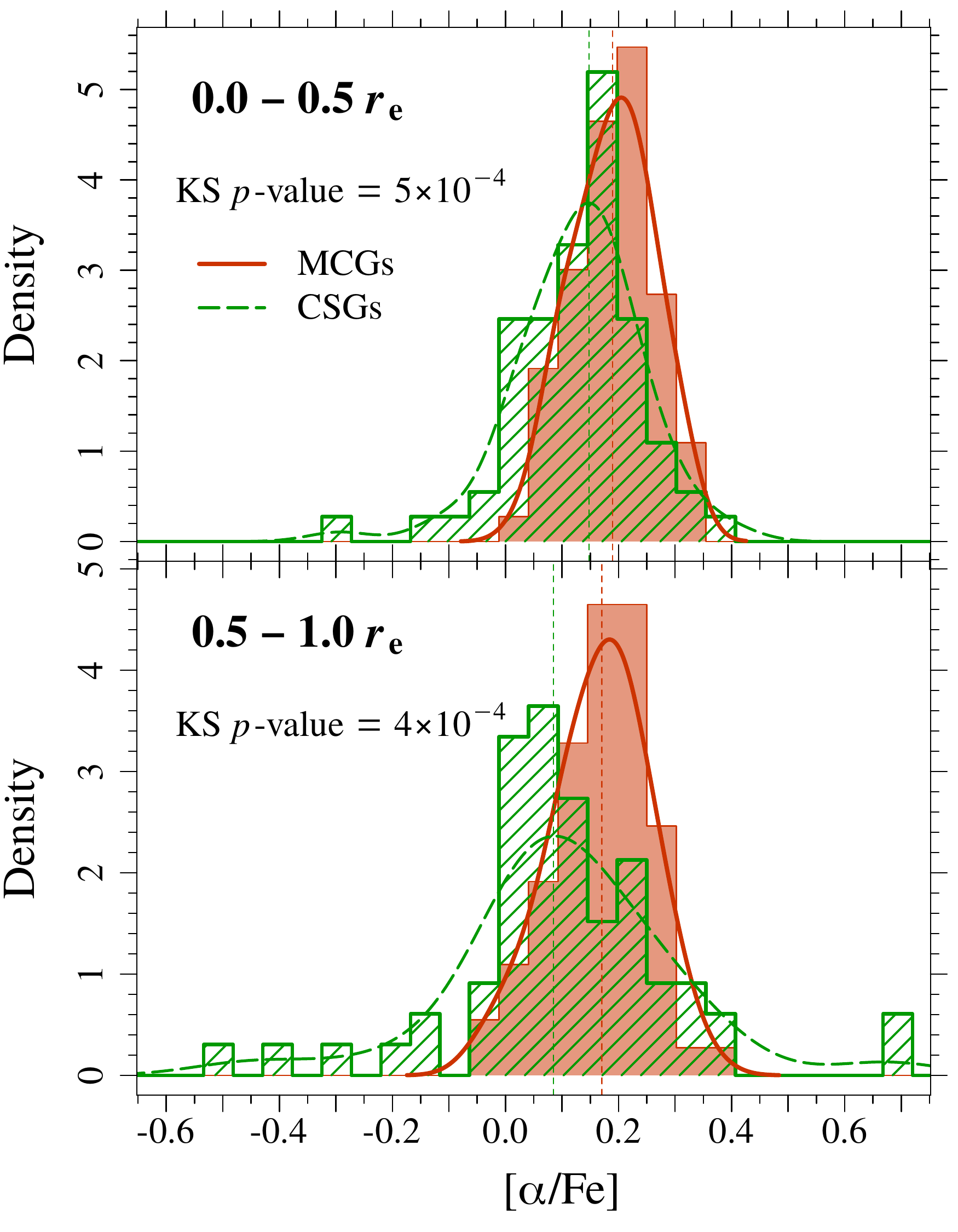} 
    \includegraphics[width=0.95\hsize]{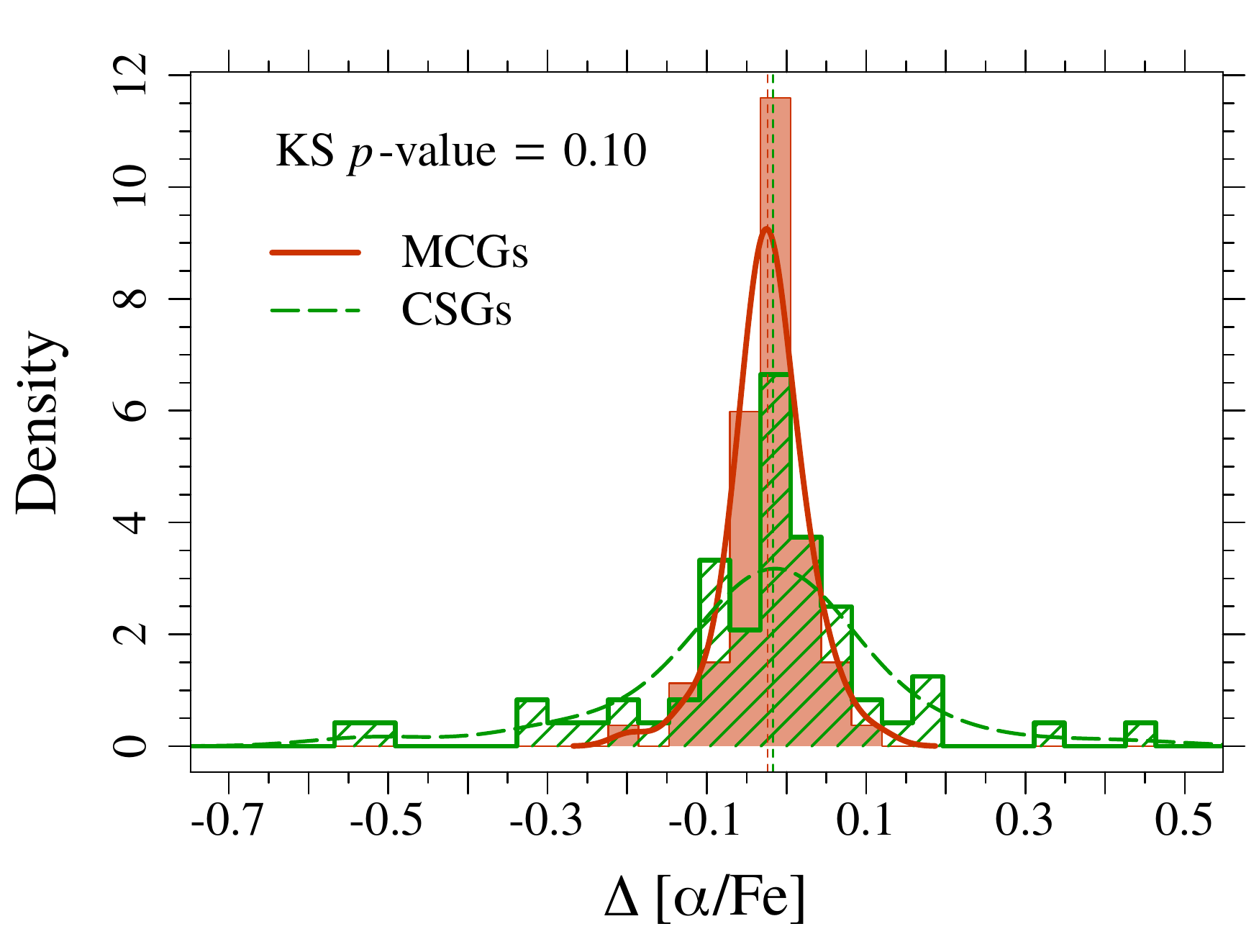} 
\caption{Distribution of [$\alpha$/Fe] inferred from the spectra integrated from $0$ to $0.5\,r_{\rm e}$ (\emph{upper panel}) and from $0.5$ to $1\,r_{\rm e}$ (\emph{middle panel}). The Distributions of the [$\alpha$/Fe] gradients of the compact and control samples are shown in the \emph{lower panel}. In all panels, the \emph{vertical dashed lines} indicate the median values. The curves are obtained by smoothing the positions of the data points using a Gaussian kernel with the standard deviation equal to half of the standard deviation of the data points. }
\label{fig:aFe}
\end{figure}

In Fig.\,\ref{fig:aFe}, we show that MCGs have higher [$\alpha$/Fe] values compared to the CSGs, and the differences between MCGs and CSGs are observed both in the inner $0.5\,r_{\rm e}$ and within $0.5-1.0\,r_{\rm e}$.
The distribution of [$\alpha$/Fe] within $0.5-1.0\,r_{\rm e}$ shows a larger scatter in CSGs compared to the MCGs [$\alpha$/Fe], which might be related to larger uncertainties in the measurements of spectral indices. However, the spectra of MCGs and CSGs within $0.5-1.0\,r_{\rm e}$ have similar signal-to-noise ratios ${\rm S/N} \gtrsim 15$. Besides, we compared the  [$\alpha$/Fe] distributions after excluding the values inferred from spectra with ${\rm S/N} \leq 20$, and the scatter in the CSG [$\alpha$/Fe] values remains larger than that of MCGs. Therefore, this scatter may indicate that the assembly histories of CSGs are more heterogeneous compared to those of MCGs.

Given the small sizes of MCGs, the spatial resolution of MaNGA observations is a concern. In Appendix\,\ref{app:seeing} , we investigate if the results on the gradients of stellar population properties change if we exclude MCGs that are not well resolved from the analysis. Although we see that the MCG age and metallicity gradients depends on how well the MCGs are spatially resolved, we get similar results when analysing only resolved MCGs.


\begin{figure}
\centering
    \includegraphics[width=0.95\hsize]{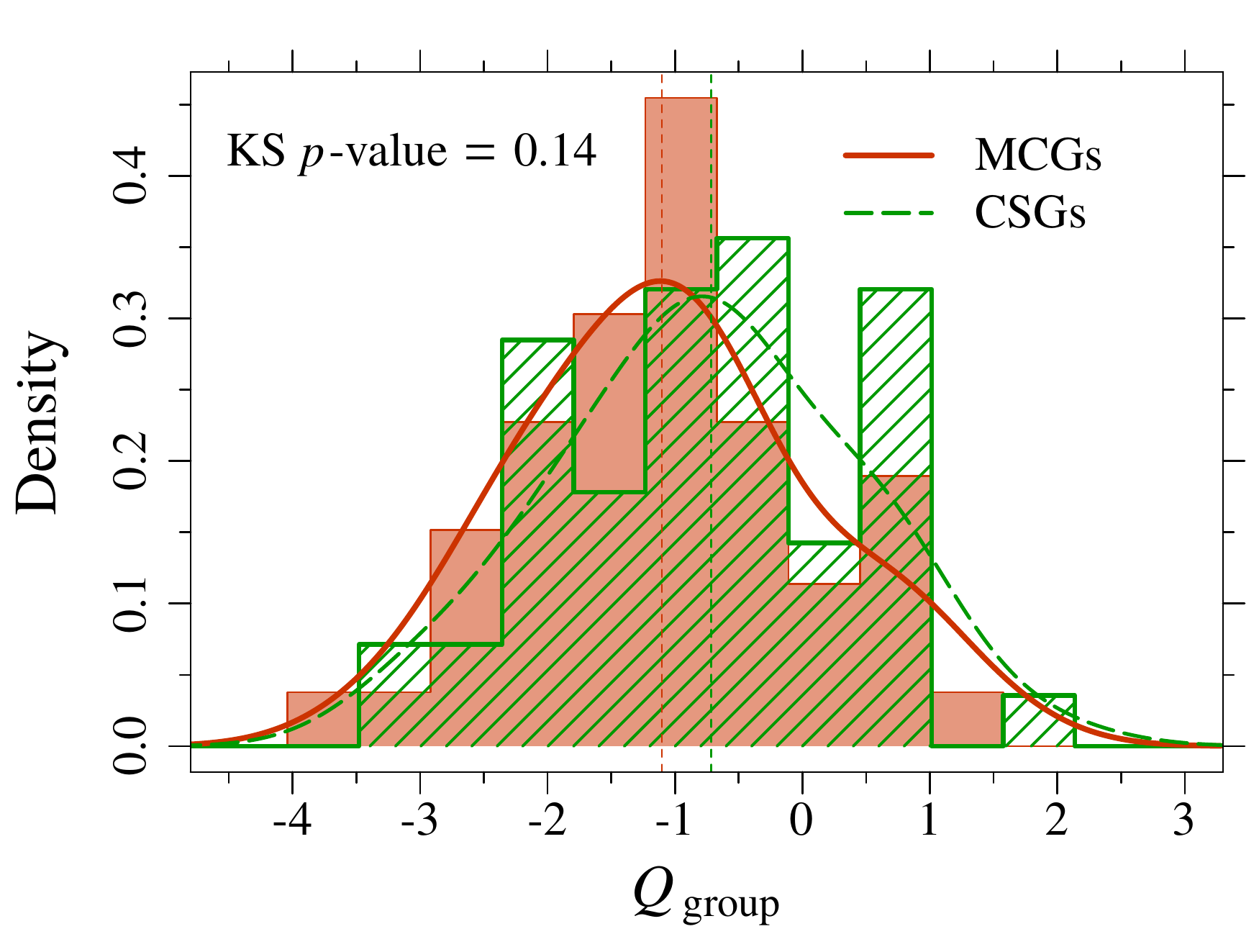}
    \caption{Distribution of the tidal strength parameter ($Q_{\rm group}$) of the compact and control samples. $Q_{\rm group}$ quantifies the local environment, as it is an estimation of the total gravitational interaction strength produced on a given galaxy by its neighbours with respect to the internal binding forces.
    The \emph{vertical dashed lines} indicate the median values. The curves are obtained by smoothing the positions of the data points using a Gaussian kernel with the standard deviation equal to half of the standard deviation of the data points.}
    \label{fig:q}
\end{figure}

\subsection{Environment}
\renewcommand{\arraystretch}{1.5}
\begin{table*}
\centering
\caption{Environment of compact and control sample galaxies. Halo masses and group membership information were extracted from an updated version of the catalogue of groups and clusters by \citet{Yang.etal:2007}.}
\begin{tabular}{c|cc|cc|cc|c}
\hline
\multicolumn{1}{c|}{} & \multicolumn{2}{c|}{$M_{\rm halo}$\,$<$\,10$^{13}$\,M$_{\odot}$} & 
\multicolumn{2}{c|}{10$^{13}$\,$\leq$\,$M_{\rm halo}$\,$<$\,10$^{14}$\,M$_{\odot}$} & 
\multicolumn{2}{c|}{$M_{\rm halo}$\,$\geq$\,10$^{14}$\,M$_{\odot}$} & Total\\
\cline{2-8}
& Centrals  & Satellites & Centrals & Satellites & Centrals & Satellites & \\ 
\hline \hline
\multirow{2}{*}{MCGs} & 43 (97.7\%) & 1 (2.3\%) & {\bf 4 (33.3\%)} & {\bf 8 (66.7\%)} & {\bf 0 (0\%)} & {\bf 14 (100\%)} & \\
    & \multicolumn{2}{c|}{\cellcolor{lightgrey} {\bf 44 (62.9\%)}} &                          \multicolumn{2}{c|}{\cellcolor{lightgrey} {\bf 12 (17.1\%)}} & 
    \multicolumn{2}{c|}{\cellcolor{lightgrey} 14 (20.0\%)} & 
    \multicolumn{1}{c}{\cellcolor{lightgrey} 70} \\  
    \hline       
\multirow{2}{*}{CSGs} & 25 (100\%) & 0 (0\%) & {\bf 30 (88.2\%)} & {\bf 4 (11.8\%)} & {\bf  5 (45.5\%)} & {\bf 6 (54.5\%)} & \\ 
    & \multicolumn{2}{c|}{\cellcolor{lightgrey} {\bf 25 (35.7\%)}} &             \multicolumn{2}{c|}{\cellcolor{lightgrey} \bf{34 (48.6\%)}} &     
    \multicolumn{2}{c|}{\cellcolor{lightgrey} 11 (15.7\%)} & 
    \multicolumn{1}{c}{\cellcolor{lightgrey} 70} \\
    \hline         

    \multicolumn{8}{c}{\bf \large Fisher's tests} \\
    \hline \hline
    \multirow{2}{*}{$p$-value} & 1.0 & 1.0 & $\mathbf{6 \times 10^{-4}}$ & $\mathbf{6 \times 10^{-4}}$ & 
        $\mathbf{9 \times 10^{-3}}$ & $\mathbf{9 \times 10^{-3}}$ & \\
    & \multicolumn{2}{c|}{\cellcolor{lightgrey} $\mathbf{2 \times 10^{-3}}$} & 
        \multicolumn{2}{c|}{\cellcolor{lightgrey} $\mathbf{1 \times 10^{-4}}$} & 
        \multicolumn{2}{c|}{\cellcolor{lightgrey} 0.7} & 
        \multicolumn{1}{c}{\cellcolor{lightgrey}}\\
    \hline
    \multicolumn{8}{c}{\bf \large Barnard's tests} \\
    \hline    \hline
    \multirow{2}{*}{$p$-value} & 0.6 & 0.6 & $\mathbf{3 \times 10^{-4}}$ & $\mathbf{3 \times 10^{-4}}$ &  
        $\mathbf{4 \times 10^{-3}}$ & $\mathbf{4 \times 10^{-4}}$ & \\
    & \multicolumn{2}{c|}{\cellcolor{lightgrey} $\mathbf{2 \times 10^{-3}}$} & 
        \multicolumn{2}{c|}{\cellcolor{lightgrey} $\mathbf{7 \times 10^{-5}}$} & 
        \multicolumn{2}{c|}{\cellcolor{lightgrey} 0.6} & 
        \multicolumn{1}{c}{\cellcolor{lightgrey}}\\
    \hline
\end{tabular}
\parbox{\hsize}{{\bf Notes:}
    The number of galaxies in haloes within different mass bins are divided in central and satellites. The last lines show the $p$-values of Fisher's and Barnard's tests comparing the fractions of compact and control sample galaxies that are centrals and satellites. 
    Results with $p$-values\,$< 0.05$ are highlighted in boldface.}
\label{table:env}
\end{table*}
\renewcommand{\arraystretch}{1}

In table\,\ref{table:env}, we present the environmental characteristics of MCGs and CSGs. We find that 86\% of CSGs are centrals (60) while 66\% (47) of MCGs are centrals. Central MCGs reside in lower mass halos than their non-compact counterparts. However, these results are a consequence of differences in the stellar mass distribution of the two samples, since it is expected that more massive galaxies reside in more massive haloes. Besides, the halo masses from the \citet{Yang.etal:2007} catalogue are obtained through abundance matching based on galaxy luminosity, and more luminous galaxies will be assigned to the more massive haloes. As we will discuss in Sect.\,\ref{sec:results:M-CSGs}, if we select a control sample with similar stellar mass distribution as that of MCGs, no difference between their environmental properties will emerge from the data.

In Fig.\,\ref{fig:q} we show the distribution of the tidal strength parameter ($Q_{\rm group}$) which quantifies the degree of isolation of a galaxy. A galaxy is considered isolated if $Q_{\rm group}$\,$\le$\,-2. According to this criteria, 33\% (23 objects) of MCGs are isolated, while 29\% (20) of CSGs are isolated. The distributions of the compact and control samples are consistent with being drawn from the same parent distribution, pointing to no significant differences in the local environment of MCGs and CSGs. 

\subsection{Comparison to a mass-matched sample}
\label{sec:results:M-CSGs}

\begin{figure}
\centering
    \includegraphics[width=0.9\hsize]{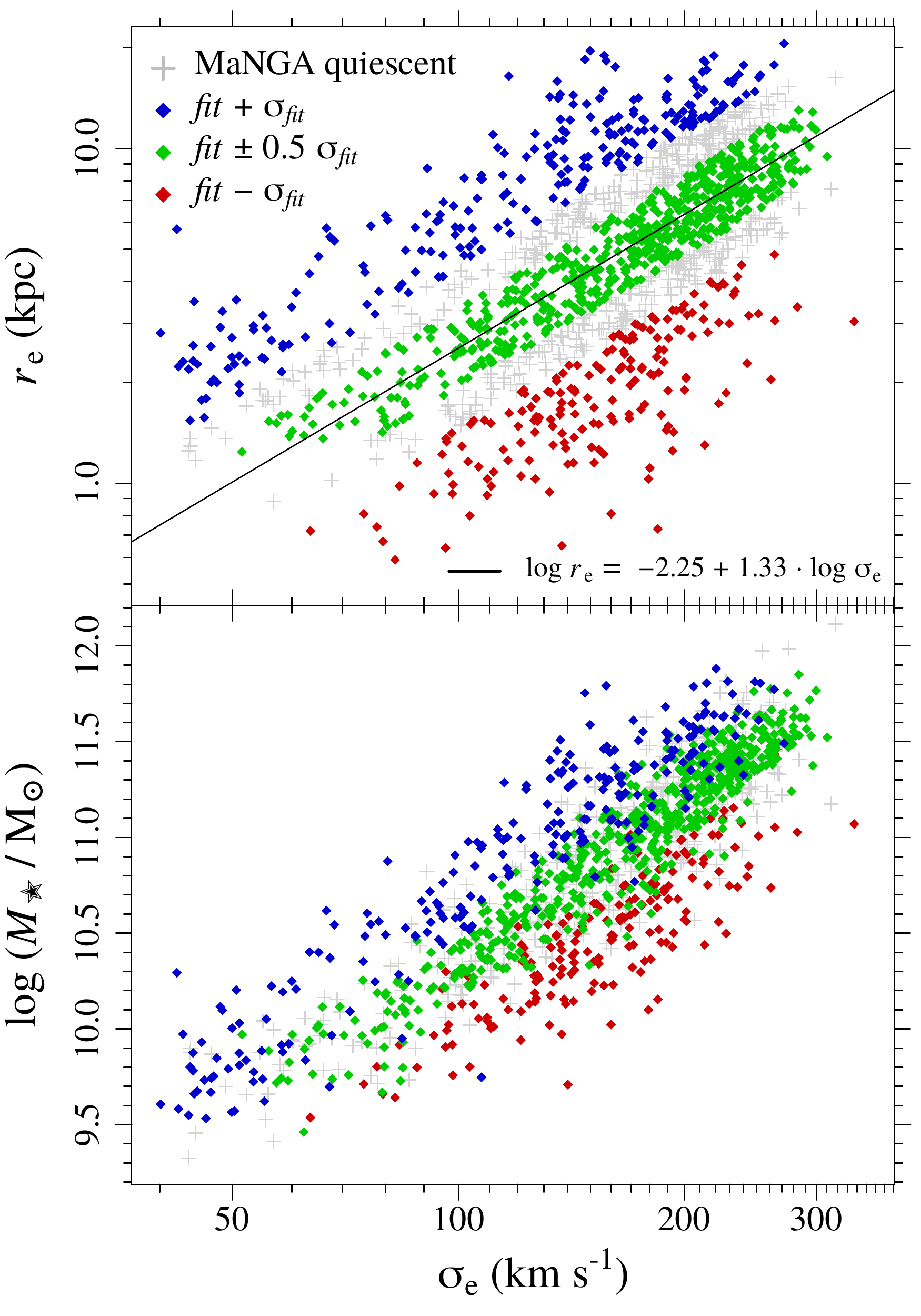} 
\caption{Relations between $r_{\rm e}$, $\sigma_{\rm e}$ and $M_{\star}$. {\bf Top:} Galaxy size as a function of velocity dispersion. The \emph{black solid line} indicates the best linear fit to the 
$\log r_{\rm e} - \log \sigma_{\rm e}$ relation of the MaNGA quiescent galaxies. The \emph{blue}, \emph{green} and \emph{red symbols} correspond to the galaxies that are above $+\sigma_{fit}$, within $\pm\, 0.5\,\sigma_{fit}$ and below $-\sigma_{fit}$ from the best-fit relation, where $\sigma_{fit}$ is the standard deviation of the residuals from the fit. {\bf Bottom:} Galaxy stellar masses vs. velocity dispersion. 
We indicate the galaxies that are above (\emph{blue symbols}), within (\emph{green}) and below (\emph{red}) the 
$r_{\rm e} - \sigma_{\rm e}$ relation as shown in the upper panel.}
\label{fig:sigma_Mstar_Re}
\end{figure}

As previously discussed in Section\,\ref{sec:control}, the stellar mass distributions of MCGs and CSGs differ significantly (see Fig.\,\ref{fig:mass}). This is expected, since the relation between $M_{\star}$ and $\sigma_{\rm e}$ depends on the size of the galaxy, as illustrated in Fig.~\ref{fig:sigma_Mstar_Re}. In this figure, we show the galaxy sizes vs. velocity dispersion for a sample of quiescent MaNGA galaxies that were selected based on their sSFRs (sSFR $\leq 10
^{-11}\,$Gyr$^{-1}$). In the bottom panel of Fig.~\ref{fig:sigma_Mstar_Re}, we show $\log M_{\star}$ vs. $\sigma_{\rm e}$ for galaxies that are above $+\sigma_{fit}$, within $\pm\, 0.5\,\sigma_{fit}$ and below $-\sigma_{fit}$ from the best fit to the $\log r_{\rm e} - \log \sigma_{\rm e}$ relation. It can be clearly seen that, at fixed $\sigma_{\rm e}$, smaller galaxies have lower stellar masses. 

\begin{figure}
\centering
    \includegraphics[width=0.9\hsize]{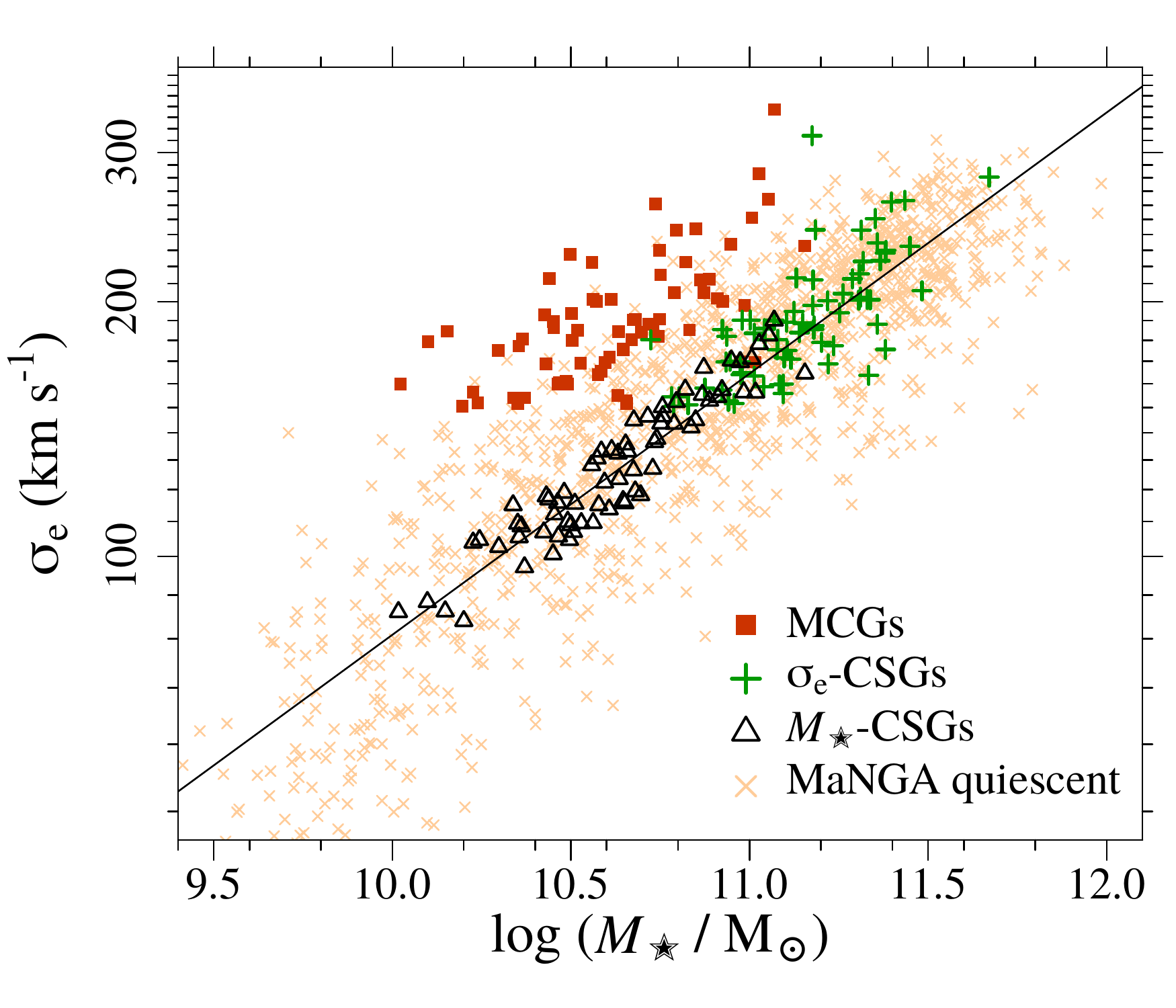} 
\caption{Velocity dispersion vs. stellar mass showing the compact (\emph{red squares}), 
CSGs (\emph{green crosses}) and $M_{\star}$-CSGs (\emph{black triangles}) samples. 
The \emph{light orange symbols} indicate the quiescent galaxies in the MaNGA catalogue. The \emph{black solid line} corresponds to the best linear fit to the $\log \sigma_{\rm e} - \log M_{\star}$ relation.}
\label{fig:sigma_Mstar}
\end{figure}

To investigate if the differences that we find between the properties of MCGs and those of CSGs are due to some dependency of the properties with stellar mass, we defined a control sample of quiescent galaxies with similar $M_{\star}$ ($M_{\star}$-CSGs). As illustrated in Fig.\,\ref{fig:sigma_Mstar}, we fitted the $\log \sigma_{\rm e}- \log M_{\star}$ relation of a general sample of MaNGA quiescent galaxies, and selected the $M_{\star}$-CSGs from a sub-sample of objects that lie within $\pm 0.5\,\sigma_{fit}$ from the best-fit relation. 

The fraction of $M_{\star}$-CSGs that are slow-rotators is 11.4\% (8 out of 70 galaxies), which is lower than the fraction of slow-rotator CSGs (28.6\%), and more compatible with the fraction that we find for MCGs (5.7\%). We find that 17 $M_{\star}$-CSGs are spirals (24.3\%), 36 are S0s (51.4\%) and 17 are classified as ellipticals (24.3\%). The fraction of $M_{\star}$-CSG spirals is 5.7 times larger than the fraction of spirals in the compact sample. 

In Table~\ref{tab:control_samples}, we summarize the results of the comparison between MCGs, CSGs and $M_{\star}$-CSGs.

\renewcommand{\arraystretch}{1.2}
\begin{table*}
    \centering
    \caption{Summary of the properties of MCGs and control sample galaxies. {\bf (1)} Galaxy property; 
    {\bf (2)} median value of the MCG property distribution; {\bf (3)} median value of the CSG property distribution; {\bf (4)} $p$-value of a KS test comparing the distributions of the MCG and CSG properties; {\bf (5)} median value of the $M_{\star}$-CSG property distribution; {\bf (6)} $p$-value of a KS test comparing the distributions of the MCG and $M_{\star}$-CSG properties.}
    \begin{tabular}{l|c|cc|cc}
    \hline
      & MCGs & 
    \multicolumn{2}{c|}{CSGs} & 
    \multicolumn{2}{c}{$M_{\star}$-CSGs} \\
    \cline{2-6}
     Property & \emph{median} & \emph{median} & $p$-value & \emph{median} & $p$-value \\
     \multicolumn{1}{c|}{(1)} & (2) & (3) & (4) & (5) & (6) \\
    \hline \hline
    $\tau$                  &  ${-0.46 \pm 0.03}$ &  $\mathbf{-0.17 \pm 0.03}$ &      $\mathbf{< 10^{-10}}$ & $\mathbf{-0.29 \pm 0.03^a}$ & $\mathbf{2 \times 10^{-5}}$ \\
    $\tau$ (fast-rotators)  &  ${-0.48 \pm 0.03}$ &  $\mathbf{-0.24 \pm 0.03}$ & $\mathbf{7 \times 10^{-9}}$& $\mathbf{-0.32 \pm 0.03^a}$ & $\mathbf{4 \times 10^{-5}}$ \\
    B/T                     &   $\ 0.69 \pm 0.06$ &          $\ 0.73 \pm 0.04$ &                       0.22 & $\mathbf{\ 0.49 \pm 0.06}$&            $\mathbf{0.01}$ \\
    
     T-type                     &   $-2.32 \pm 0.04$ &          $-2.20 \pm 0.06$ &     0.05 & $\mathbf{-1.71 \pm 0.25}$&            $\mathbf{1 \times 10^{-6}}$ \\
     
     $p_{\rm S0}$ (T-type < 0)                     &   $\ 0.63 \pm 0.05$ &          $\mathbf{\ 0.23 \pm 0.05}$ &     $\mathbf{5 \times 10^{-7}}$ & $\ 0.80 \pm 0.03$&            $0.10$ \\
    
    age$_{\rm LW}\ ({\rm Gyr}, 0 \leq R < 0.5 r_{\rm e})$  &  ${\ 8.26 \pm 0.23}$ & 
       ${\ 7.57 \pm 0.26}$ & ${0.08}$ & 
       ${\ 7.95 \pm 0.29}$ & $0.36$\\
    age$_{\rm LW}\ ({\rm Gyr}, 0.5 \leq R < 1 r_{\rm e})$  &  ${\ 8.72 \pm 0.19}$ & 
       $\mathbf{\ 7.37 \pm 0.31}$ & $\mathbf{1 \times 10^{-3}}$ & 
       $\mathbf{\ 7.11 \pm 0.48}$ & $\mathbf{1 \times 10^{-4}}$\\
    $\Delta\,\log ($age$_{\rm LW}$ / Gyr)  &  $\ 0.02 \pm 0.01$ & 
       $\mathbf{-0.02 \pm 0.02}$&    $\mathbf{3 \times 10^{-3}}$ &     
       $\mathbf{-0.02 \pm 0.02}$&    $\mathbf{1 \times 10^{-3}}$ \\
    
    ${\rm [Z/H]}_{\rm LW}\ (0 \leq R < 0.5 r_{\rm e})$  &  ${+0.13 \pm 0.02}$ & $\mathbf{+0.03 \pm 0.01}$ & $\mathbf{5 \times 10^{-5}}$ & $\mathbf{-0.06 \pm 0.02}$ & $\mathbf{< 10^{-10}}$\\
    ${\rm [Z/H]}_{\rm LW}\ (0.5 \leq R < 1 r_{\rm e})$   &  ${+0.02 \pm 0.02}$ & $\mathbf{-0.07 \pm 0.01}$ & $\mathbf{2 \times 10^{-5}}$ & $\mathbf{-0.15 \pm 0.02}$ & $\mathbf{2 \times 10^{-10}}$\\
    $\Delta {\rm [Z/H]}_{\rm LW}$  & ${-0.11 \pm 0.01}$ & ${-0.11 \pm 0.01}$& ${0.88}$ & $-0.09 \pm 0.01$ & 0.36 \\
    
    ${\rm [\alpha/Fe]}_{\rm LW}\ (0 \leq R < 0.5 r_{\rm e})$  & ${+0.19 \pm 0.01}$ & $\mathbf{+0.15 \pm 0.01}$ & $\mathbf{5 \times 10^{-4}}$ & $\mathbf{\ 0.04 \pm 0.01}$ & $\mathbf{< 10^{-10}}$\\
    ${\rm [\alpha/Fe]}_{\rm LW}\ (0.5 \leq R < 1 r_{\rm e})$   & ${+0.17 \pm 0.01}$ & $\mathbf{+0.08 \pm 0.02}$ & $\mathbf{4 \times 10^{-4}}$ & $\mathbf{\ 0.03 \pm 0.01}$ & $\mathbf{< 10^{-10}}$\\
    $\Delta {\rm [\alpha/Fe]}_{\rm LW}$   & ${-0.02 \pm 0.01}$ & ${-0.02 \pm 0.01}$ &  $0.10$  & $-0.02 \pm 0.01$ & 0.14\\

    $Q_{\rm group}$         & $-1.10 \pm 0.22$ & $-0.72 \pm 0.17$ & 0.14 & $-0.94 \pm 0.20$ & 0.59\\
    \hline
    \end{tabular}
    \parbox{\hsize}{{\bf Notes:}
    Results with $p$-values\,$< 0.05$ are highlighted in boldface. The errors of the median values were estimated by bootstrapping the samples 1000 times.
    \\ {\bf{$^a$:}} 32 of the 70 $M_{\star}$-CSGs have more than 30\% of the spaxels with velocity dispersion lower than 80\,km/s. As these spaxels are excluded from the computation of $\tau$ and the remaining spaxels are usually located near the center, where the fraction of z-tube orbits is lower (see Fig.\,7 of \citealt{rottgers14}), $\tau$ in $M_{\star}$-CSGs might be biased against small values.}
    \label{tab:control_samples}
\end{table*}
\renewcommand{\arraystretch}{1}

\section{Discussion}\label{sec:discussion}

 We compared the kinematics, morphology, stellar populations properties and environment of MCGs with those of a control sample of average-sized quiescent galaxies with similar velocity dispersion. We find that MCGs have, on average, higher $\lambda_{\rm e}$ and show a stronger anti-correlation between $h_3$ and $V/\sigma$. The morphology of MCGs and  CSGs is also distinct: two thirds of the MCGs are classified as S0s or early (red) spirals, while about 70\% of the CSGs are classified as ellipticals. The stellar population analysis revealed that MCGs have similar ages to CSGs of the same velocity dispersion, but they are more metal-rich and $\alpha$-enhanced than CSGs. MCGs have flatter age gradients compared to the CSGs, but no significant difference is observed between their metallicity and [$\alpha/$Fe] gradients.  Finally, we find that CSGs are more likely to be centrals and they tend to reside in more massive haloes, but there are no differences in the local environment of MCGs and  CSGs. 

\subsection{Efficiency and duration of the last episode of star formation in MCGs}\label{sec:starpops}

As shown in Figs.\,\ref{fig:metal} and \ref{fig:aFe}, MCGs have higher metallicities and [$\alpha/$Fe] than CSGs, indicating that the last episode of star formation in the inner (within $0.5\,r_{\rm e}$) and outer (between $0.5\,r_{\rm e}$ and $1.0\,r_{\rm e}$) parts of MCGs was efficient and occurred in short timescales. 
The age gradients of MCGs are shallow and their [$\alpha/$Fe] gradients are flat, suggesting similar star-formation timescales throughout the MCG up to $1.0\,r_{\rm e}$. The short timescales and high efficiency of star-formation in MCGs suggest they might have gone through a post-starburst phase, as we discuss in Sect.\,\ref{sec:discussion:psbs}.

The negative metallicity gradients in MCGs can indicate either that star-formation efficiency was higher in the inner $0.5\,r_{\rm e}$ of MCGs compared to within $0.5\,r_{\rm e}-1.0\,r_{\rm e}$; or that the galaxy  was already more metal-rich in the inner $0.5\,r_{\rm e}$ prior to the last episode of star formation. The shallow slope can be a consequence of an efficient radial mixing of the material caused by interactions or strong stellar feedback \citep{ma17}. Alternatively, if the quenching of the star formation is fast, and occurs approximately at the same time inside $1.0\,r_{\rm e}$ (as the age and [$\alpha/$Fe] gradients suggest), there would be no time to build a steep metallicity gradient. Nonetheless, observations of high redshift galaxies suggest gas metallicity gradients were flatter in the past (see \citealt{maiolino19} for a review), which is consistent with the stellar metallicity gradients we measured for both MCGs and CSGs.

\subsection{Constrains on the formation of compact galaxies}

 In hydrodynamical simulations, compact galaxies are formed when large amounts of gas are driven to the center of a galaxy due to major mergers, violent disk instabilities or counter-rotating streams, fueling a nuclear starburst \citep{wuyts10,ceverino15,zolotov15}. The star-formation is quenched soon after due to a combination of gas consumption and feedback \citep{tacchela16}. We note that violent disk instabilities and counter-rotating streams are phenomena restricted to the z\,$\gtrsim$\,1 universe, as disks stabilize at lower redshifts \citep{cacciato12}.
 
 Hydrodynamical simulations of binary gas-rich mergers show that during the interaction, gas within some characteristic radius of the primary galaxy loses angular momentum and flows inward, fueling a nuclear starburst which grows the bulge. A disk is reformed after the merger from gas at large radii that was unable to lose angular momentum efficiently \citep{hopkins09a}. The bulge to total ratio of the remnant is a function of the gas mass fraction ($f_{\rm gas}$) of the progenitors and their mass ratio \citep{hopkins10b}. The remnants of very gas-rich major mergers are disk dominated systems, as gas loses angular momentum less efficiently at high $f_{\rm gas}$ (because there are less stars to transfer angular momentum to).  
 
 Simulations of gas-rich major mergers, however, predict that, although the mixing of stars formed prior to the merger flattens the metallicity gradient, centrally concentrated star-formation generates a strong negative gradient \citep{hopkins09}. Strong AGN feedback is necessary to prevent the steepening of the metallicity gradient \citep{taylor17}. A link between compaction and AGN activity is supported by observations: compact star-forming galaxies at $z$\,$\sim$\,2 are more likely to host an AGN than extended star forming galaxies of similar mass \citep{kocevski17}. Similar results are found at $z$\,$\sim$\,0, where star-forming galaxies with compact cores have a higher probability of hosting WISE selected AGN \citep{woo19}. Nonetheless, evidence of AGN-driven winds suppressing star formation is still elusive \citep{schulze19,shangguan20,stacey20}, and observations of post-starburst galaxies show that, although AGN are more common in these objects than in star-forming galaxies, there is a delay between the peaks of starburst and AGN activity \citep{yesuf14}, so AGN feedback is unlikely to play a prominent role in the quenching of starbursts \citep{sell14}. The increased fraction of AGN in compact star-forming and post-starburst galaxies might just be indicating an increased amount of gas available to fuel the central black hole \citep{greene20}. 

A recent study of the molecular gas properties in a massive compact star-forming galaxy at $z = 2.234$ by \citet{spilker19} presented evidence on how these objects quench. Molecular gas fractions and gas depletion times were found to be lower in the inner 1\,--\,2\,kpc than at 3\,--\,4\,kpc, suggesting the star formation quenches inside-out. The depletion time at the center, however, is just 10--15\,Myr, only increasing by a factor of $\sim 2$ at large radii. In another recent study, \citet{Jafariyazani20} reported the detection of flat age and [Mg/Fe] gradients and a shallow negative [Fe/H] gradient in a compact quiescent galaxy at $z = 1.98$. Although the age distribution of MCGs is inconsistent with the bulk of them being the descendants of $z \gtrsim 2$ compact star forming galaxies, the formation scenario proposed by \citet{spilker19} and the gradients reported by \citet{Jafariyazani20} are consistent with what we infer based on the fossil record imprinted in stellar populations of MCGs. This suggests that MCGs form in a similar manner to z\,$\sim$\,2 compact quiescent galaxies, but under less extreme conditions. For example, while the progenitors of $z \sim 2$ compact quiescent galaxies experienced a strong  (SFR\,$\gtrsim$\,1000\,M$_\odot$\,yr$^{-1}$) and short ($\sim$\,50\,Myr) burst of star formation before quenching \citep{valentino20}, the [$\alpha$/Fe] distribution of MCGs points to longer star formation timescales and less intense bursts.

\subsection{Did MCGs go through a post-starburst phase in the past?}
\label{sec:discussion:psbs}

MCGs have a broad age distribution, with luminosity-weighted ages ranging from $\sim$\,4\,Gyr to 10\,Gyr in the inner 1.0\,$r_{\rm e}$ (a formation epoch of $z$\,$\sim$\,0.4\,--\,2.0). While in the inner $0.5\,r_{\rm e}$ the ages of CSGs and MCGs are consistent, CSGs are younger at larger radii. Furthermore, MCGs are more metal rich and alpha enhanced in their centers than CSGs. We argue that these differences cannot be attributed to a difference in accretion histories. They are, instead, a result of the progenitors of compact galaxies and CSGs following distinct quenching routes. We base our interpretation on a study of a large sample of quiescent galaxies at $z\,\sim$\,0.7 by \citet{wu18}, which suggests that there exists a fast and a slow path to quiescence. They argue that galaxies following the fast path experience significant structural change and evolve into compact post-starbursts, while galaxies following the slow path have their star formation gradually shut off and do not necessarily change their structure.

The star forming progenitors of MCGs experiencing a centrally concentrated starburst followed by an abrupt quenching of the star formation can explain the differences in ages, metallicity and [$\alpha$/Fe] compared to CSGs. Typically, intermediate redshift PSBs have either flat or shallow positive age gradients \citep{hunt18,deugenio20,setton20}, in agreement with the age and [$\alpha/$Fe] gradients of MCGs. Moreover, observations of $z$\, $\sim1-1.4$ compact PSBs indicate that these objects present higher stellar velocity dispersion than passive galaxies of the same mass \citep{maltby19}, as is the case of MCGs (see Fig.\,\ref{fig:sigma_Mstar}). Our findings are also in agreement with \citet{zahid16}, which argued that MCGs and intermediate redshift post-starburst galaxies are linked based on a comparison of their stellar ages and central velocity dispersions.


\subsection{The accretion history of MCGs and  CSGs}\label{sec:assembly}

Constraints on the accretion history of MCGs can be gathered from their kinematic properties. Hydrodynamical simulations show that slow- and fast-rotators that experienced late gas-poor mergers show either a weak $h_3$\,--\,$V/\sigma$ anti-correlation or no anti-correlation at all \citep{hoffman09,naab14,rottgers14}. In constrast, fast rotators which experienced late gas-rich mergers present a strong $h_3$\,--\,$V/\sigma$ anti-correlation. Thus, the strong  $h_3$\,--\,$V/\sigma$ anti-correlation observed in MCGs suggests that these systems have had a quiet accretion history, experiencing only few, if any, dry minor mergers since the last star formation episode.  On the other hand, most of the CSGs show a weak anti-correlation, pointing to an increased importance of dry mergers in the mass assembly of these objects. The relative importance of dry major and minor mergers cannot be constrained from kinematics, as simulations show that repeated dry minor mergers can also effectively reduce the rotational support of a galaxy producing an elliptical remnant, provided the accreted mass is of the order of 40\% of the mass of the initial disk galaxy \citep{bournaud07}. 

Dry major and minor mergers affect the stellar population gradients in a distinct manner. Dry major mergers lead to a flattening of gradients \citep{Kobayashi:2004}, but the amount of flattening depends on the differences between the gradients of the progenitors \citep{dimatteo09}. In contrast, dry minor mergers steepen pre-existing metallicity gradients \citep{Hirschmann.etal:2015}, although gradients become flat at large radii as accreted stars become dominant \citep{cook16,oyarzun19,zibetti20}. Our observations cover only the inner 1\,$r_{\rm e}$, however, where simulations point to the in-situ population being dominant (see Fig.\,3 of \citealt{cook16}), which explains why we do not observe significant differences in the metallicity gradients of MCGs and CSGs, despite their distinct accretion histories. 

The differences in accretion histories of MCGs and CSGs might be driven, at least in part, by their environments. MCGs are more likely to be satellites in massive halos, where the large velocity dispersion of the galaxies inhibit merging. Moreover, recent work suggests that galaxies in denser environments experience significant minor merging earlier \citep{cole20}; MCGs tend to be central galaxies of lower mass halos, where significant merging might not have yet occurred.

In summary, while the accreted stellar mass fraction and the stellar population properties of accreted starts cannot be constrained based on observations of the inner $\sim$\,1\,$r_e$, where in-situ stars dominate, simulations points to dry mergers significantly affecting the orbital distribution of these central stars, lowering the rotational support of the galaxy. Considering compact galaxies display elevated rotational support, these objects must have had a quiet accretion history since they quenched.

\subsection{Relation with S0s}

The fraction of S0 galaxies in the MCG sample is very high (62\%) compared to that of the CSGs (16.7\%), but it is more similar to the fraction of $M_{\star}$-CSGs that are S0s (51.4\%). The bulge to total light ratios and velocity dispersion of MCG S0s tend to be larger than those of $M_{\star}$-CSG S0s; so, \emph{could MCGs be S0s with large bulges and high-$\sigma_{\rm e}$?}

The differences in the stellar population properties of the S0s in the MCG and $M_{\star}$-CSG samples indicate that they have had distinct assembly histories. MCG S0s are older, more metal-rich and $\alpha$-enhanced than $M_{\star}$-CSG S0s.

Recent studies of large samples of early-type galaxies based on integral field spectroscopic data reported flat or shallow negative age and mildly negative metallicity gradients. Typical light weighted metallicity gradients are of the order of $\Delta [{\rm Z/H}]_{\rm LW}$\,$\sim$\,--0.1\,dex/$r_e$, while light weighted age gradients are slightly flatter, varying between $\Delta$\,log(age$_{LW}$)\,$\sim$\,--0.03 and $\sim$\,--0.08\,dex/$r_e$ \citep{gonzalez-delgado15,goddard17,zheng17,li18,lacerna20}. When analysed separately, S0s and slow- and fast-rotating ellipticals show significant differences in their stellar population gradients \citep{lacerna20,krajnovic20}. In particular, massive S0s tend to have steeper age gradients and shallower metallicity gradients than fast-rotating ellipticals of the same luminosity and velocity dispersion \citep{dominguez-sanchez20}. In this regard, MCGs are unlike typical S0s, as their metallicity and age gradients resemble those of fast-rotating ellipticals.

In conclusion, MCGs appear to be an homogeneous class of objects in terms of their kinematics, morphology and stellar population properties, but their characteristics point to a population in between S0s and elliptical galaxies. Therefore, it is unlikely that MCGs simply represent the high-$\sigma_{\rm e}$, high-B/T end of the S0 population.

\subsection{Relation with red spirals}

In the local universe, there exists a rare population of massive red spiral galaxies with old metal rich centers \citep{masters10}. Spatially resolved spectroscopic observations of the inner 1\,$r_{\rm e}$ of these massive red spiral revealed stellar populations properties more similar to elliptical galaxies than to blue spirals: shallow negative age gradients, flat metallicity gradients and alpha enhanced centers \citep{robaina12,hao19}. These are broadly similar to what we observe in compact galaxies, and considering that red spirals harbour compact cores \citep{guo20}, one might question if MCGs and red spirals share a common origin. For example, considering a significant fraction of massive red spirals exhibit signatures of interactions \citep{guo20}, it is plausible that red spirals are rejuvenated compact galaxies which have since quenched again. Star forming galaxies with old quenched bulges have been observed at intermediate redshifts \citep{mancini19}, and there are blue spirals with compact cores in the local universe \citep{fang13,guo20}, supporting a rejuvenation scenario. Another possibility is that, similarly to compact galaxies, red spirals were formed in major mergers at intermediate and high redshift, although with significantly larger gas fractions \citep{hao19} such that gas inflows are inhibited and consequently the remnants are disk dominated systems instead of compact galaxies. In a recent work, \citet{peschken20} identified the formation of disk dominated remnants of very gas rich mergers ($f_{\rm gas}$\,$\sim$\,0.7) in the Illustris simulation. They found that star formation continued in the remnants for a long period, fueled by gas scattered to the halo during the merger which is slowly accreted as it cools.

\section{Summary and Conclusions} \label{sec:conclusions}

We characterized the 2D kinematics, morphology and stellar population properties of a sample of 70 massive compact quiescent galaxies with integral field spectroscopic data publicly available as part of the MaNGA survey. We compared the properties of the massive compact galaxies to a control sample, matched in effective velocity dispersion, of 70 massive quiescent galaxies with median sizes. We found that:

\begin{itemize}
    \item MCGs are fast rotators (96$\%$), while  CSGs are composed of a mix of fast (71$\%$) and slow rotators (29$\%$);
    
    \item Compared to fast-rotating CSGs, MCGs are more likely to have $V/\sigma$ and $h_3$ anti-correlated. This is evidence of increased rotational support;
    
    \item 62$\%$ of MCGs are S0s, 34$\%$ are ellipticals. In contrast, 72$\%$ of CSGs are ellipticals and 14$\%$ are S0s;
    
    \item MCGs and CSGs have similar ages inside $0.5\,r_{\rm e}$, but MCGs are more metal-rich and $\alpha$-enhanced; 
    
    \item Both MCGs and CSGs have shallow negative age gradients and flat [$\alpha$/Fe] gradients. MCGs and CSGs have flat age gradients, but the distribution of CSG $\Delta\,\log($age) have a significantly larger tail towards negative values;
    
    \item The high MCG metallicities and [$\alpha$/Fe] suggest that these objects experienced a short and efficient burst of star formation at $z \sim 0.4 - 2.0$ followed by a post-starburst phase;

    \item The sizes of compact galaxies are well reproduced by hydrodynamical simulations of gas rich binary major mergers. These simulations predict steep negative metallicity gradients due to centrally concentrated star formation, in contrast to the approximately flat gradients we observe; 
    
    \item The MCG kinematics, morphology and stellar population properties are very homogeneous and they appear to be a population in between S0s and elliptical galaxies.

\end{itemize}

%

\section*{Acknowledgements}
The authors thank the anonymous referee for their comments and suggestions, which led to an improved version of the manuscript.
ASM acknowledges the financial support from the Brazilian national council for scientific and technological development (CNPq). MT thanks the support of CNPq (process \#307675/2018-1) and the program L'Or\'eal UNESCO ABC \emph{Para Mulheres na Ci\^encia}. RR thanks CNPq, CAPES and FAPERGS. ACS acknowledges support from CNPq-403580/2016-1, L'Or\'eal UNESCO ABC \emph{Para Mulheres na Ci\^encia}, CNPq-311153/2018-6,  PqG/FAPERGS-17/2551-0001 and FAPERGS/CAPES 19/2551-0000696-9. CF acknowledges the financial support form CNPQ (processes 433615/2018-4 e 311032/2017-6. TVR also acknowledges CNPq (process 306790/2019-0). We acknowledge the use of SDSS data (\url{http://www.sdss.org/collaboration/credits.html}). This research made use of Marvin, a core Python package and web framework for MaNGA data (\url{https://dr16.sdss.org/marvin/}).

\section*{Data availability}
The data underlying this article are available at
https://www.sdss.org/surveys/manga/ for the MaNGA Survey
DR15; and at https://classic.sdss.org/dr14/
for the SDSS DR14. Additional data
generated by the analyses in this work are available upon request
to the corresponding author.





\bibliography{bibliografia}






\bsp	
\label{lastpage}
\appendix

\section{Spatial resolution of MaNGA data}
\label{app:seeing}

The resolution of MaNGA observations is a concern when analysing the spatially-resolved properties of galaxies with small sizes like MCGs. Therefore, here we investigate if our results and conclusions are affected by this issue. 

\begin{figure}
\centering
    \includegraphics[width=0.95\hsize]{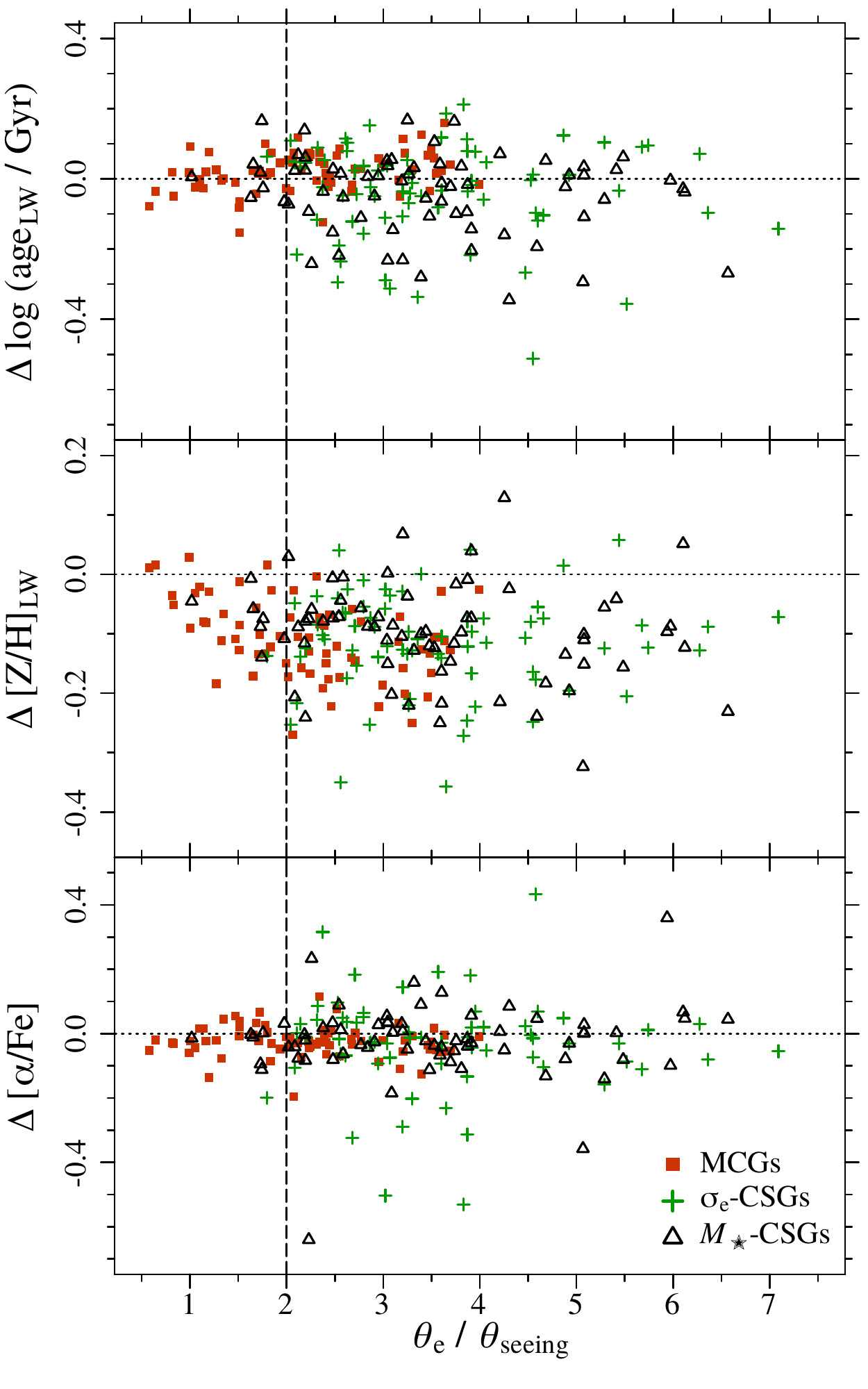}
    \caption{Gradients of age (\emph{upper panel}), metallicity (\emph{middle panel}), and $\alpha-$enhacement (\emph{bottom panel}) as a  function of the $\theta_{\rm e} / \theta_{\rm seeing}$ ratio. The \emph{vertical dashed lines} indicate $\theta_{\rm e} / \theta_{\rm seeing} = 2$.}
    \label{fig:grad_seeing_app}
\end{figure}

In Fig.\,\ref{fig:grad_seeing_app}, we show how the gradients of the stellar populations properties of compact and control sample galaxies vary with the $\theta_{\rm e}/\theta_{\rm seeing}$ ratio, where $\theta_{\rm seeing}$ is the average seeing during the observations of our galaxies and $\theta_{\rm e}/2$ is half of their effective radius in arcseconds. Only a few CSGs have $\theta_{\rm e} / \theta_{\rm seeing} < 2$ (one CSG and 7 $M_{\star}-$CSGs), but 40\% of the MCGs are not resolved within $\theta_{\rm e}/2$ (28 MCGs). 

\begin{figure}
\centering
    \includegraphics[width=0.95\hsize]{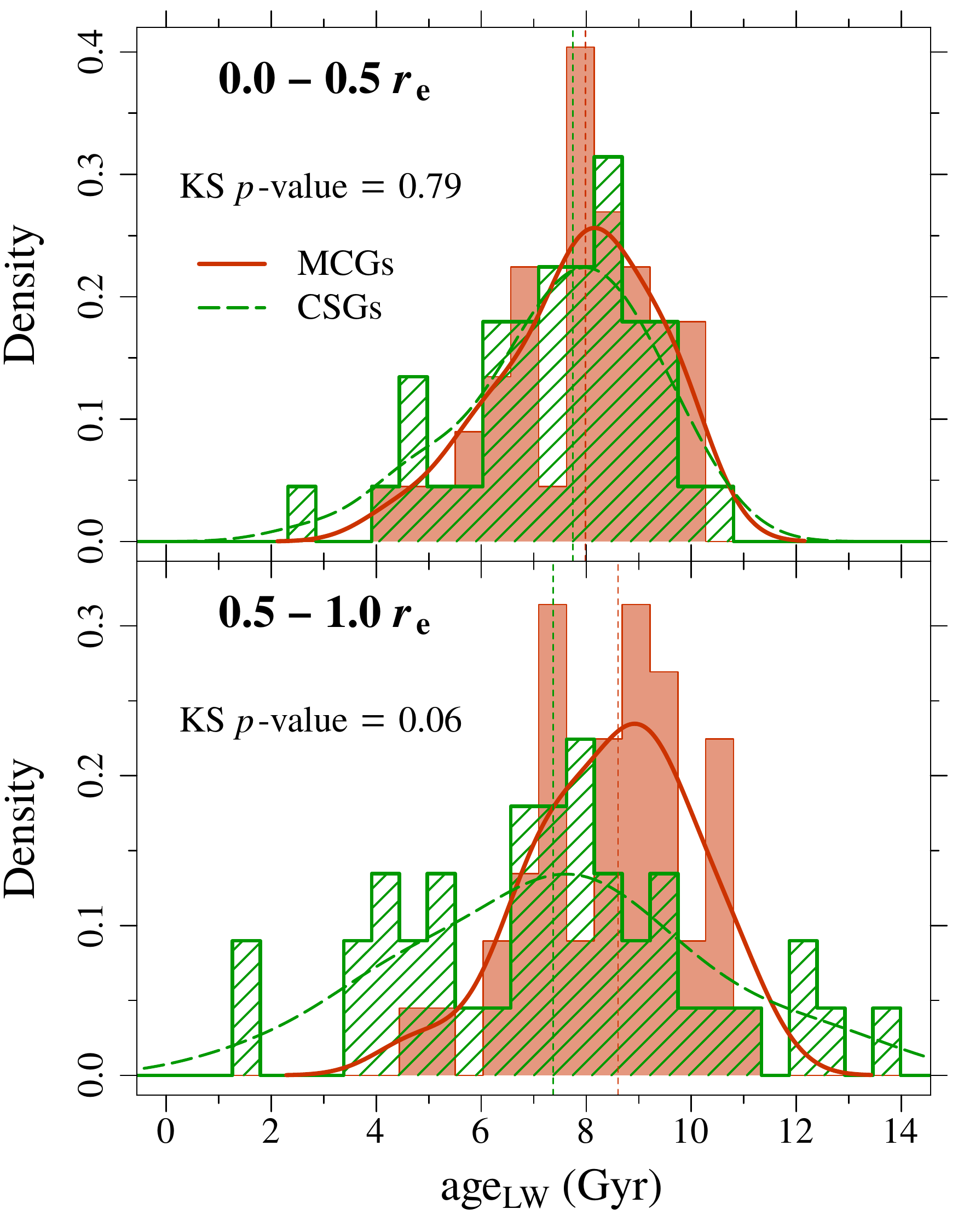}
    \includegraphics[width=0.95\hsize]{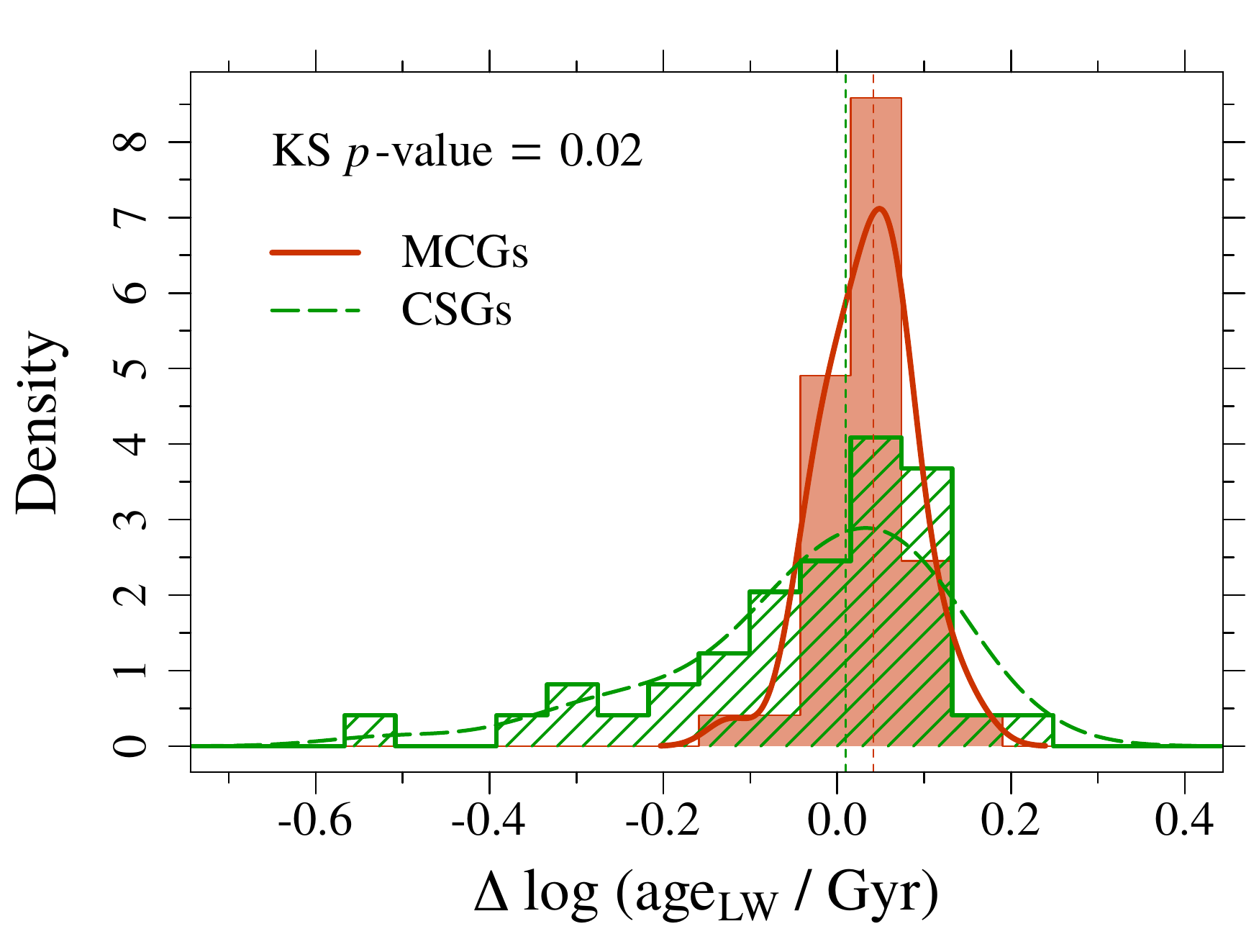}
    \caption{Same as in Fig.~\ref{fig:age}, but for 42 compact galaxies that have $\theta_{\rm e} / \theta_{\rm seeing} > 2$. The control sample was redefined to match the velocity-dispersion distribution of the spatially-resolved compact galaxies.}
    \label{fig:age_app}
\end{figure}

\begin{figure}
\centering
    \includegraphics[width=0.95\hsize]{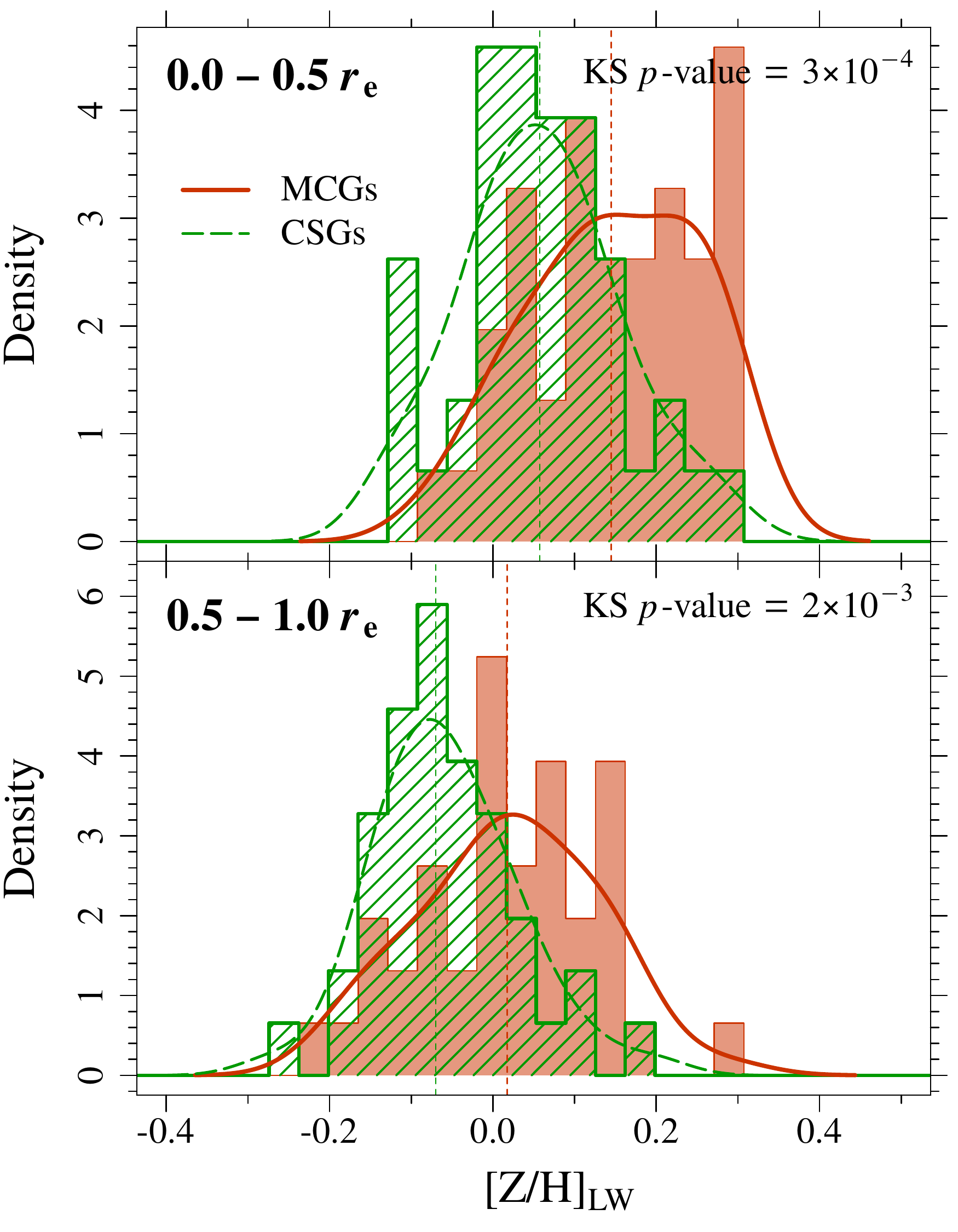}
    \includegraphics[width=0.95\hsize]{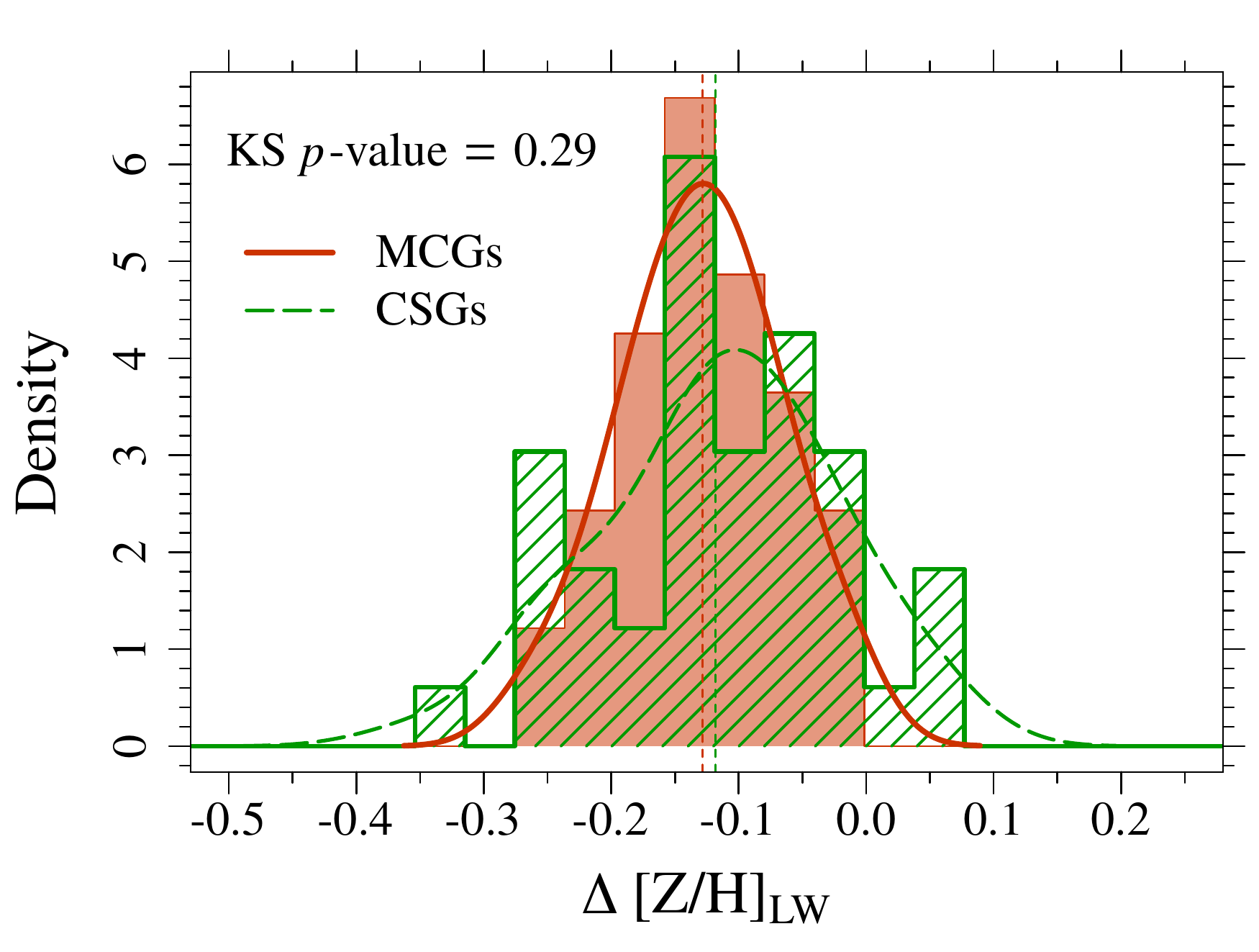}
    \caption{Same as in Fig.~\ref{fig:metal}, but for 42 compact galaxies that have $\theta_{\rm e} / \theta_{\rm seeing} > 2$. The control sample was redefined to match the velocity-dispersion distribution of the spatially-resolved compact galaxies.}
    \label{fig:metal_app}
\end{figure}

\begin{figure}
\centering
    \includegraphics[width=0.95\hsize]{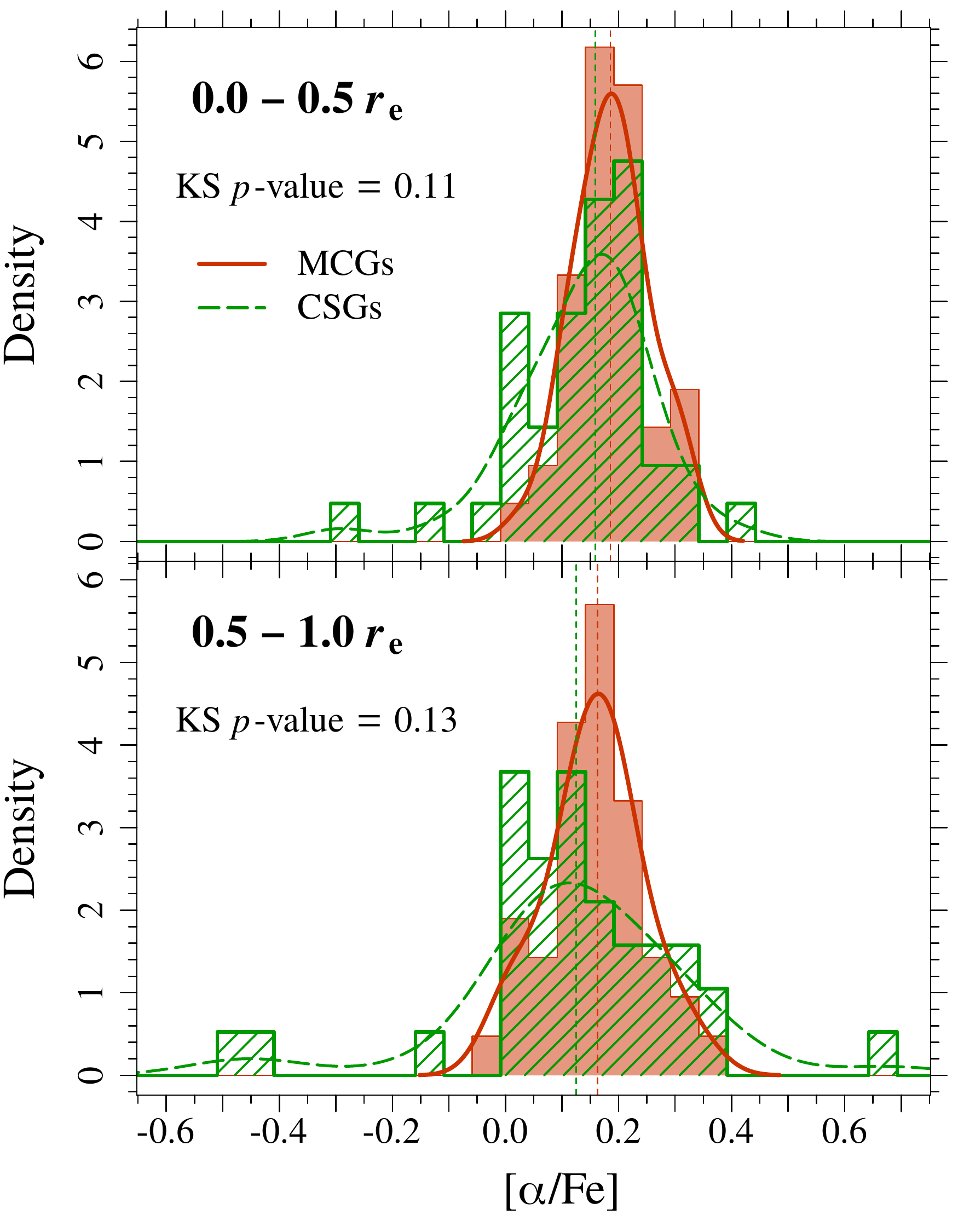} 
    \includegraphics[width=0.95\hsize]{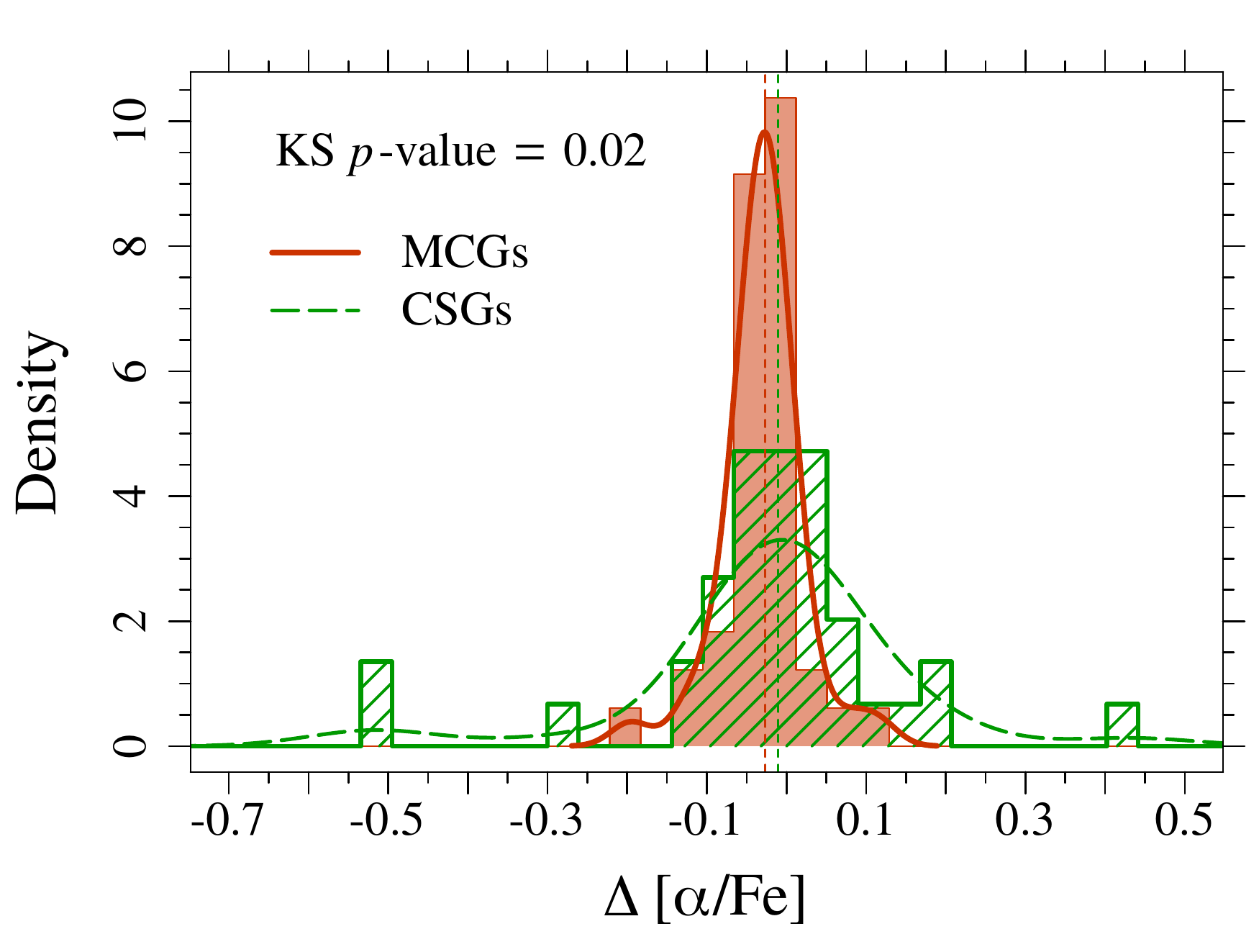} 
    \caption{Same as in Fig.~\ref{fig:aFe}, but for 42 compact galaxies that have $\theta_{\rm e} / \theta_{\rm seeing} > 2$. The control sample was redefined to match the velocity-dispersion distribution of the spatially-resolved compact galaxies.}
\label{fig:aFe_app}
\end{figure}

As shown in Table~\ref{tab:corr_tests}, Pearson and Kendall correlation tests indicate that the MCG age and metallicity gradients have a correlation with $\theta_{\rm e}/\theta_{\rm seeing}$. To investigate if this dependency affects our results, we excluded the MCGs with  $\theta_{\rm e} / \theta_{\rm seeing} < 2$ and analysed the gradients the resulting sub-sample of 42 resolved MCGs. To compare the gradients of the resolved MCG sub-sample with those of the CSGs, we selected 42 galaxies among CSGs that have similar velocity dispersions. The results of the comparisons are presented in Figs.\,\ref{fig:age_app}, \ref{fig:metal_app}, and \ref{fig:aFe_app}. It can be seen that, after removing the unresolved MCGs from the sample, we get similar results, with the exception of the [$\alpha$/Fe] gradients which are now marginally different. This difference is due to a change in the distribution of [$\alpha$/Fe] gradients of CSGs, which now show a tail extending to positive values, while previously the tails were symmetric. The distribution of [$\alpha$/Fe] gradients of MCGs is unchanged.

\begin{table*}
    \centering
    \caption{Pearson and Kendall correlation coefficients, $\rho$ and $\tau$, of the relations between stellar-population gradients and the $\theta_{\rm e}/\theta_{\rm seeing}$ ratio. 
    Results with $p$-values$\,< 0.05$ are highlighted in boldface}
    \begin{tabular}{lcccccc}
    \hline
       & \multicolumn{2}{c}{MCGs} & \multicolumn{2}{c}{CSGs} & \multicolumn{2}{c}{$M_{\star}-$CSGs} \\
    \hline\hline
    \multirow{2}{*}{$\Delta \log({\rm age}_{\rm LW}/{\rm Gyr})$}  
           & $\mathbf{\rho = 0.32}$ & $p$-value$\,= \mathbf{0.01}$ & $\rho = -0.05$ & $p$-value$\,= 0.67$ & $\rho = -0.16$ & $p$-value$\,= 0.18$ \\
           & $\mathbf{\tau = 0.20}$ & $p$-value$\,= \mathbf{0.01}$ & $\tau = -0.03$ & $p$-value$\,= 0.76$ & $\tau = -0.12$ & $p$-value$\,= 0.14$ \\
    \hline
    \multirow{2}{*}{$\Delta [{\rm Z/H}]_{\rm LW}$}  
           & $\mathbf{\rho = -0.41}$ & $p$-value$\,= \mathbf{4 \times 10^{-4}}$ & $\rho =  0.03$ & $p$-value$\,= 0.83$ & $\rho = -0.22$ & $p$-value$\,= 0.07$ \\
           & $\mathbf{\tau = -0.29}$ & $p$-value$\,= \mathbf{4 \times 10^{-4}}$ & $\tau = -0.02$ & $p$-value$\,= 0.79$ & $\tau = -0.16$ & $p$-value$\,= 0.05$ \\
    \hline
    \multirow{2}{*}{$\Delta [\alpha/{\rm Fe}]$}  
           & $\rho = -0.10$ & $p$-value$\,= 0.40$ & $\rho = -0.03$ & $p$-value$\,= 0.83$ & $\rho = 0.01$ & $p$-value$\,= 0.93$ \\
           & $\tau = -0.08$ & $p$-value$\,= 0.30$ & $\tau = -0.09$ & $p$-value$\,= 0.28$ & $\tau = 0.02$ & $p$-value$\,= 0.84$ \\
 
    \hline
    \end{tabular}
    \label{tab:corr_tests}
\end{table*}


\section{Kinematic and stellar population parameters} \label{app:ap_table}

Stellar population properties and kinematic parameters of compact and control galaxies are provided in table\,\ref{tab:general}. A full version of the table is available online as supplementary data.
\begin{table*}
    \centering
    \caption{Properties of the compact and control samples. References: $^1$:\citet{salim18},$^2$:\citet{simard11}.}
    \resizebox{\linewidth}{!}{%
    \begin{tabular}{lccccccccccccc}
    \hline
MaNGAID & Sample & $z$ & $\log M_\star/M_\odot$$^1$ & $\sigma_{\rm e}$ & $r_{\rm e}$$^2$ & $\lambda$ & $\tau$ & age$_{{\rm LW}-0.5r_{\rm e}}$ & age$_{{\rm LW}-1r_{\rm e}}$ & [Z/H]$_{{\rm LW}-0.5r_{\rm e}}$ & [Z/H]$_{{\rm LW}-1r_{\rm e}}$ & [$\alpha$/Fe]$_{0.5r_{\rm e}}$ & [$\alpha$/Fe]$_{1r_{\rm e}}$ \\
& & & & (km/s) & (kpc) & & & (Gyr) & (Gyr) & & & & \\
\hline \hline 
1-133987 & MCG & 0.01954 & 10.24 & 152 & 1.15 & 0.27 & -0.45 & 9.45 & 8.86 & 0.15 & 0.00 & 0.21 & 0.20 \\
1-135193 & MCG & 0.03089 & 10.68 & 191 & 1.48 & 0.12 & -0.14 & 6.89 & 7.24 & 0.27 & 0.16 & 0.13 & 0.20 \\
1-145995 & MCG & 0.04088 & 10.20 & 150 & 1.29 & 0.26 & -0.46 & 10.56 & 10.50 & -0.10 & -0.12 & 0.25 & 0.27 \\
1-149193 & MCG & 0.02762 & 10.79 & 243 & 2.28 & 0.35 & -0.64 & 5.82 & 7.08 & 0.27 & 0.10 & 0.30 & 0.23 \\
1-149878 & MCG & 0.02652 & 10.58 & 164 & 2.18 & 0.31 & -0.35 & 9.78 & 10.36 & -0.04 & -0.20 & 0.17 & 0.10 \\
\hline
  \end{tabular}}
    \parbox{\hsize}{{\bf Notes:}
   This table is available in its entirety in machine-readable form as supplementary material.}
    \label{tab:general}
\end{table*}

\section{Stellar velocity, velocity dispersion and Gauss-Hermite moment $h_3$ maps} \label{app:ap_maps}

\begin{figure*}
\centering
    \includegraphics[width=0.95\hsize]{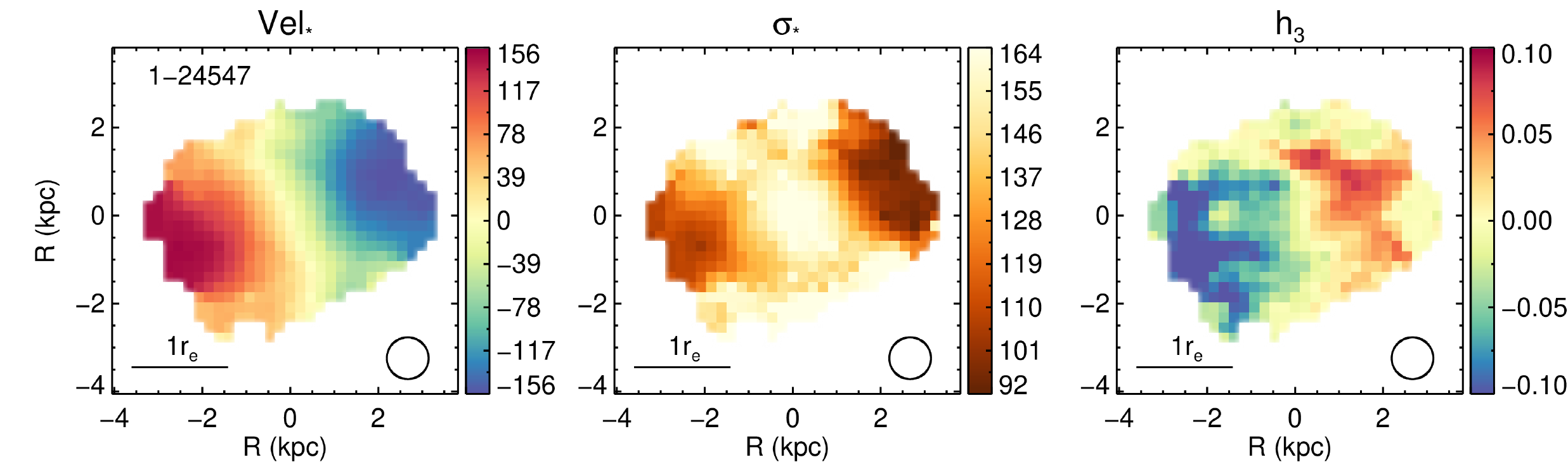}
    \includegraphics[width=0.95\hsize]{maps/1-25819.pdf}
    \includegraphics[width=0.95\hsize]{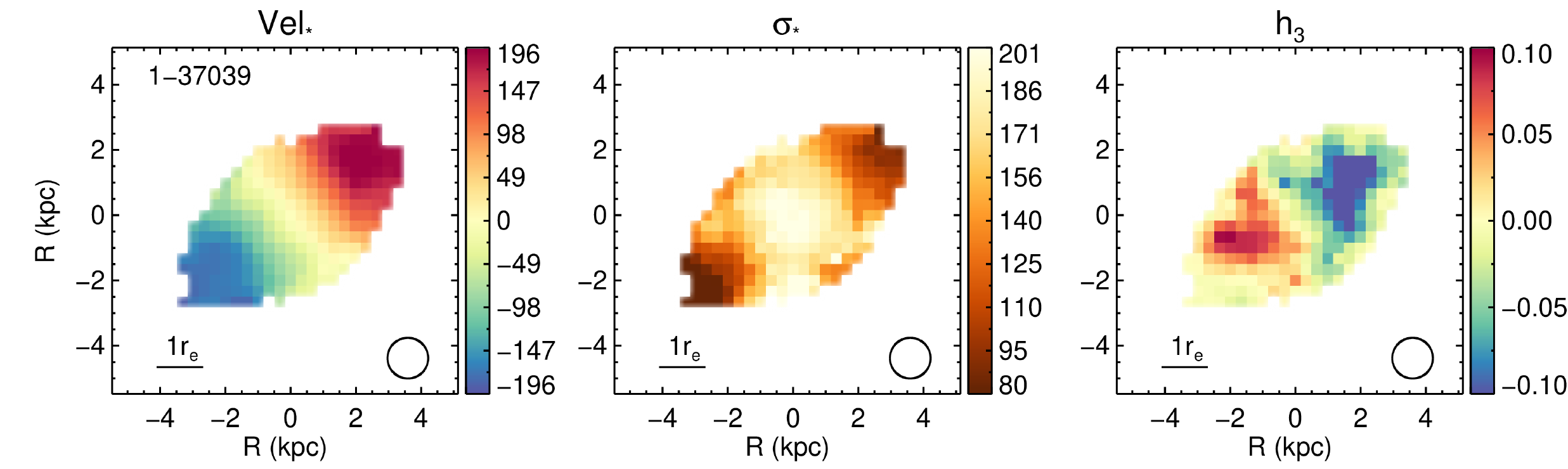}
    \includegraphics[width=0.95\hsize]{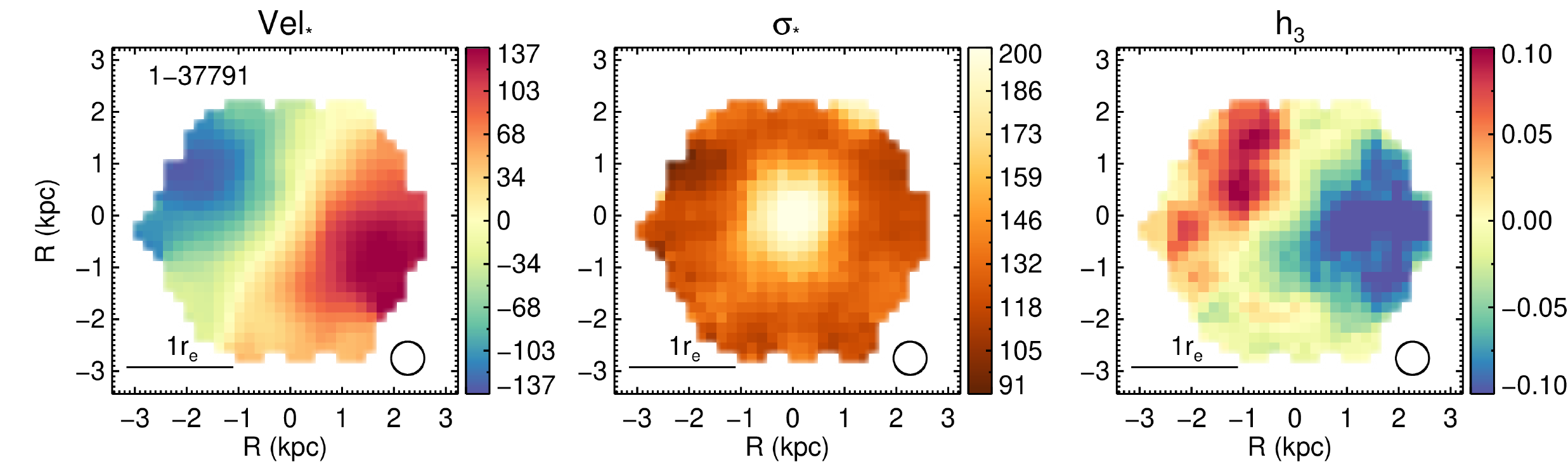}
    \caption{Stellar velocity (km\,s$^{-1}$), velocity dispersion (km\,s$^{-1}$) and Gauss-Hermite moment $h_3$ maps of MCGs. The MaNGA ID of each object is shown on the top left corner of the leftmost panel. The black circle shows the angular resolution of the observations.}
    \label{fig:kinematics_apend}
\end{figure*}

In Fig.\,\ref{fig:kinematics_apend} we show the stellar velocity (km\,s$^{-1}$), stellar velocity dispersion (km\,s$^{-1}$) and Gauss-Hermite moment $h_3$ maps for all 70 MCGs. 

\begin{figure*}
\centering
    \includegraphics[width=0.95\hsize]{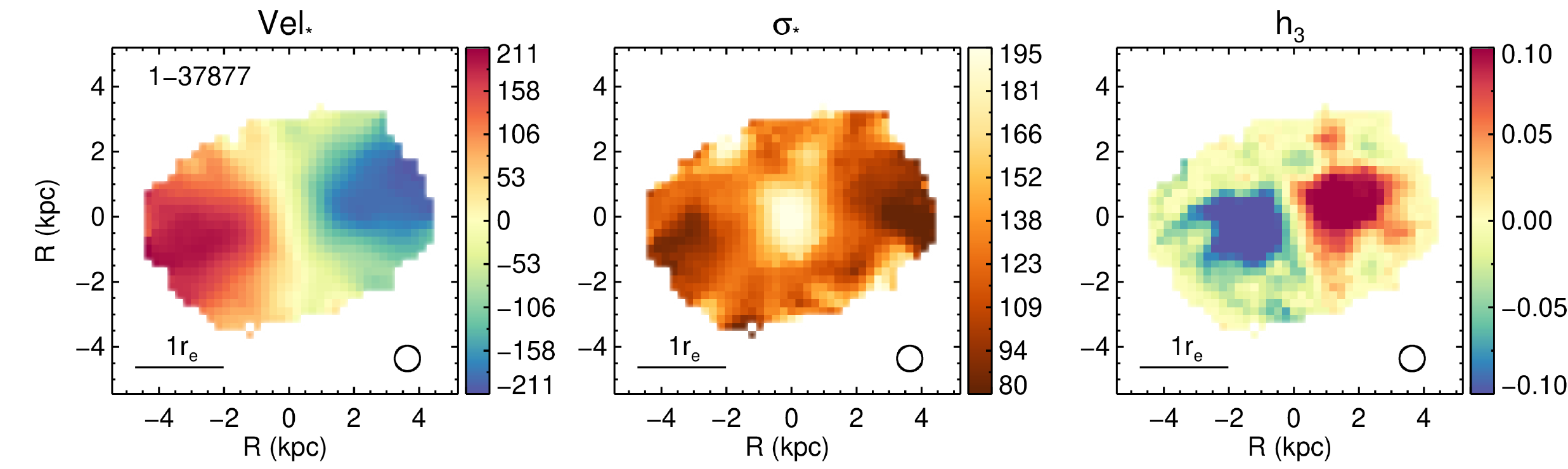}
    \includegraphics[width=0.95\hsize]{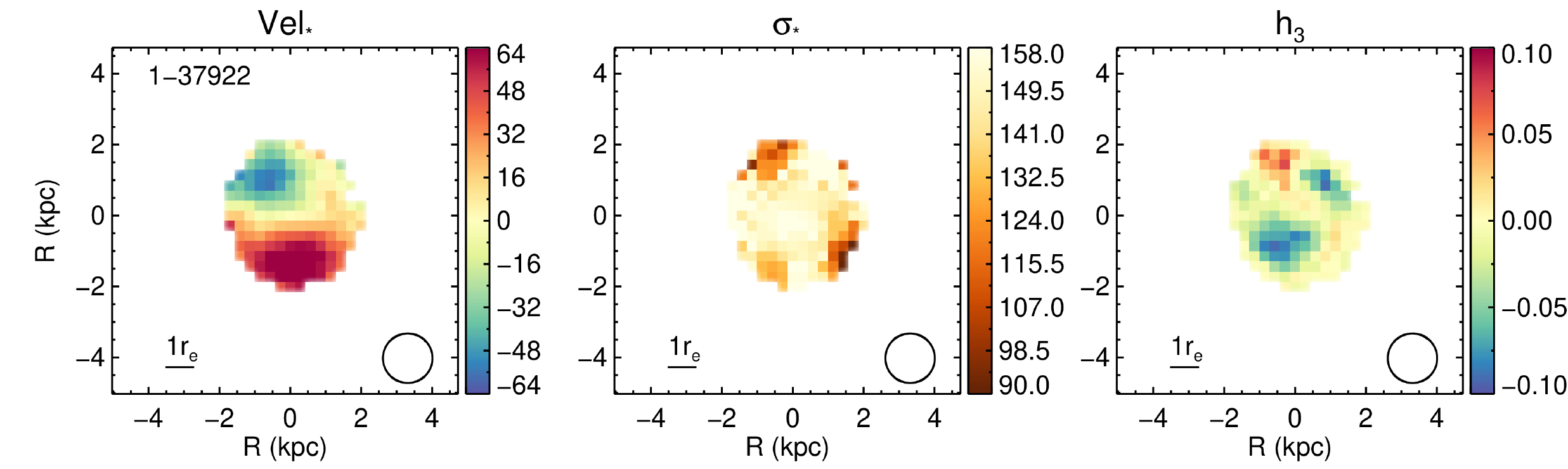}
    \includegraphics[width=0.95\hsize]{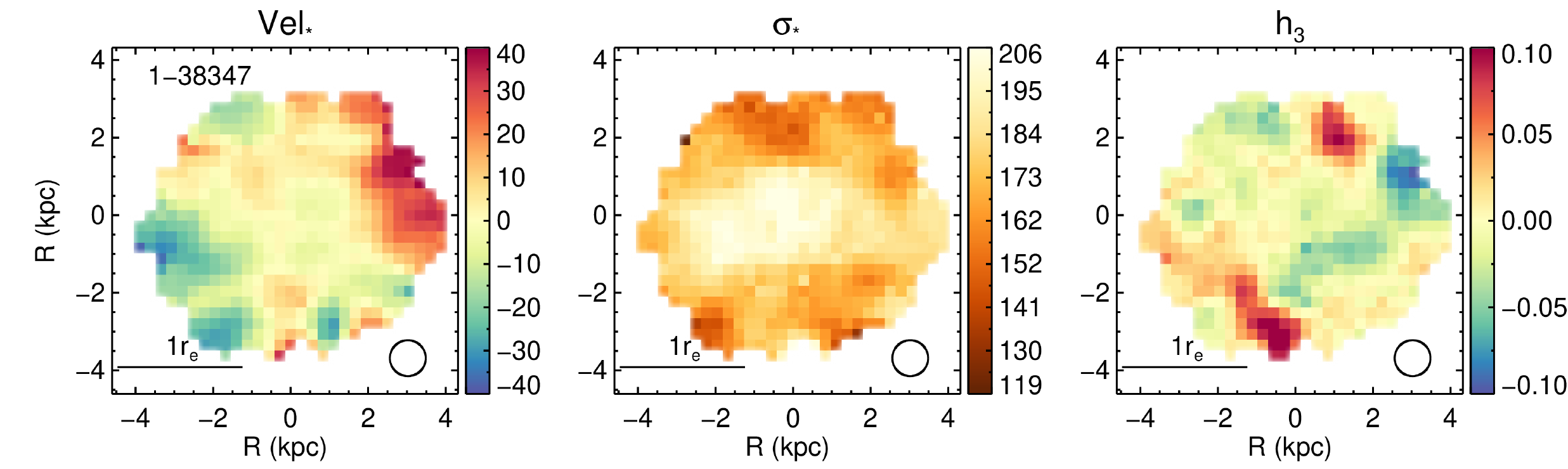}
    \includegraphics[width=0.95\hsize]{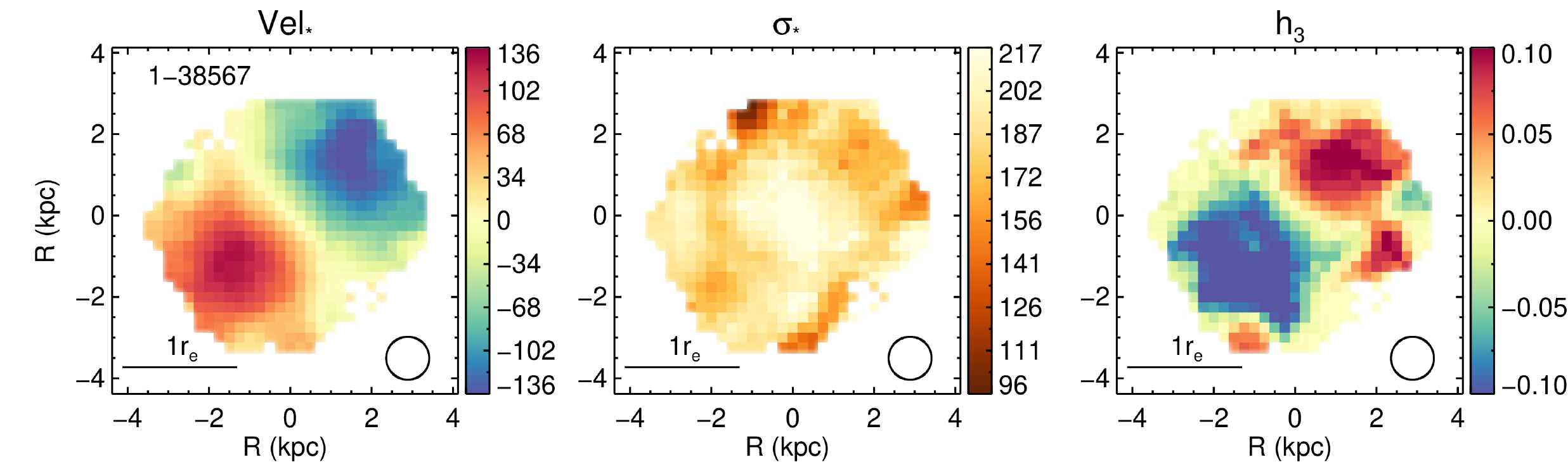}
    \caption{Fig.\,\ref{fig:kinematics_apend} continued.}
\end{figure*}

\begin{figure*}
\centering
    \includegraphics[width=0.95\hsize]{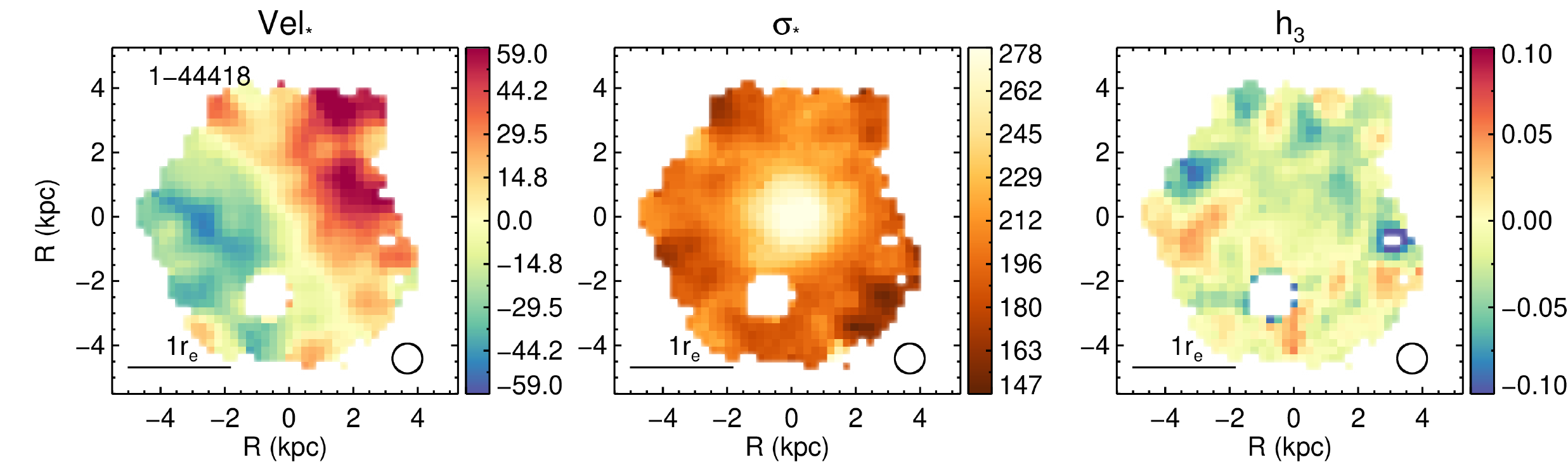}
    \includegraphics[width=0.95\hsize]{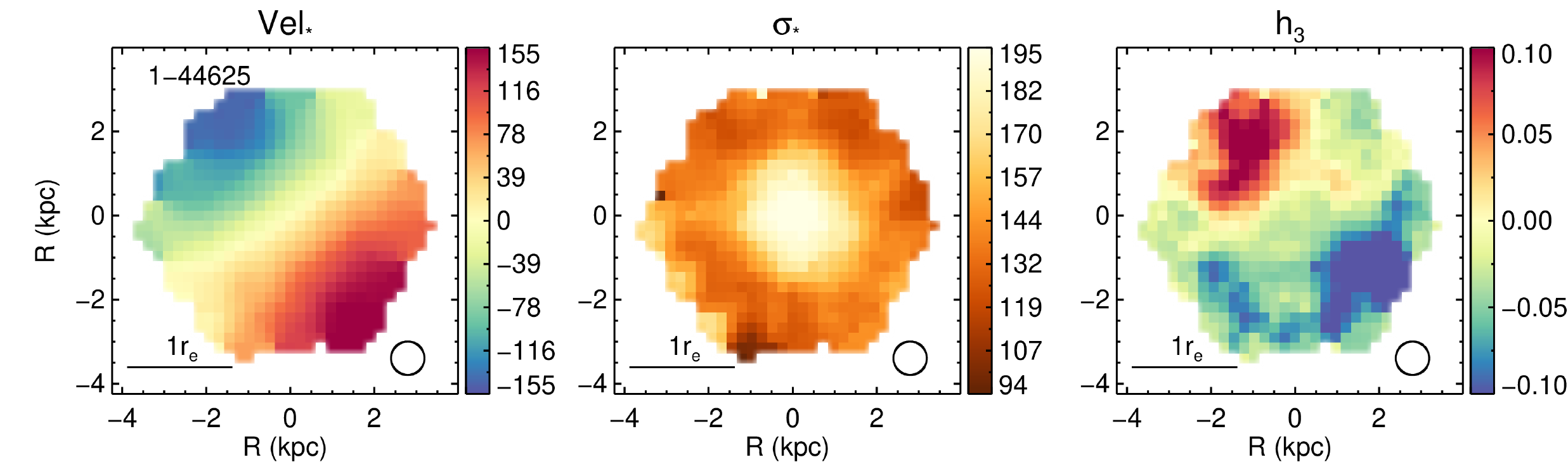}
    \includegraphics[width=0.95\hsize]{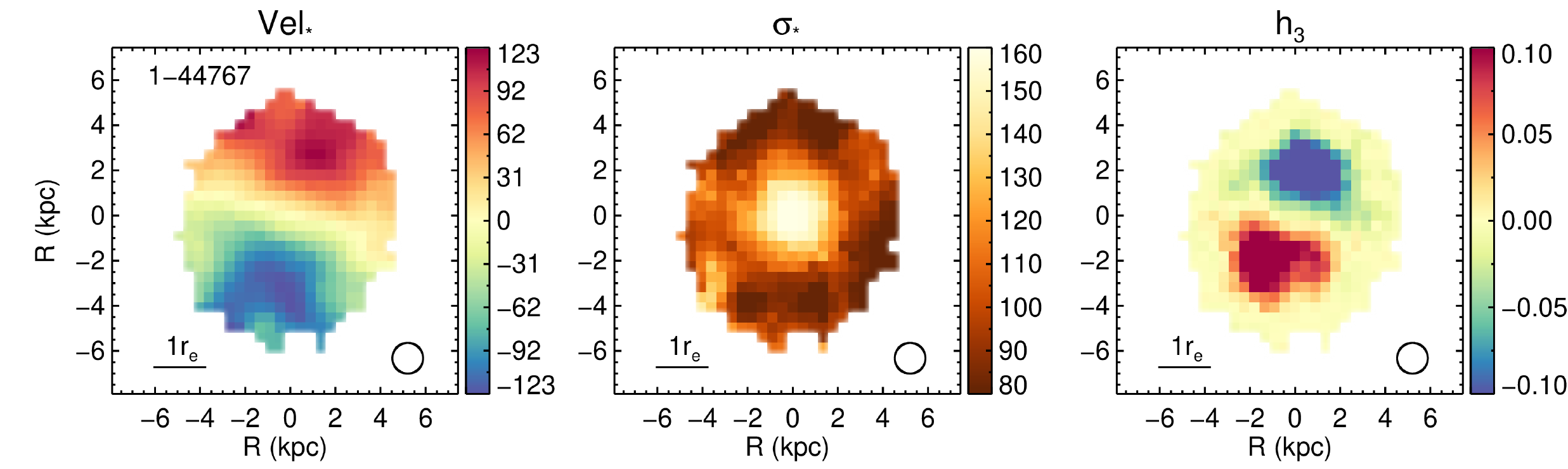}
    \includegraphics[width=0.95\hsize]{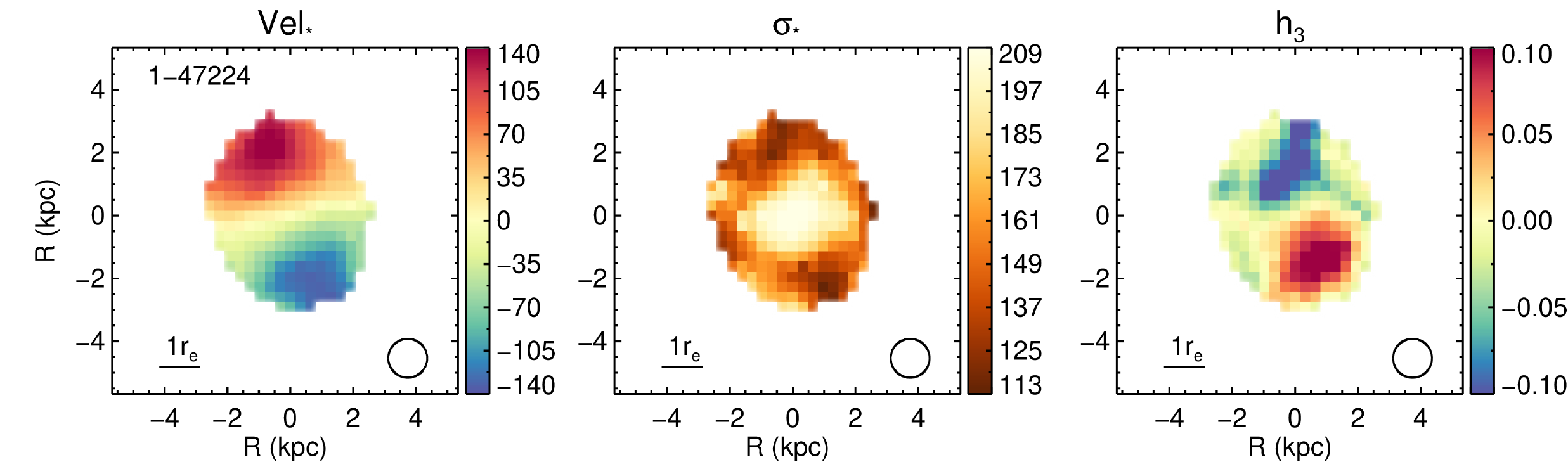}
    \caption{Fig.\,\ref{fig:kinematics_apend} continued.}
\end{figure*}

\begin{figure*}
\centering
    \includegraphics[width=0.95\hsize]{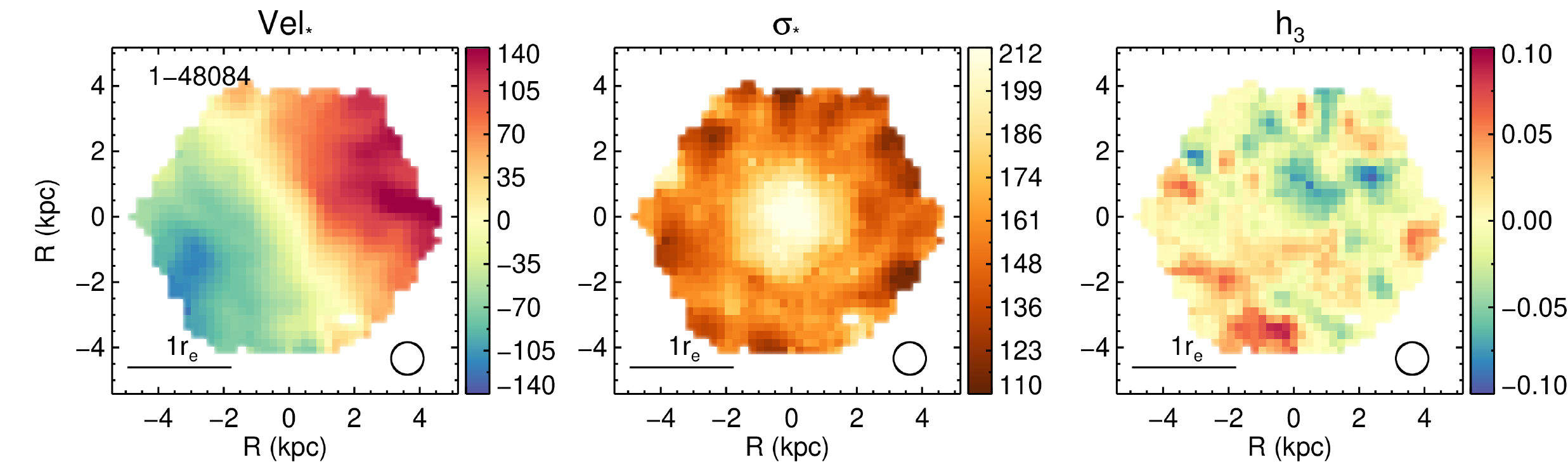}
    \includegraphics[width=0.95\hsize]{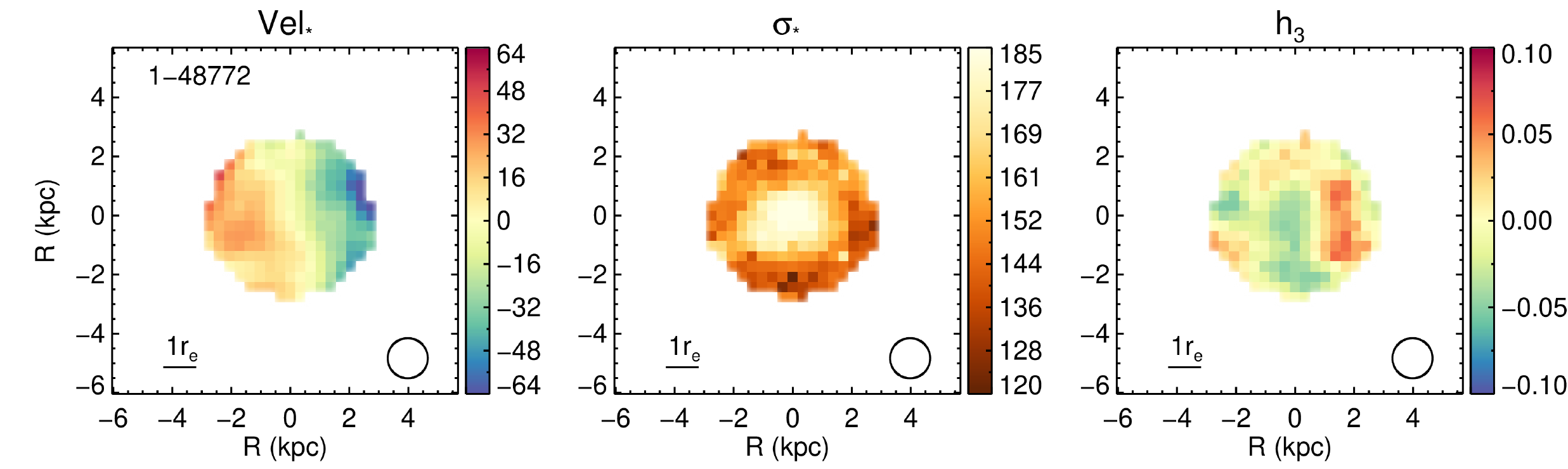}
    \includegraphics[width=0.95\hsize]{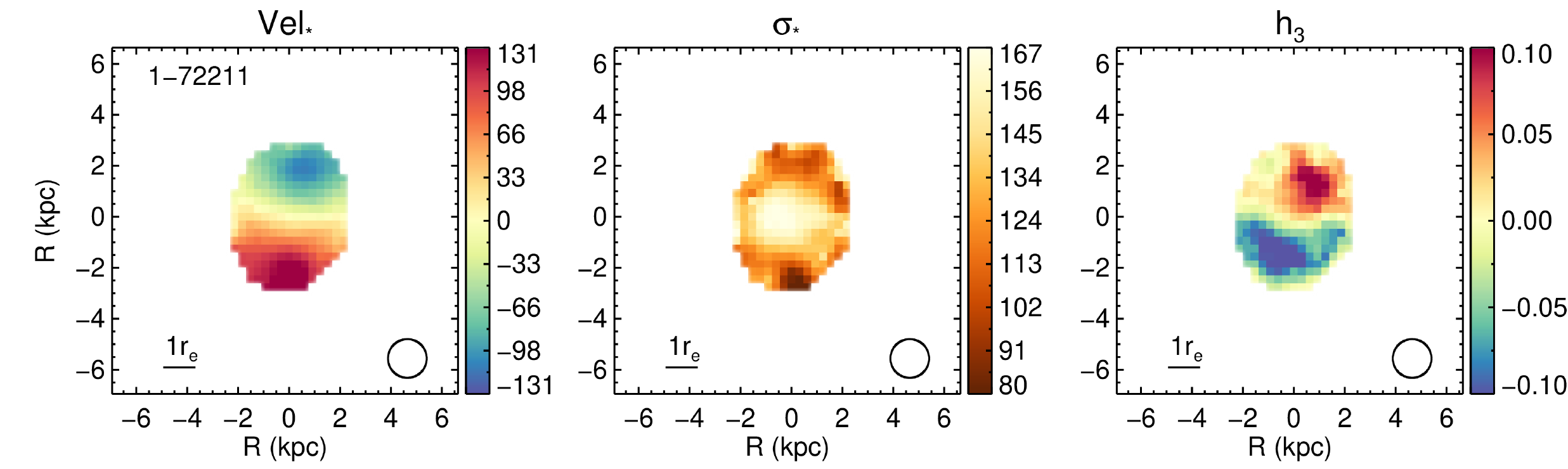}
    \includegraphics[width=0.95\hsize]{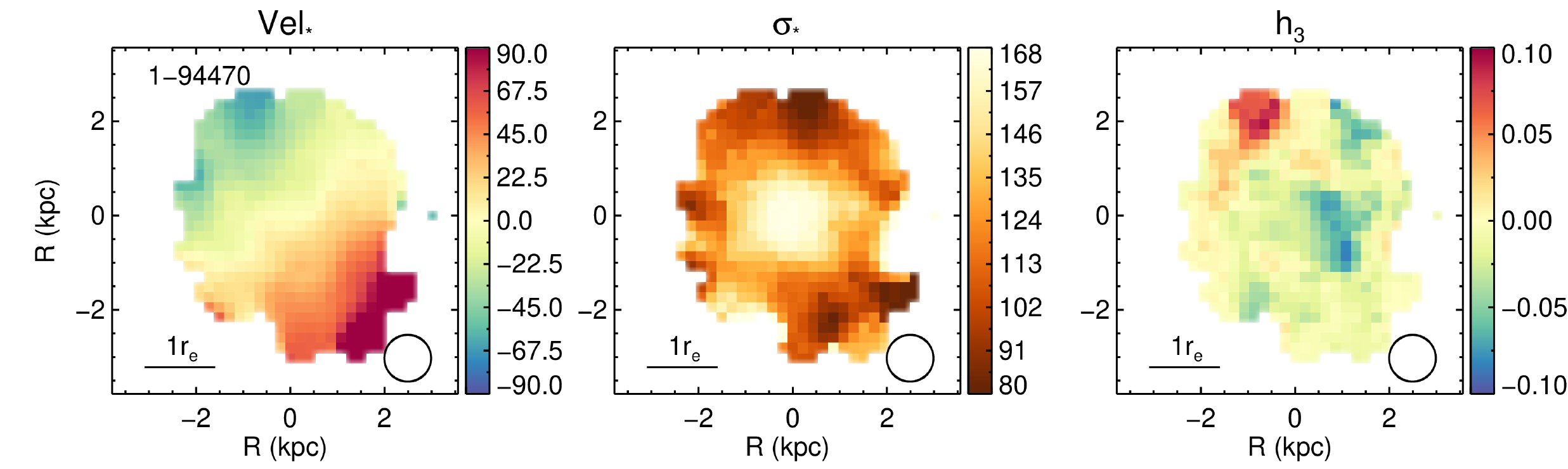}
    \caption{Fig.\,\ref{fig:kinematics_apend} continued.}
\end{figure*}

\begin{figure*}
\centering
    \includegraphics[width=0.95\hsize]{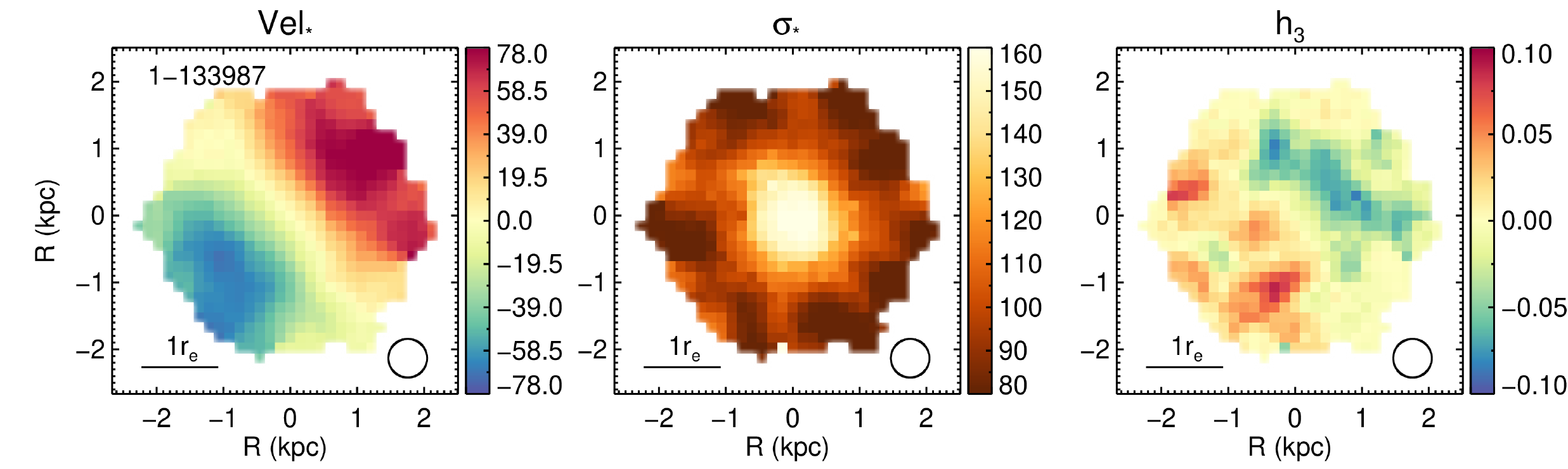}
    \includegraphics[width=0.95\hsize]{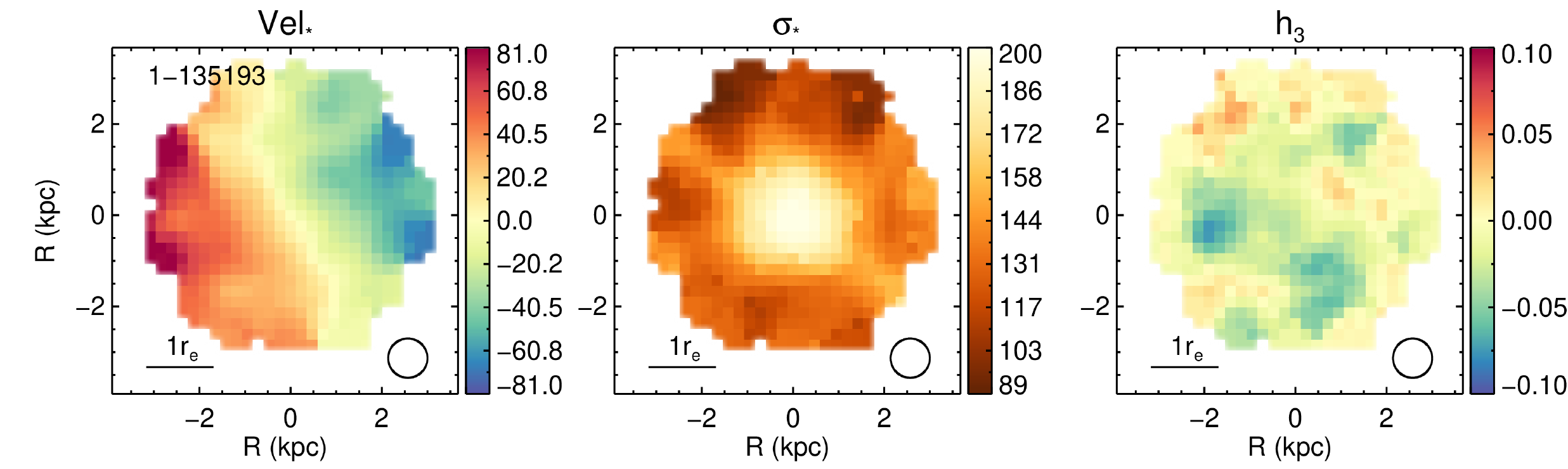}
    \includegraphics[width=0.95\hsize]{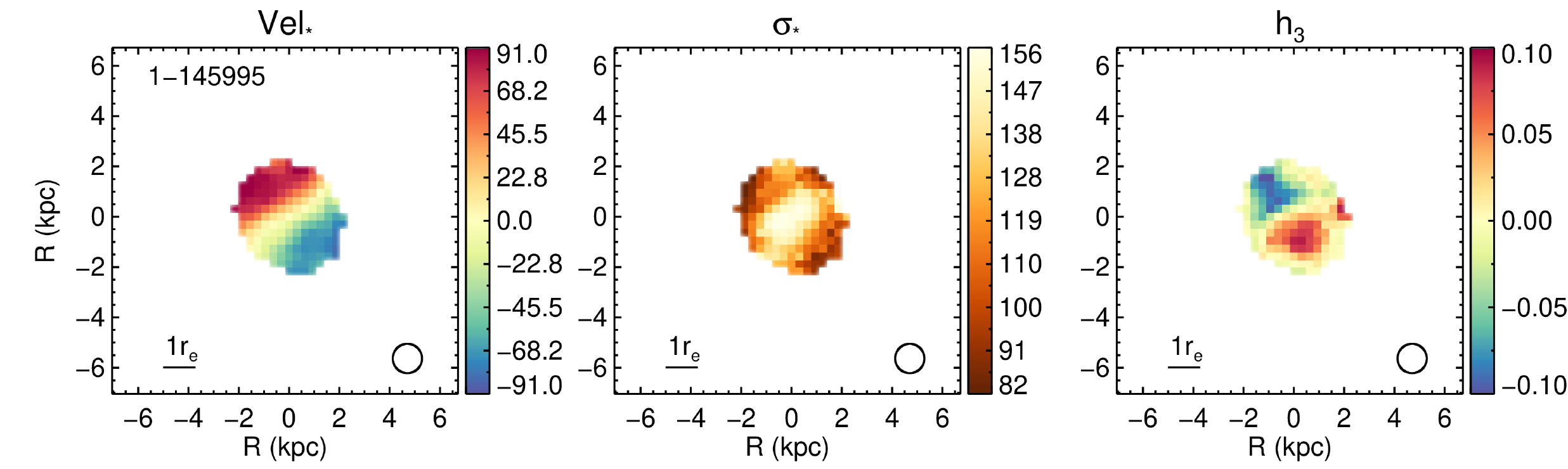}
    \includegraphics[width=0.95\hsize]{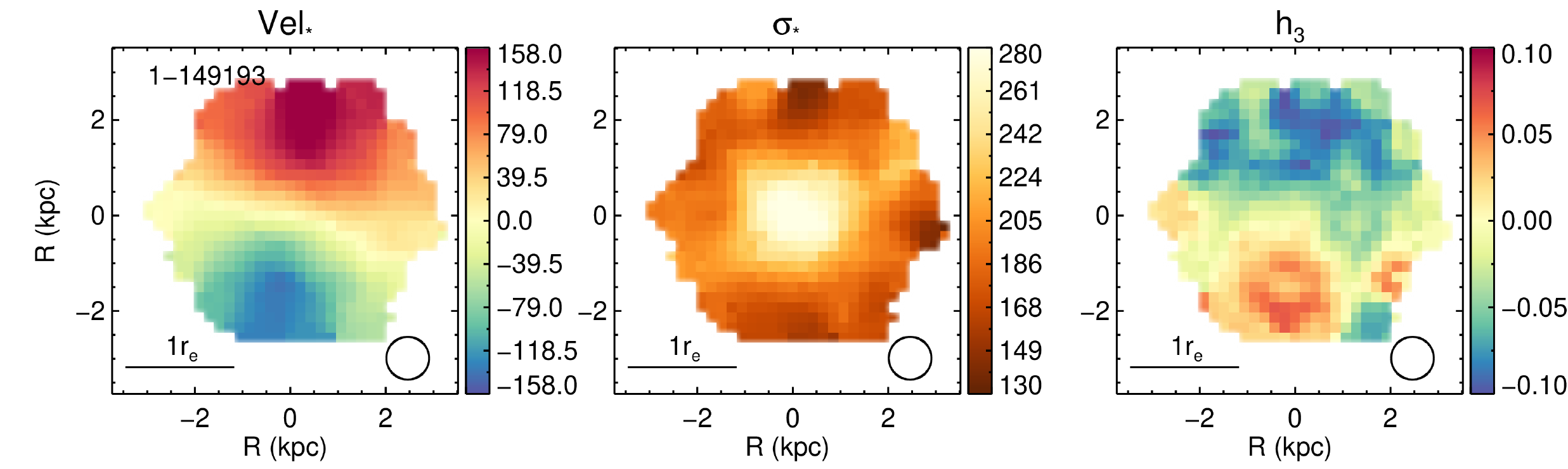}
    \caption{Fig.\,\ref{fig:kinematics_apend} continued.}
\end{figure*}

\begin{figure*}
\centering
    \includegraphics[width=0.95\hsize]{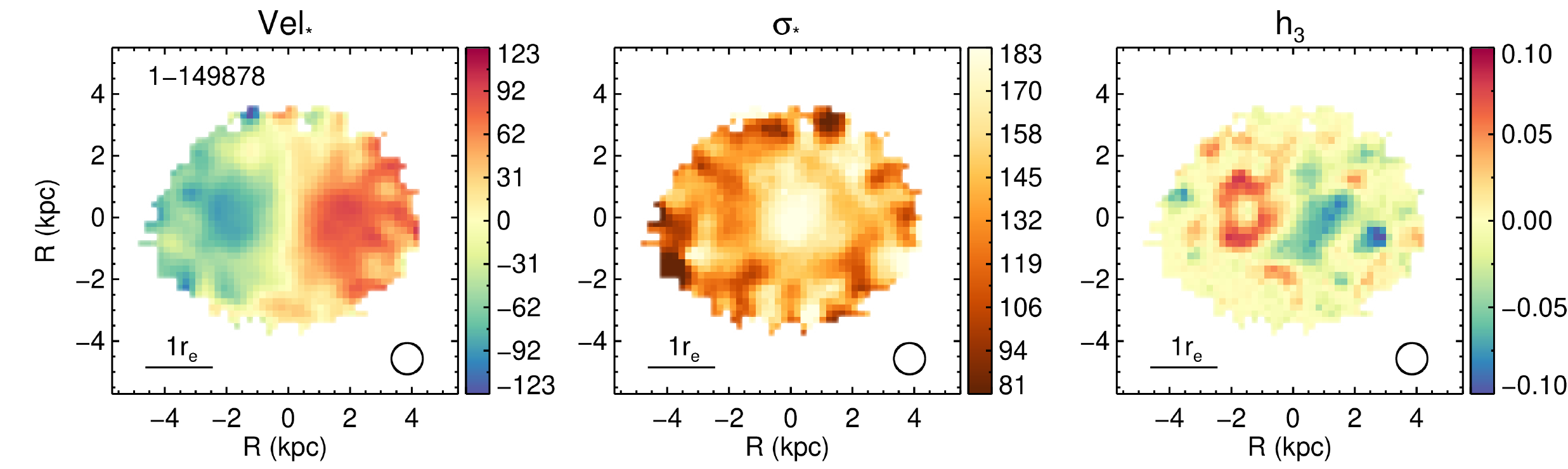}
    \includegraphics[width=0.95\hsize]{maps/1-174435.pdf}
    \includegraphics[width=0.95\hsize]{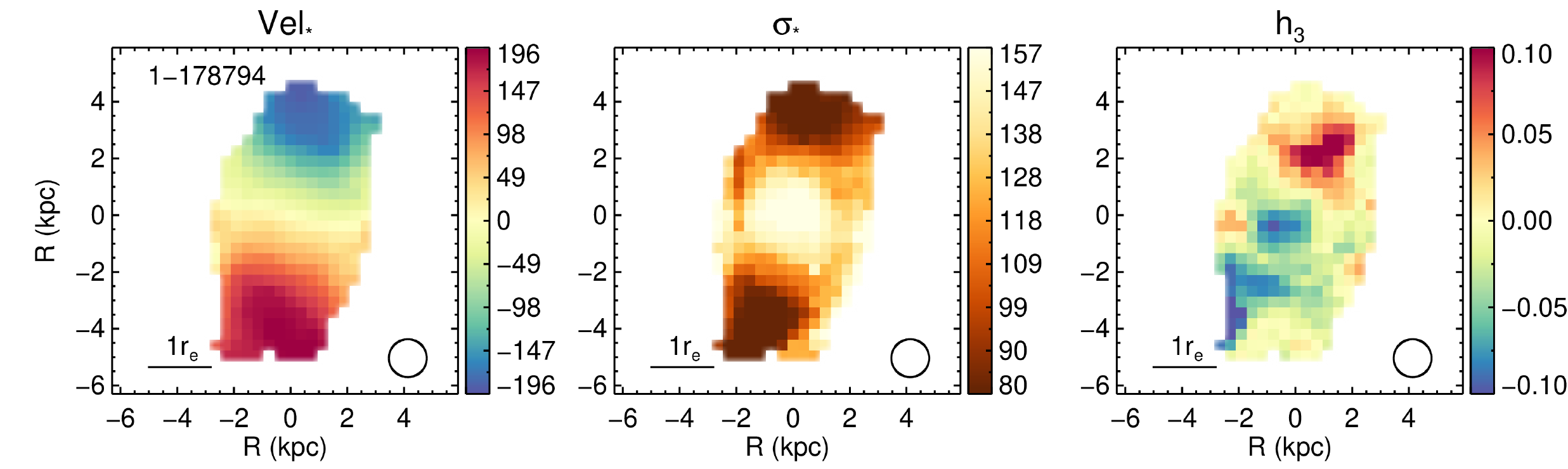}
    \includegraphics[width=0.95\hsize]{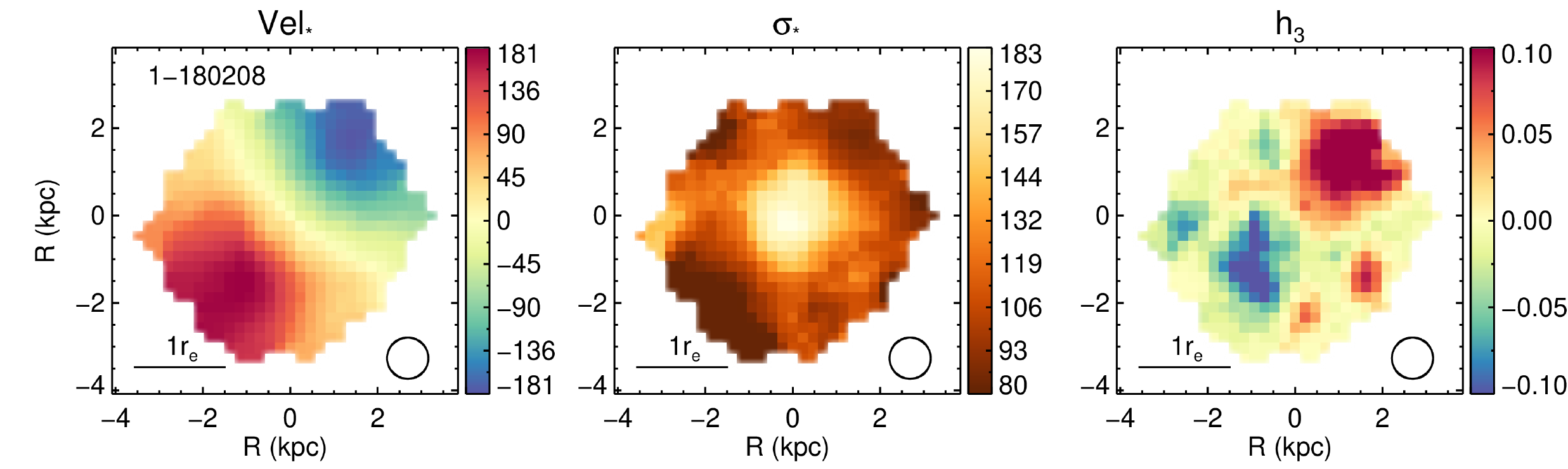}
    \caption{Fig.\,\ref{fig:kinematics_apend} continued.}
\end{figure*}

\begin{figure*}
\centering
    \includegraphics[width=0.95\hsize]{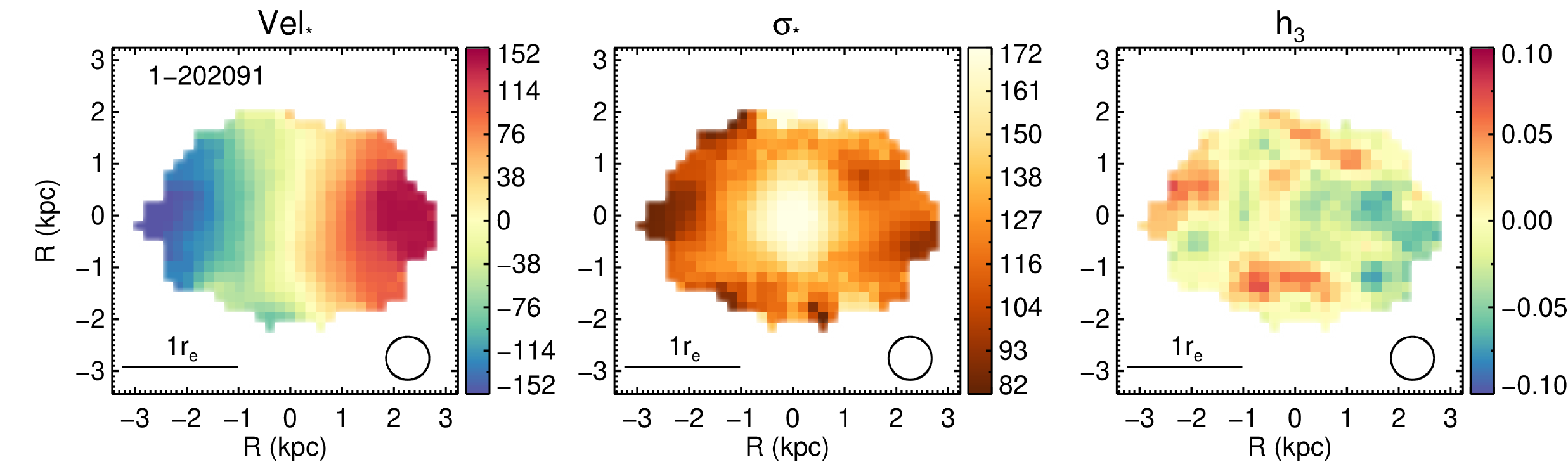}
    \includegraphics[width=0.95\hsize]{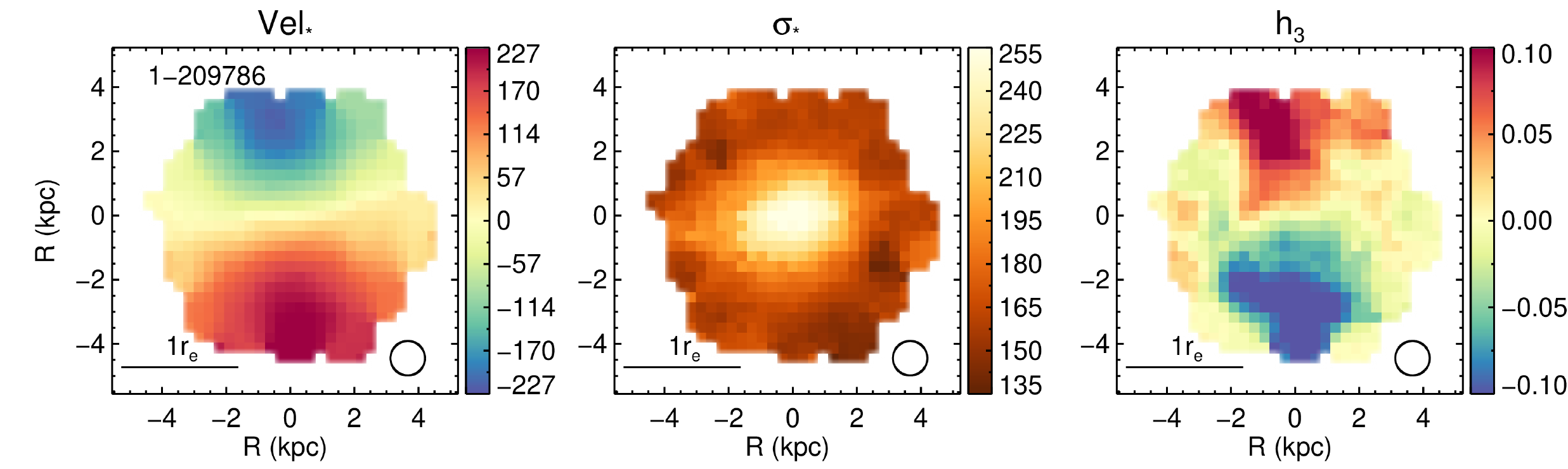}
    \includegraphics[width=0.95\hsize]{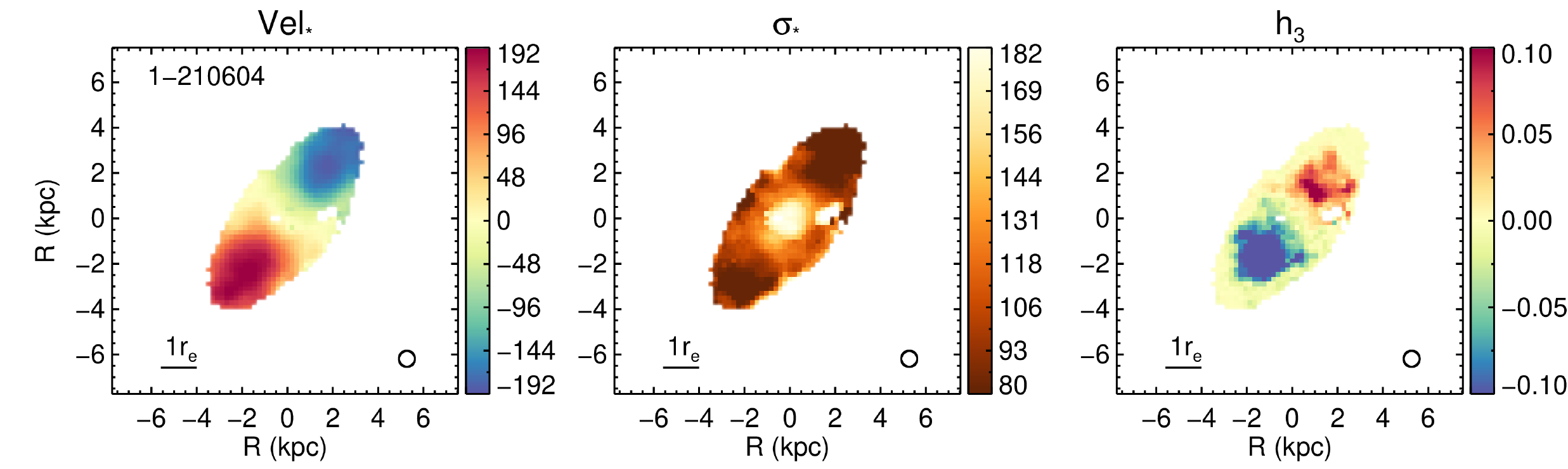}
    \includegraphics[width=0.95\hsize]{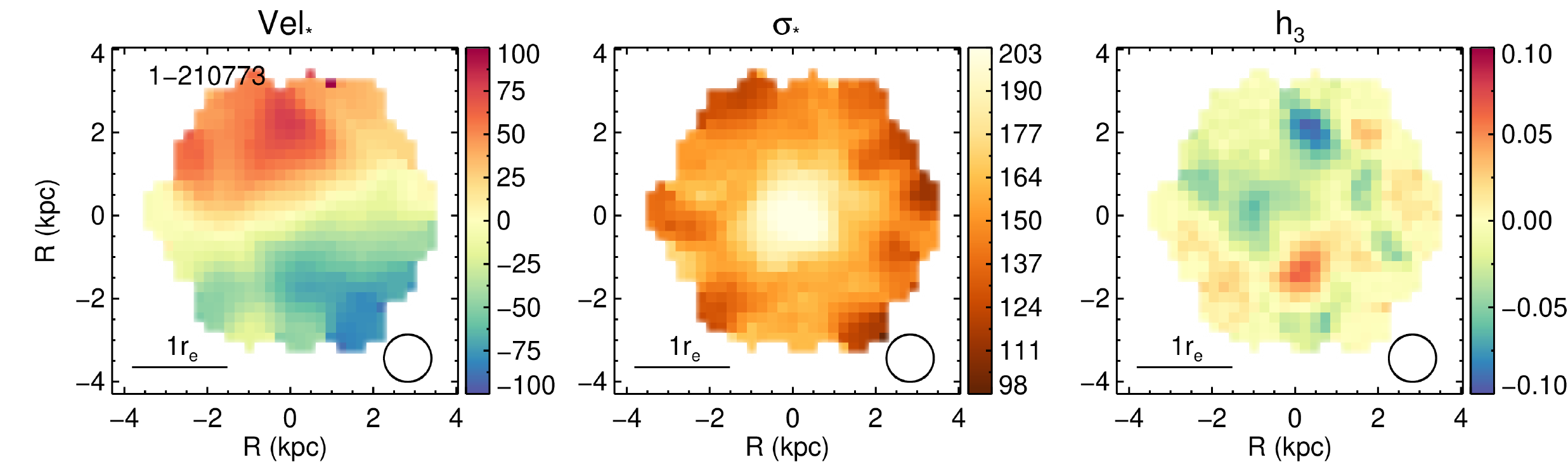}
    \caption{Fig.\,\ref{fig:kinematics_apend} continued.}
\end{figure*}

\begin{figure*}
\centering
    \includegraphics[width=0.95\hsize]{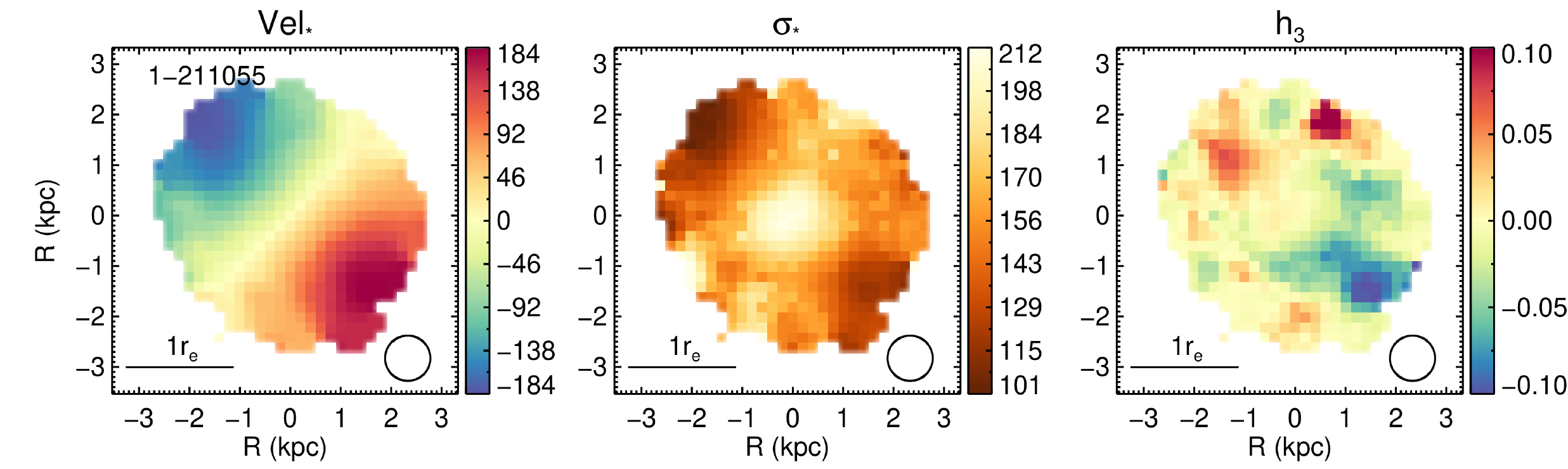}
    \includegraphics[width=0.95\hsize]{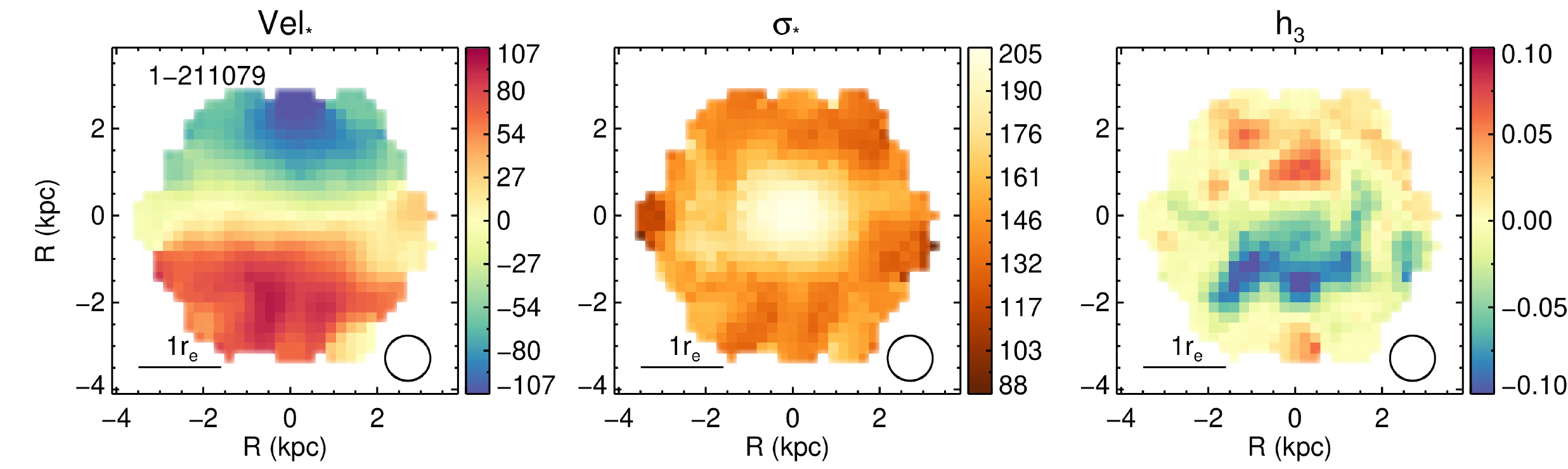}
    \includegraphics[width=0.95\hsize]{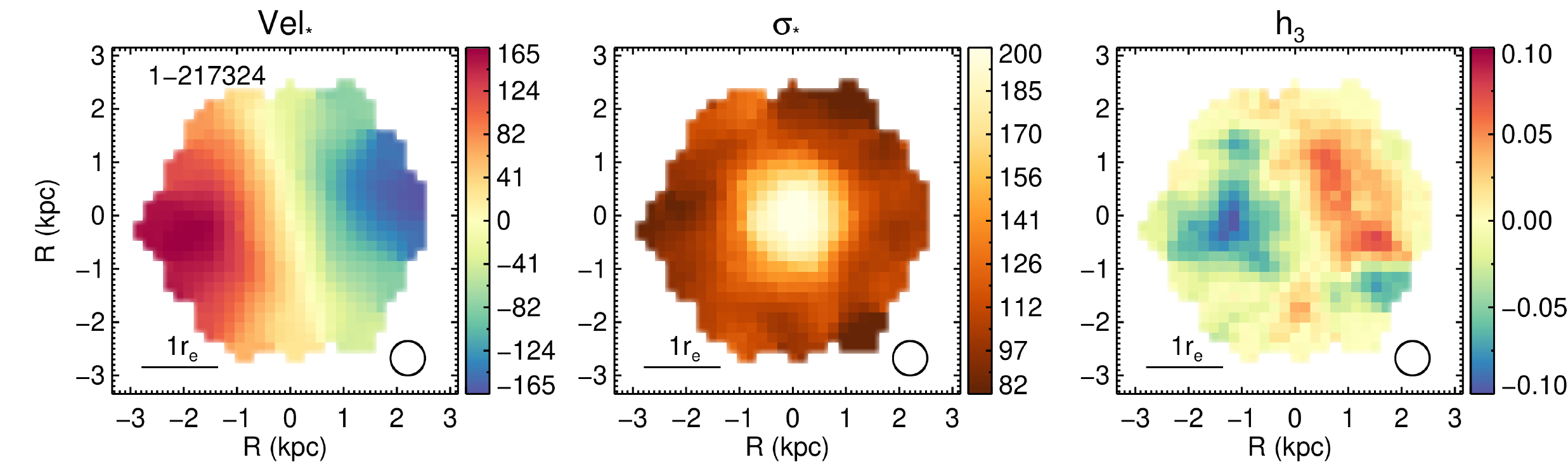}
    \includegraphics[width=0.95\hsize]{maps/1-229012.pdf}
    \caption{Fig.\,\ref{fig:kinematics_apend} continued.}
\end{figure*}

\begin{figure*}
\centering
    \includegraphics[width=0.95\hsize]{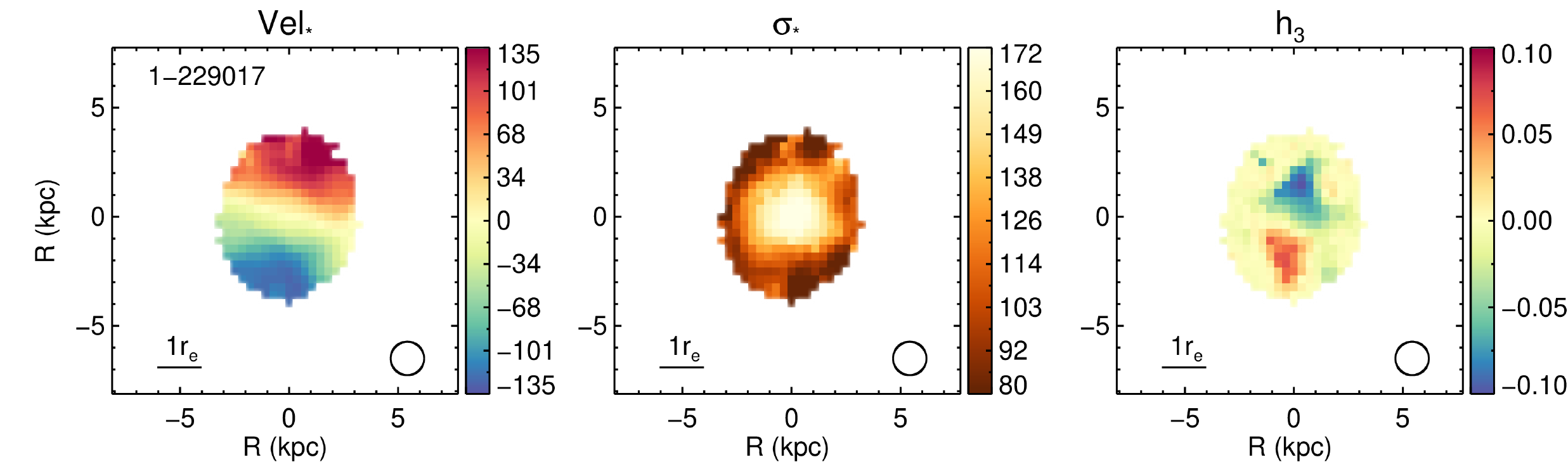}
    \includegraphics[width=0.95\hsize]{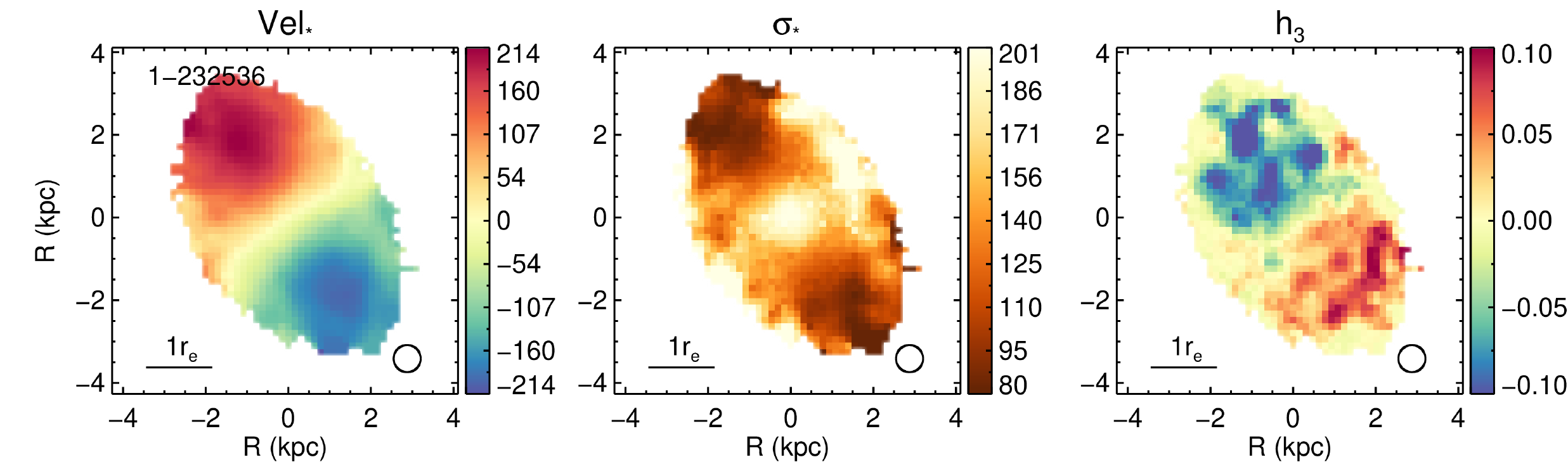}
    \includegraphics[width=0.95\hsize]{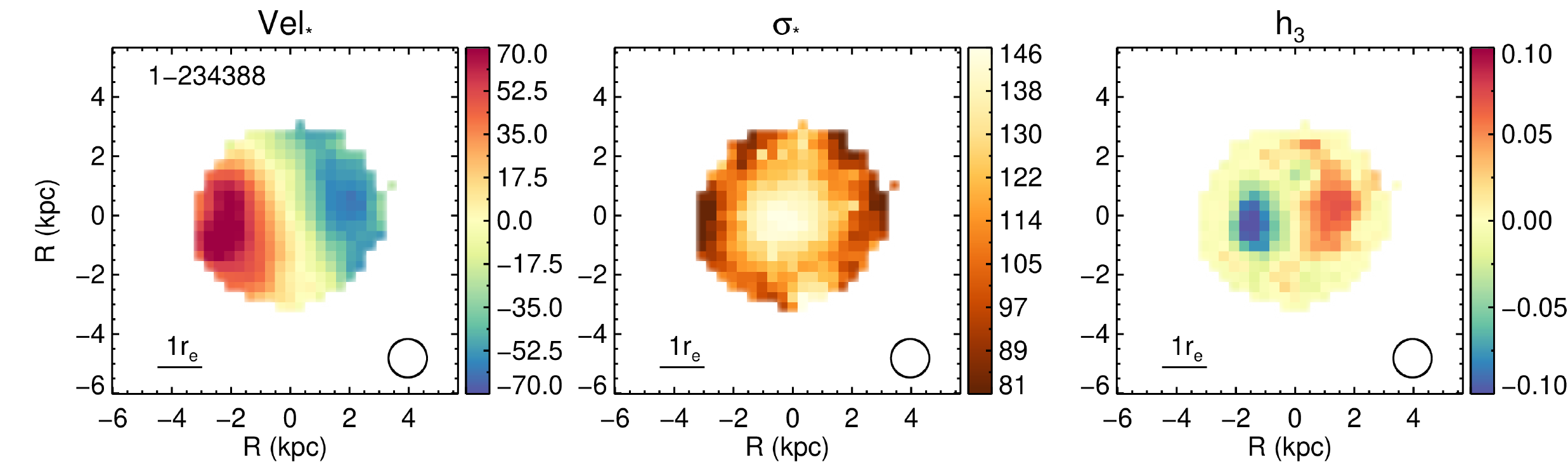}
    \includegraphics[width=0.95\hsize]{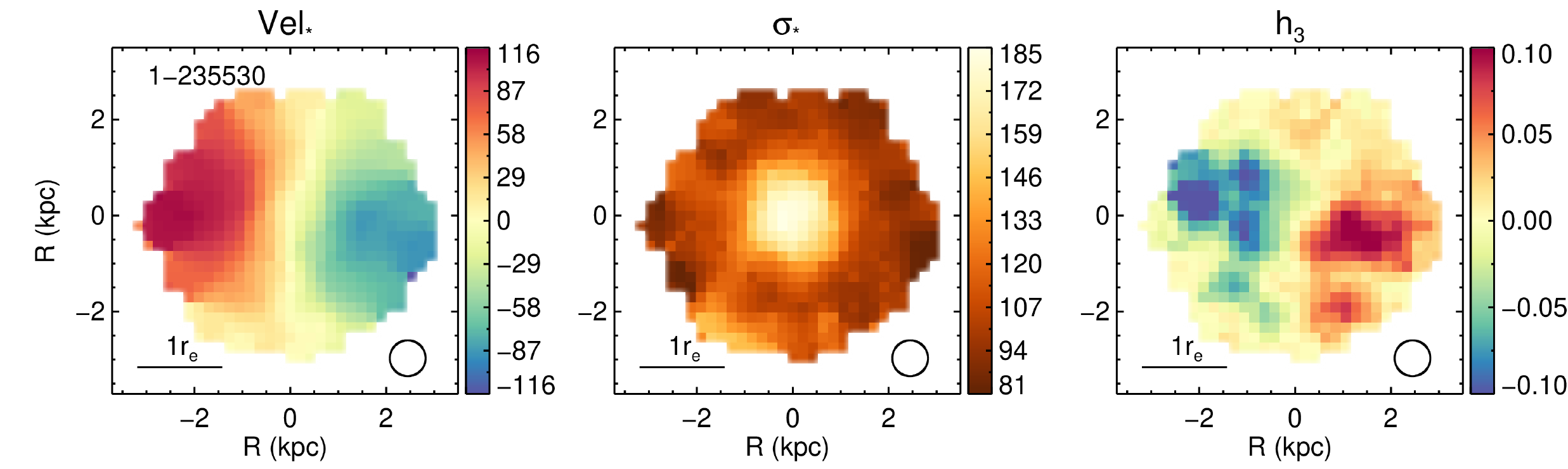}
    \caption{Fig.\,\ref{fig:kinematics_apend} continued.}
\end{figure*}

\begin{figure*}
\centering
    \includegraphics[width=0.95\hsize]{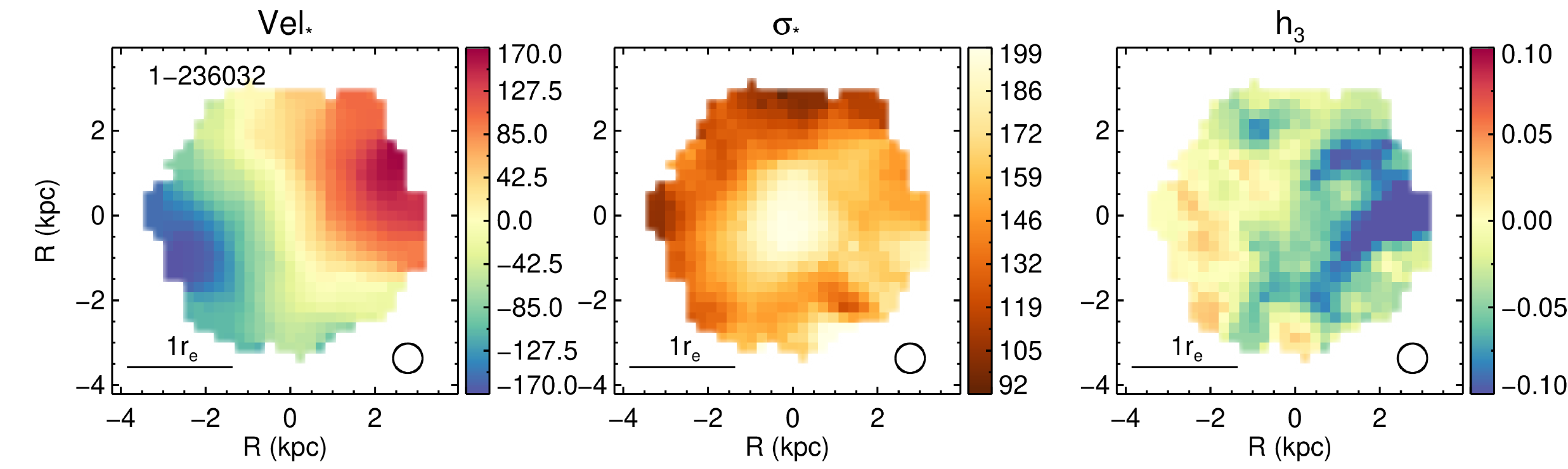}
    \includegraphics[width=0.95\hsize]{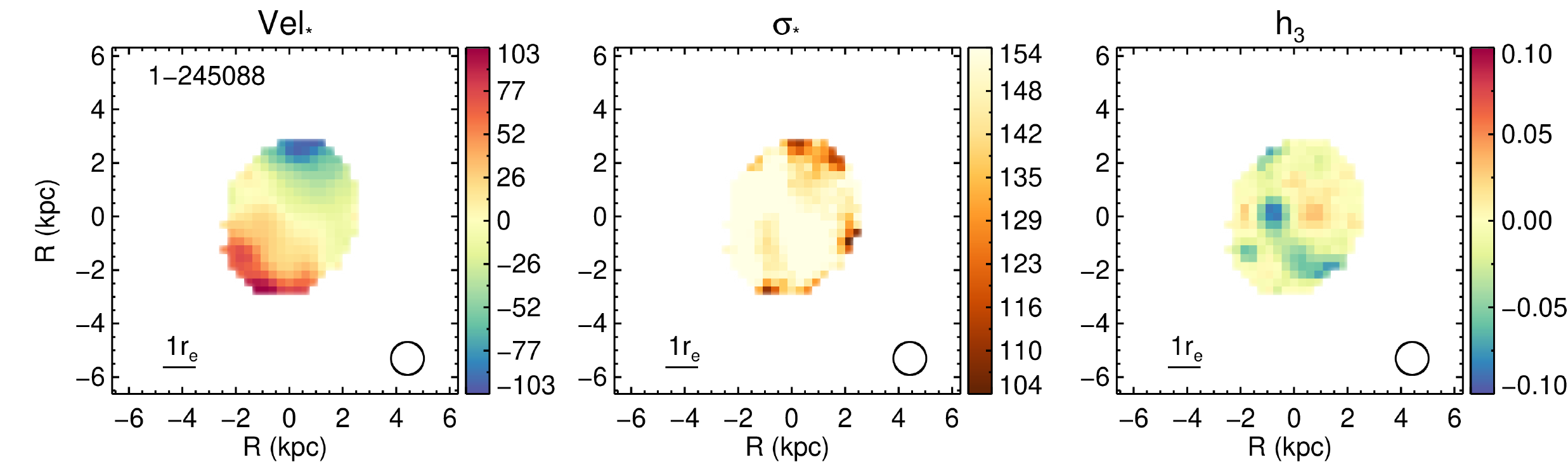}
    \includegraphics[width=0.95\hsize]{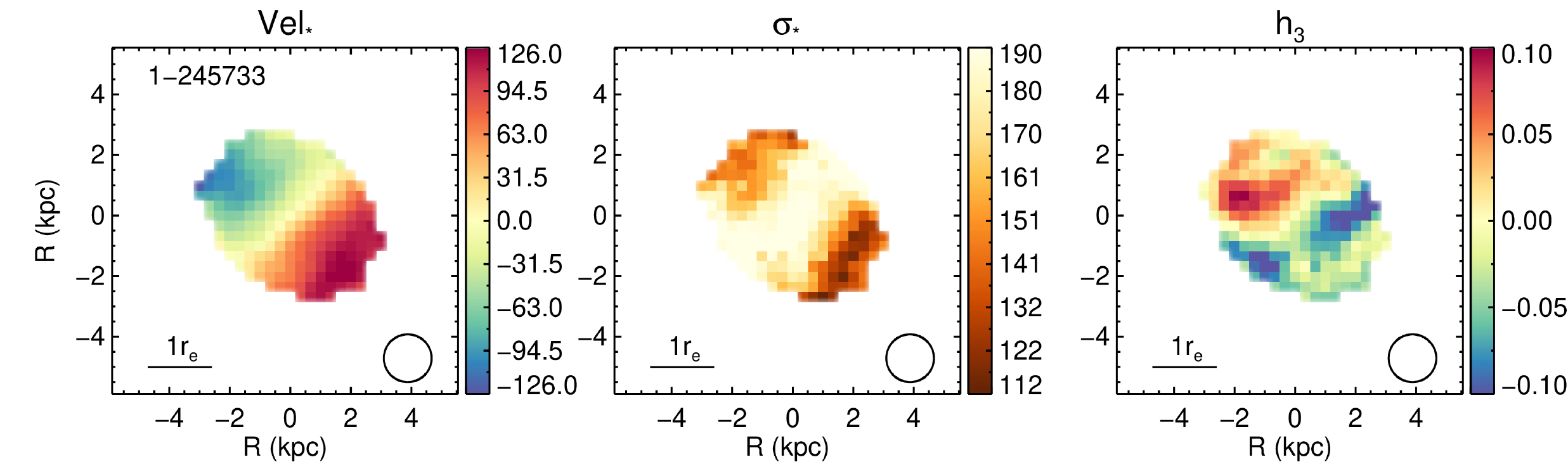}
    \includegraphics[width=0.95\hsize]{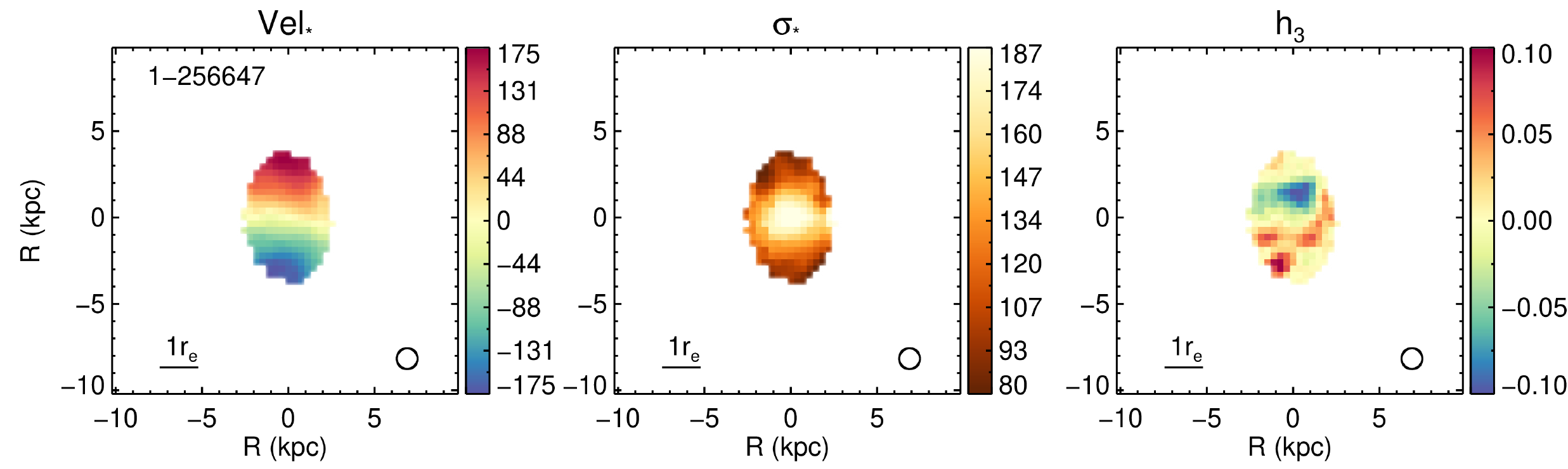}
    \caption{Fig.\,\ref{fig:kinematics_apend} continued.}
\end{figure*}

\begin{figure*}
\centering
    \includegraphics[width=0.95\hsize]{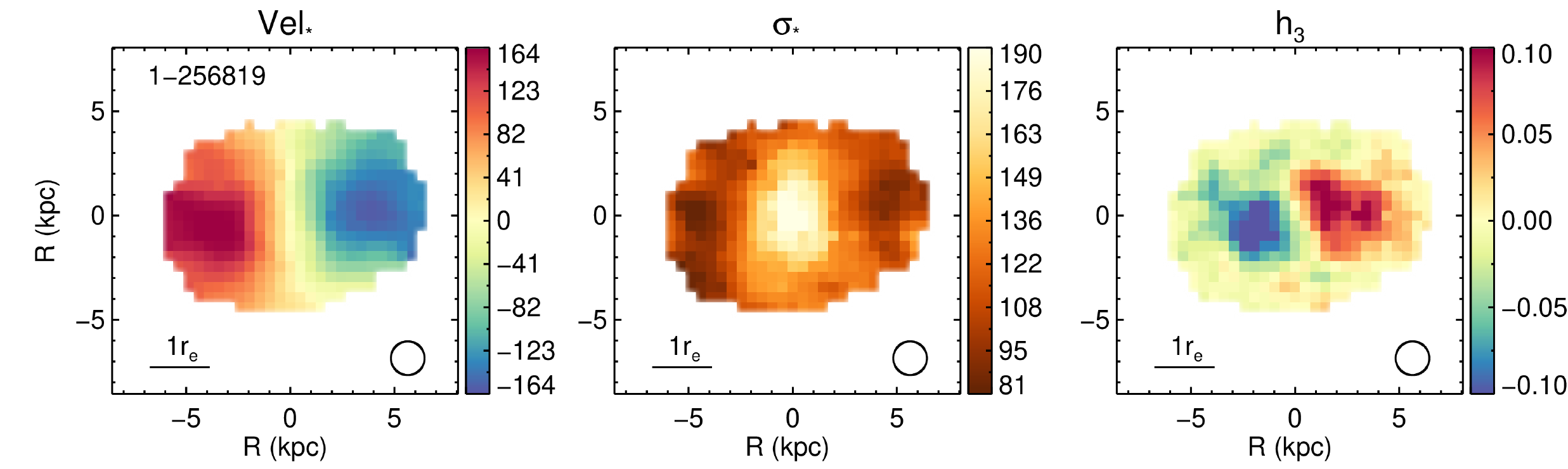}
    \includegraphics[width=0.95\hsize]{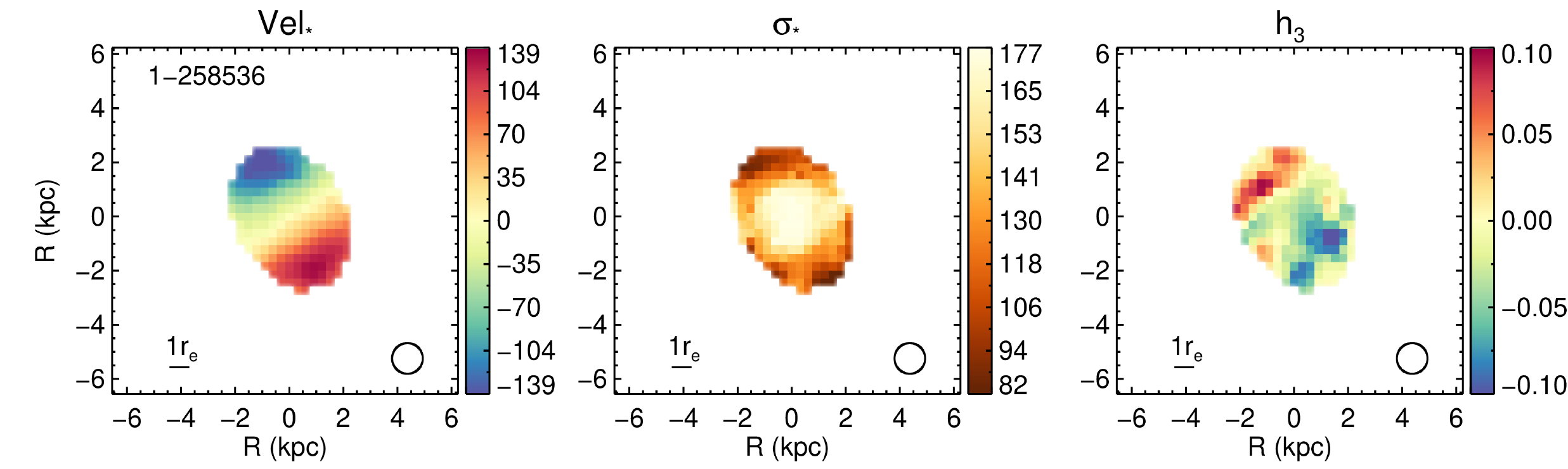}
    \includegraphics[width=0.95\hsize]{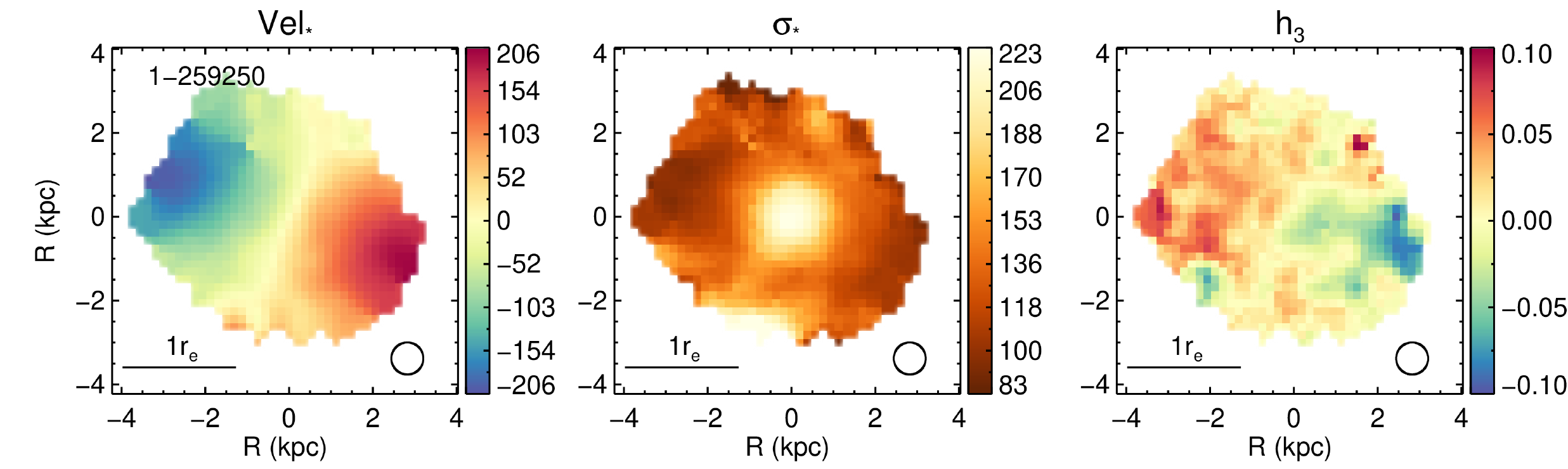}
    \includegraphics[width=0.95\hsize]{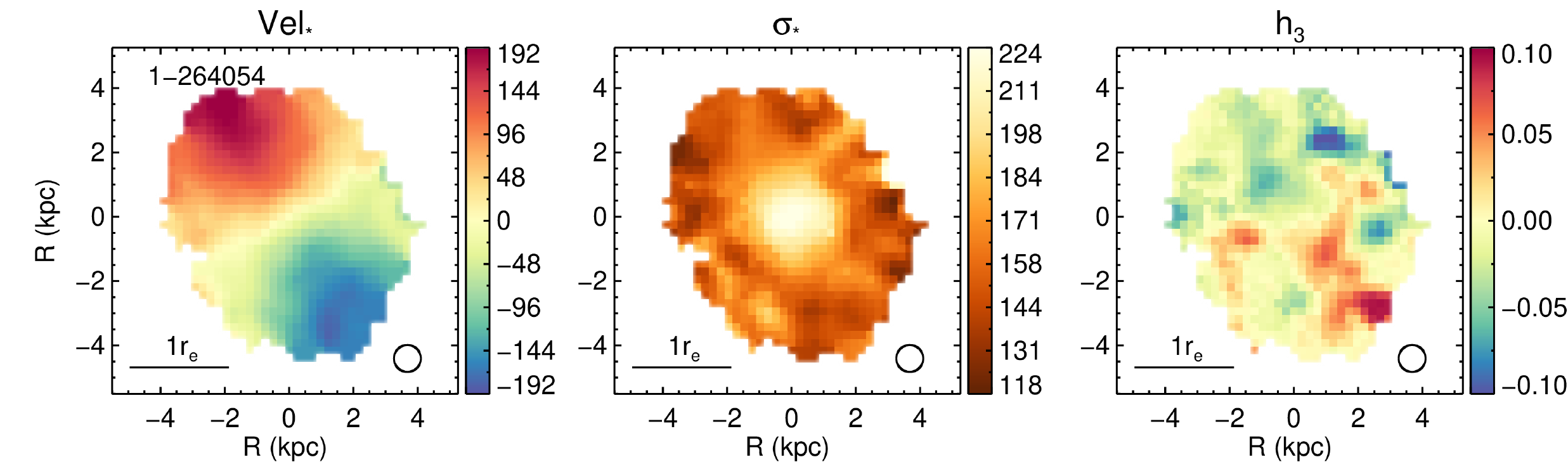}
    \caption{Fig.\,\ref{fig:kinematics_apend} continued.}
\end{figure*}

\begin{figure*}
\centering
    \includegraphics[width=0.95\hsize]{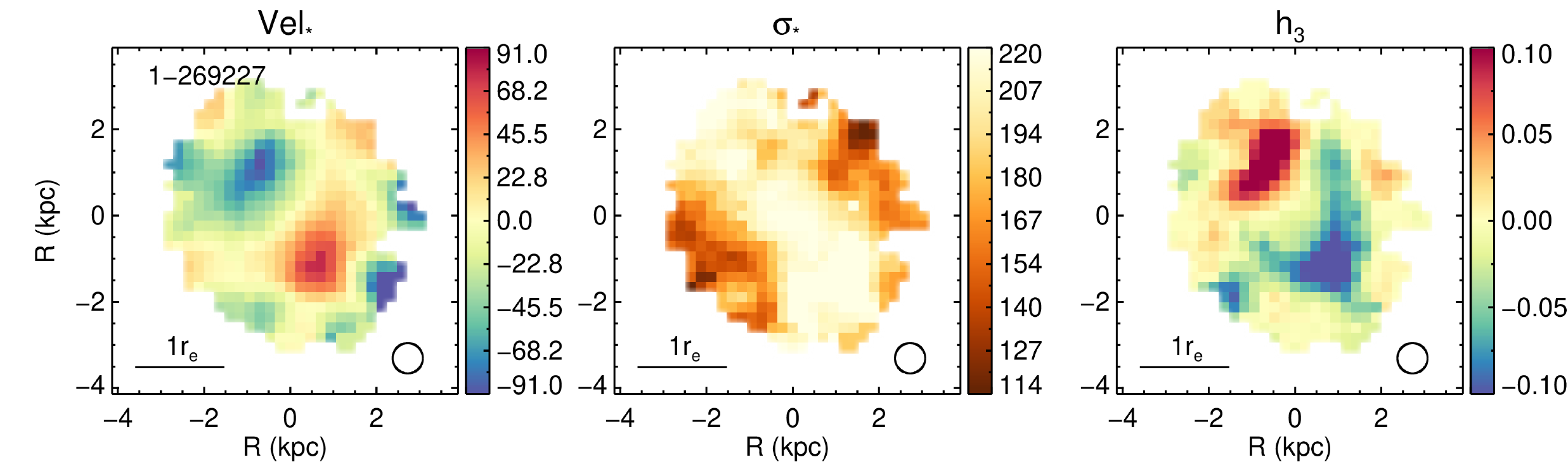}
    \includegraphics[width=0.95\hsize]{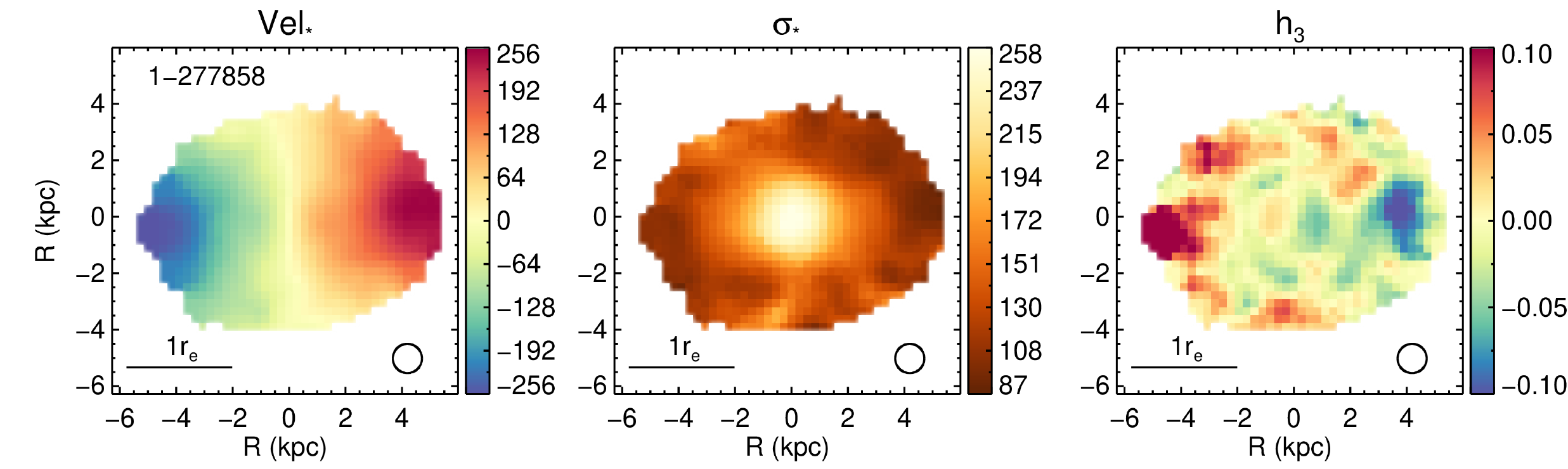}
    \includegraphics[width=0.95\hsize]{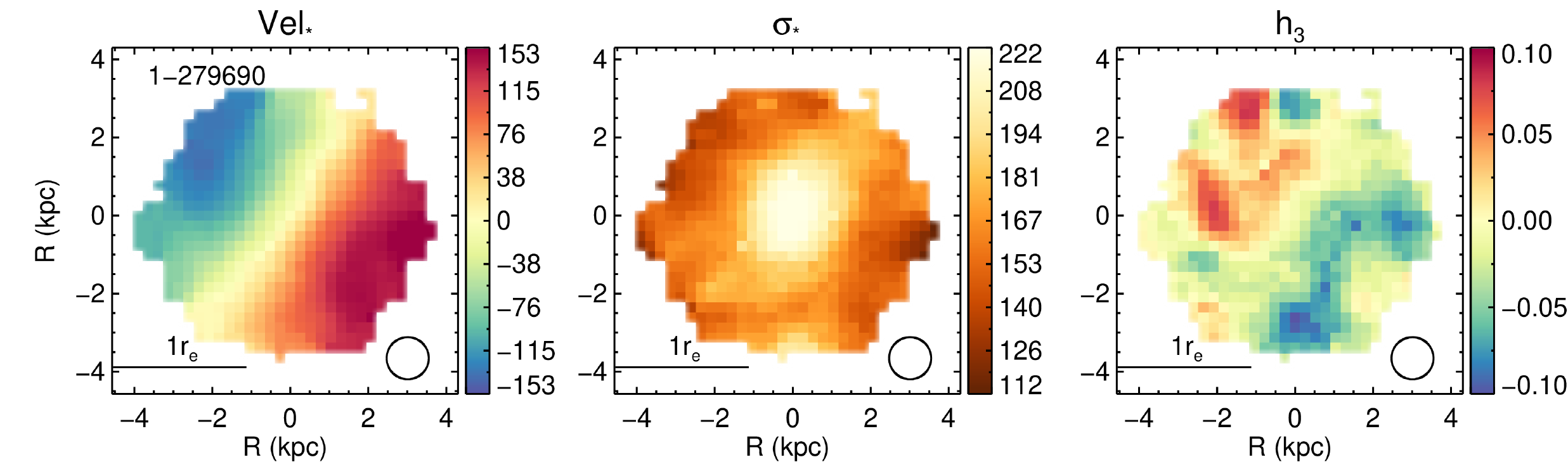}
    \includegraphics[width=0.95\hsize]{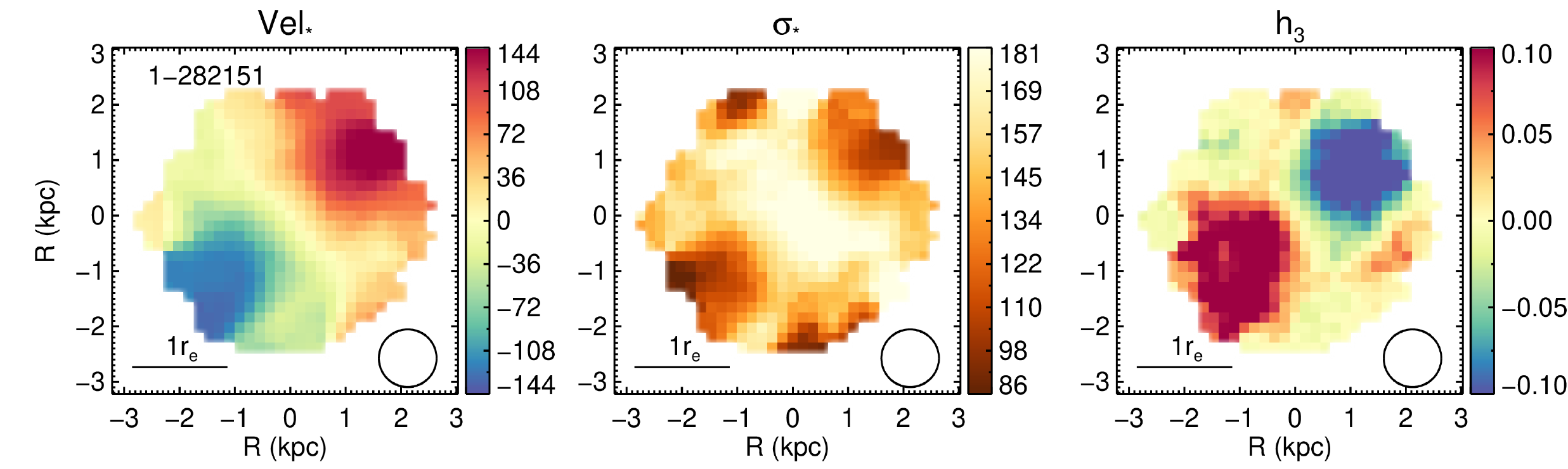}
    \caption{Fig.\,\ref{fig:kinematics_apend} continued.}
\end{figure*}

\begin{figure*}
\centering
    \includegraphics[width=0.95\hsize]{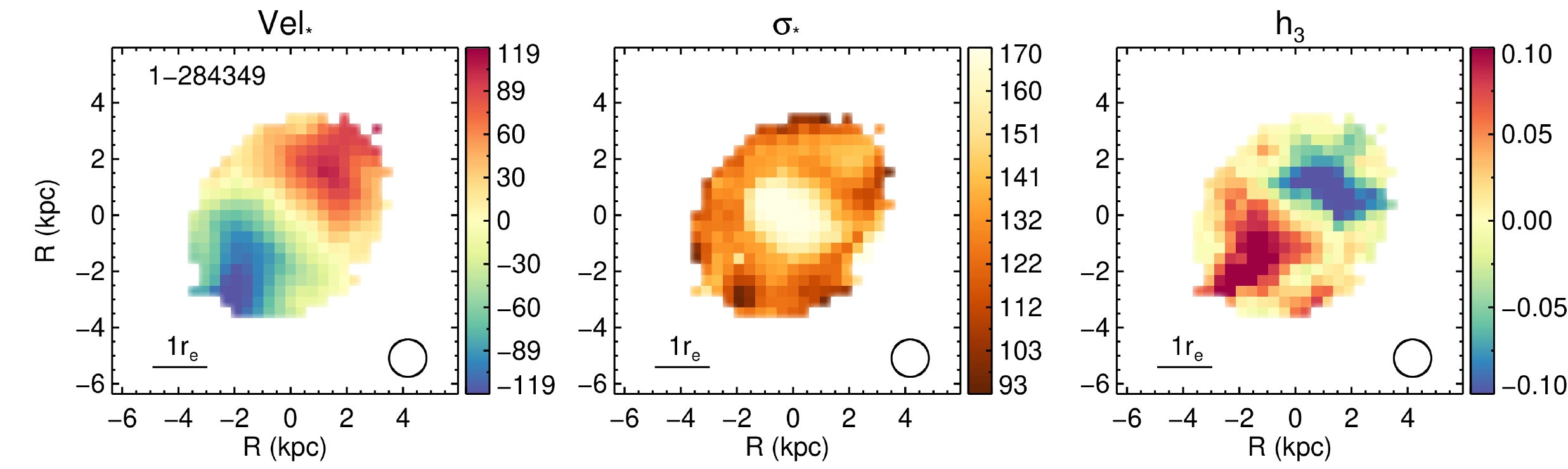}
    \includegraphics[width=0.95\hsize]{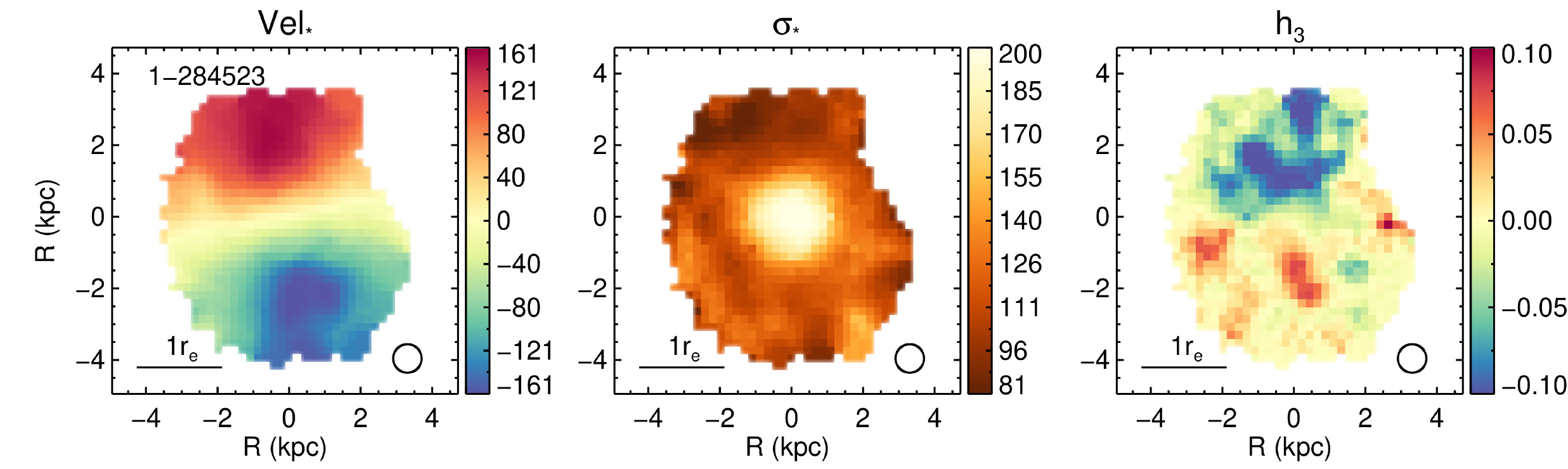}
    \includegraphics[width=0.95\hsize]{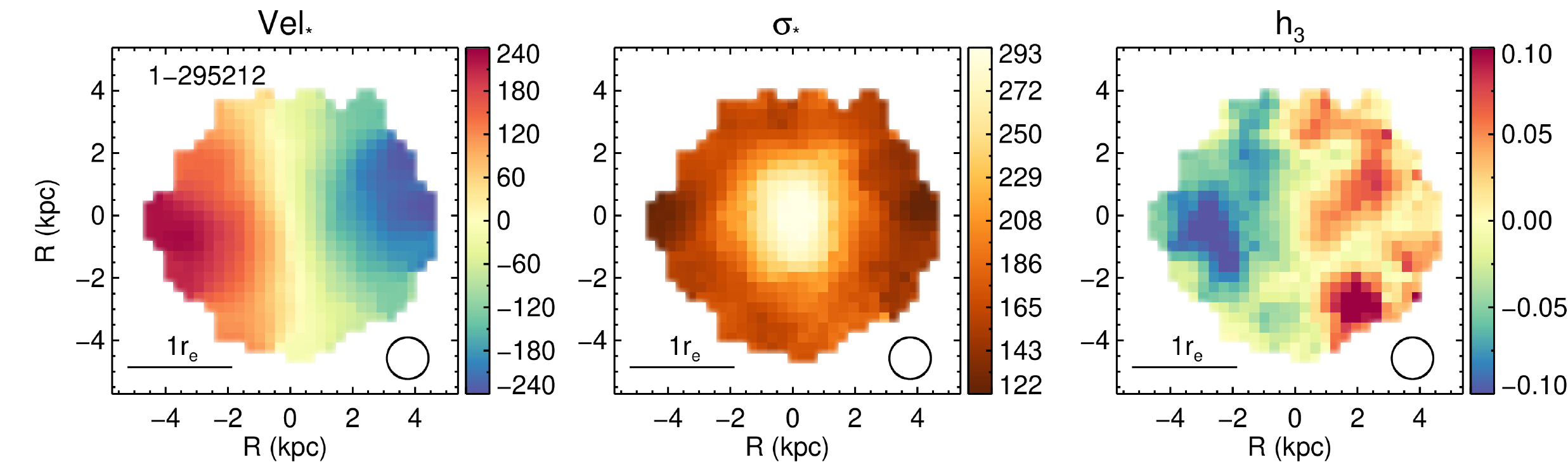}
    \includegraphics[width=0.95\hsize]{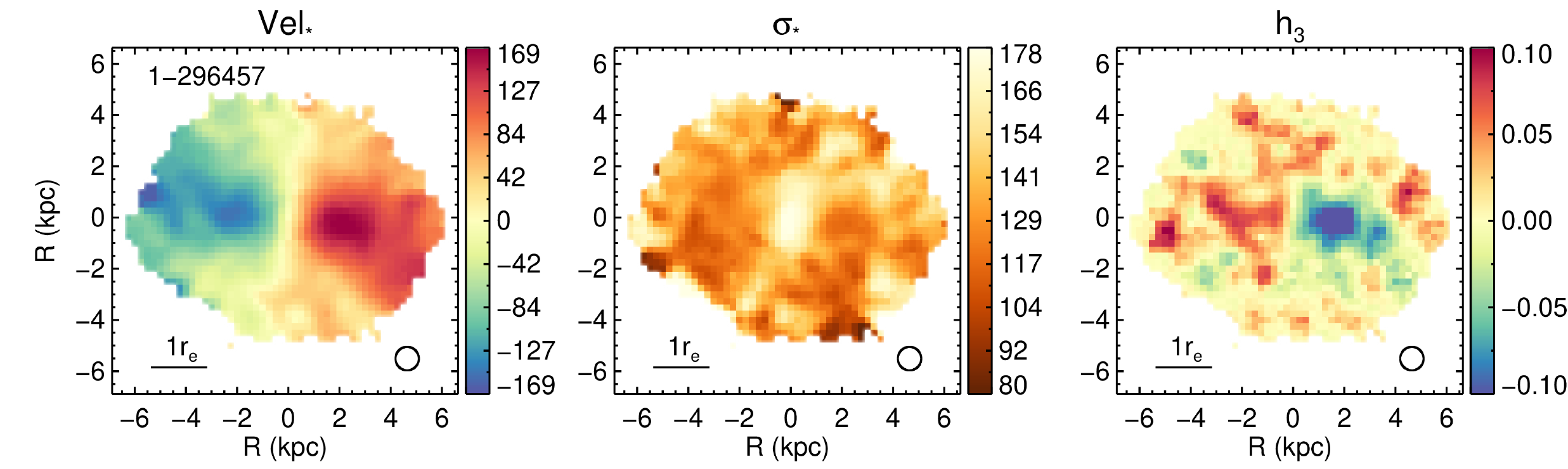}
    \caption{Fig.\,\ref{fig:kinematics_apend} continued.}
\end{figure*}

\begin{figure*}
\centering
    \includegraphics[width=0.95\hsize]{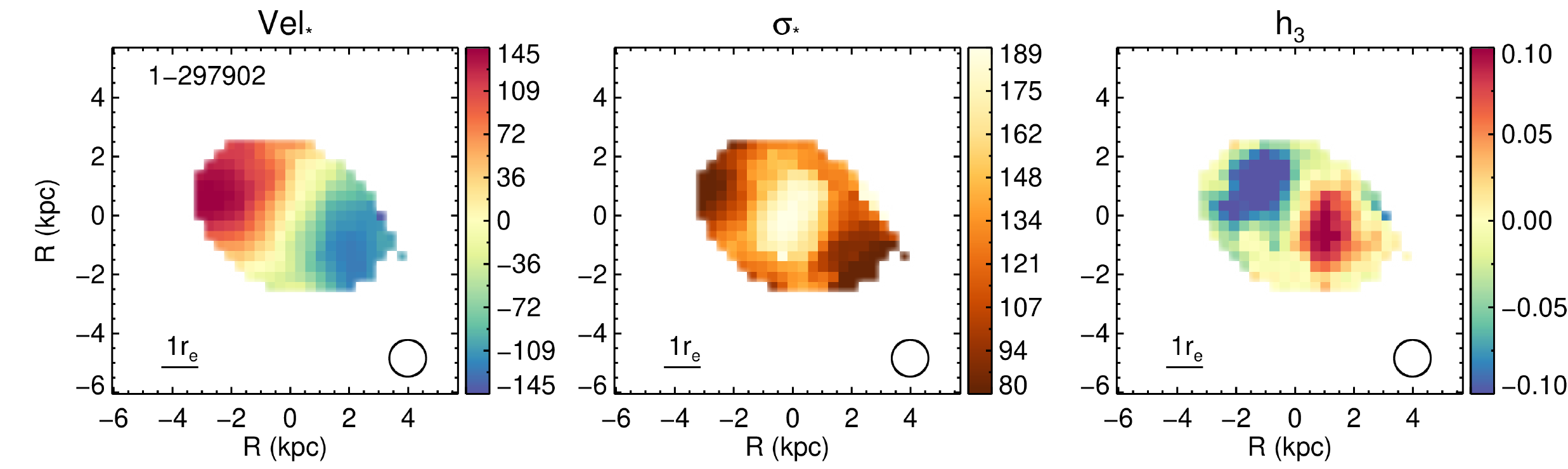}
    \includegraphics[width=0.95\hsize]{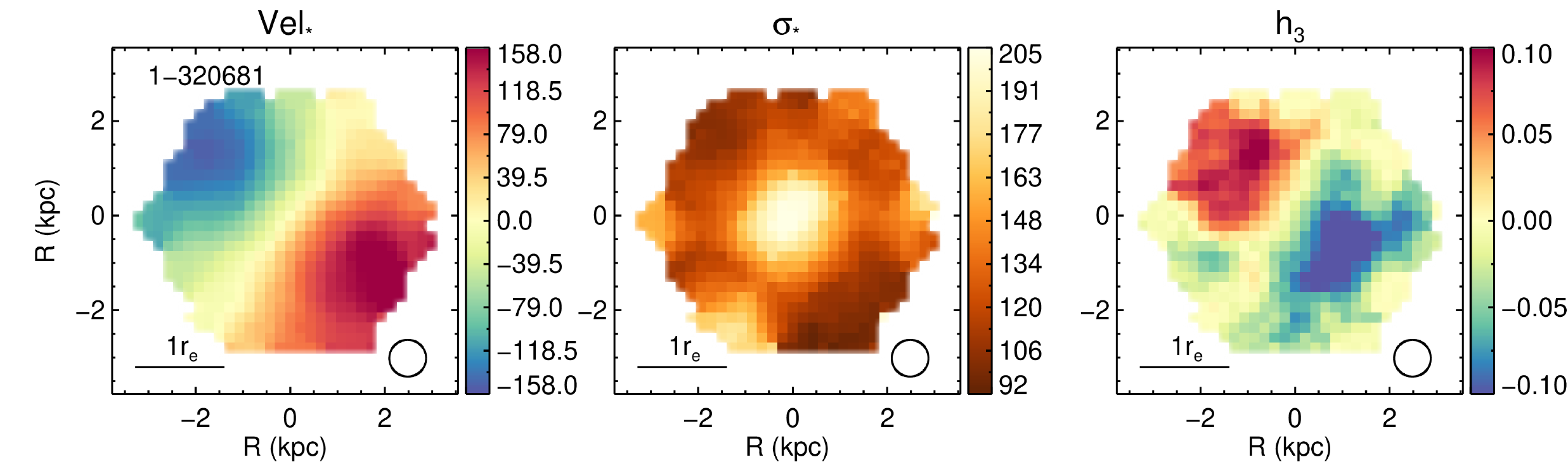}
    \includegraphics[width=0.95\hsize]{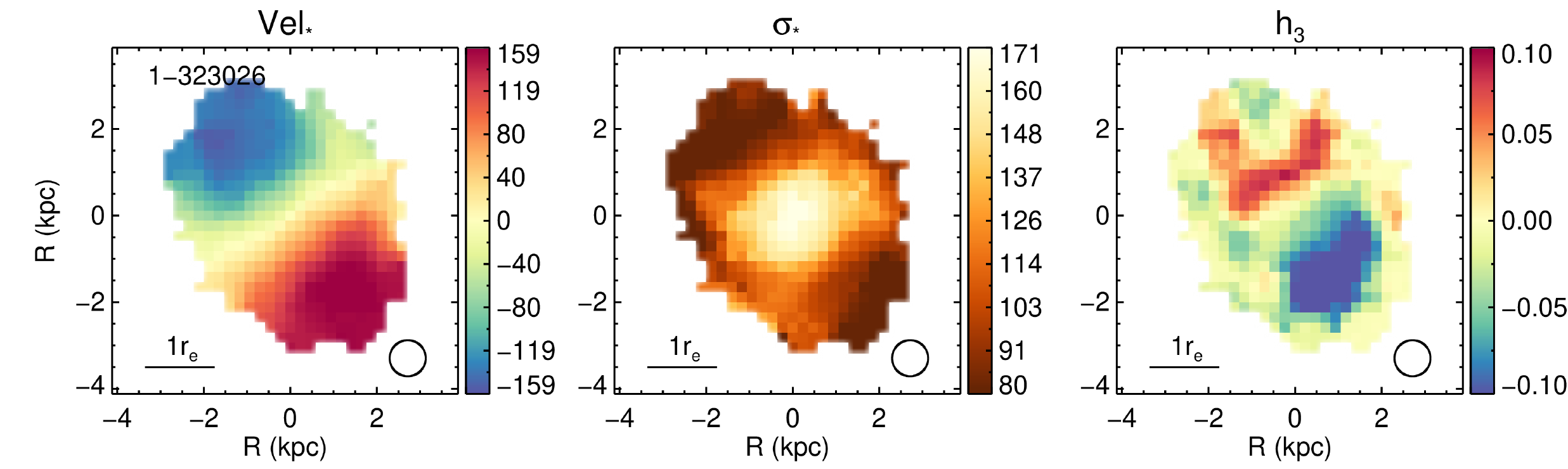}
    \includegraphics[width=0.95\hsize]{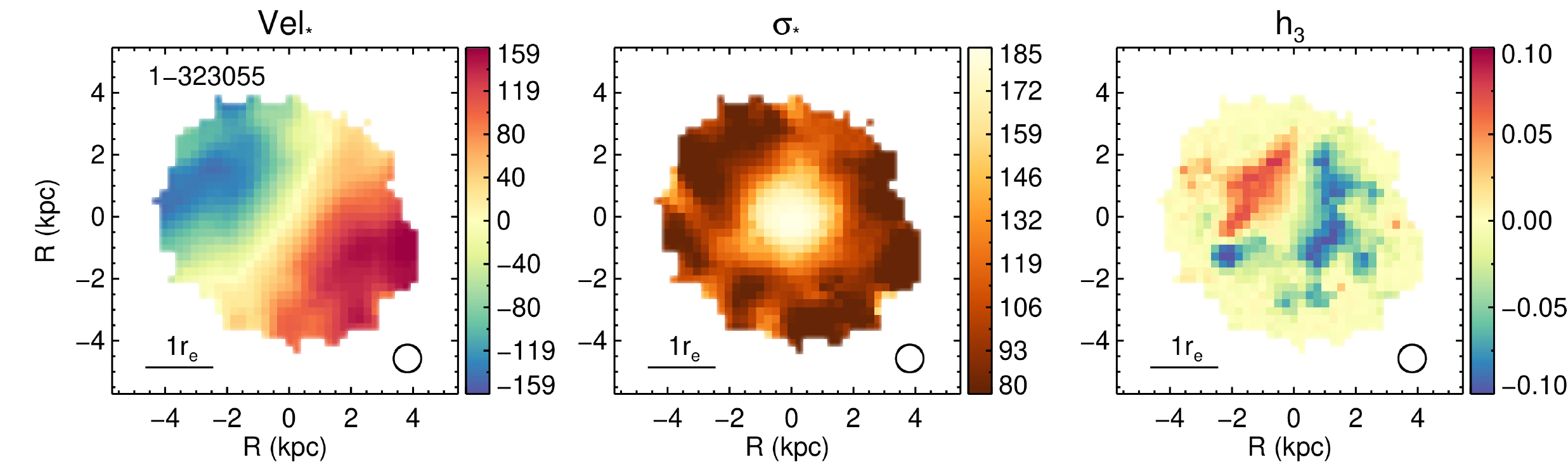}
    \caption{Fig.\,\ref{fig:kinematics_apend} continued.}
\end{figure*}

\begin{figure*}
\centering
    \includegraphics[width=0.95\hsize]{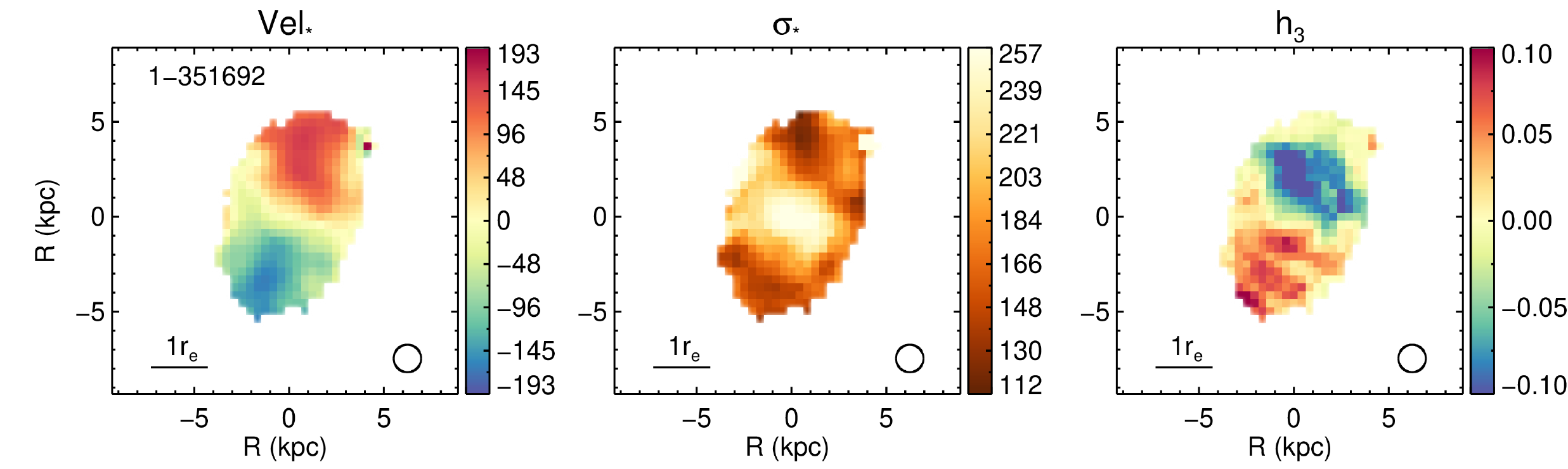}
    \includegraphics[width=0.95\hsize]{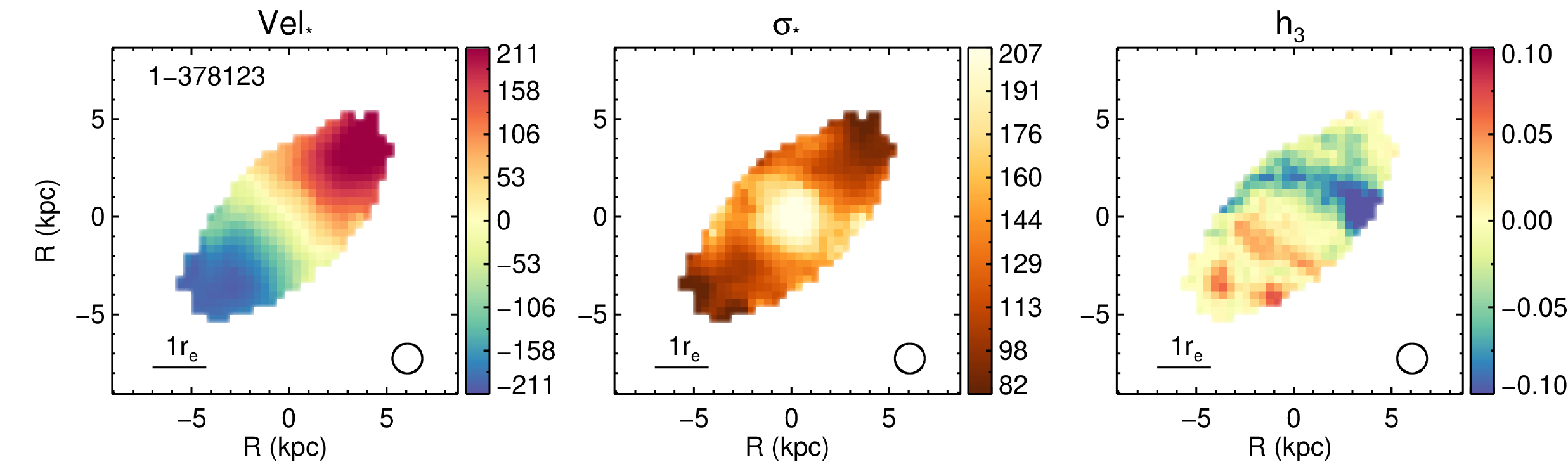}
    \includegraphics[width=0.95\hsize]{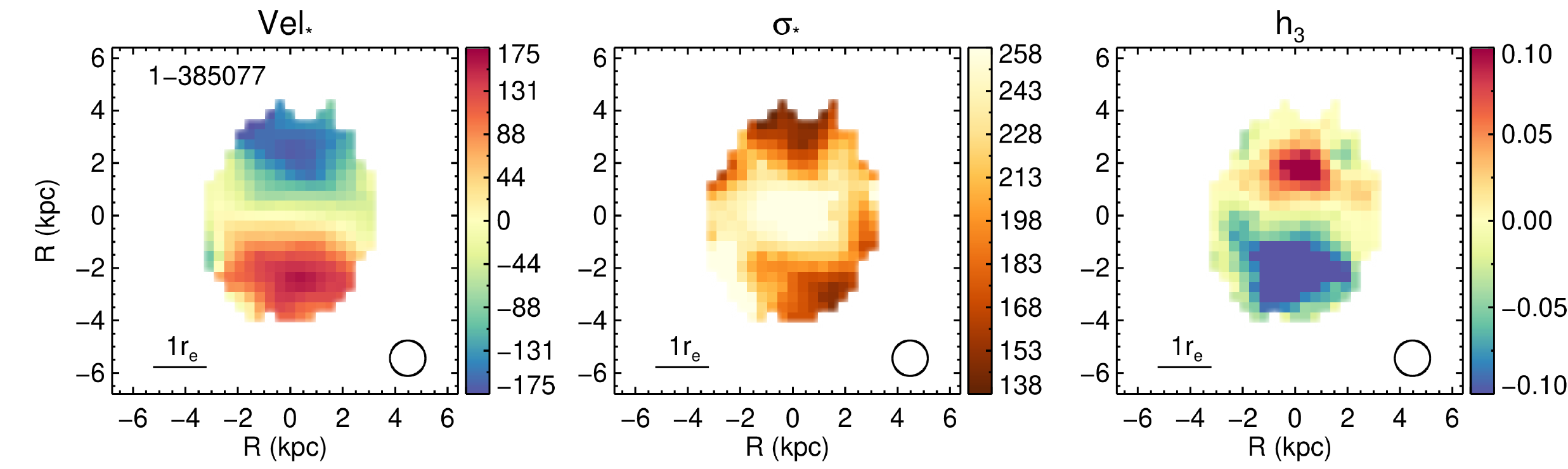}
    \includegraphics[width=0.95\hsize]{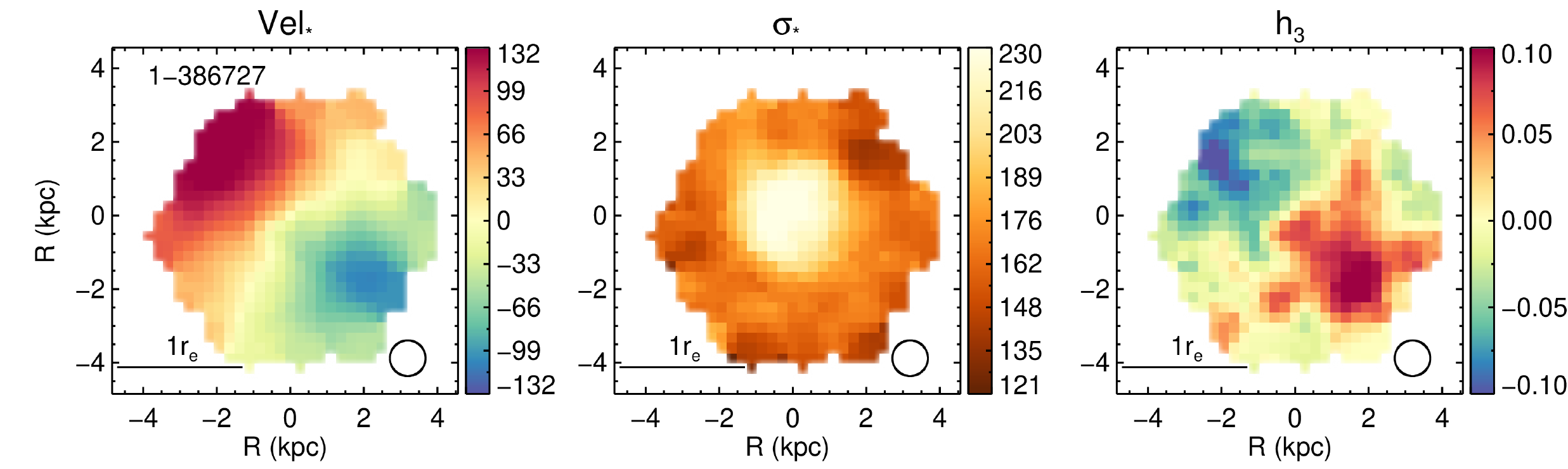}
    \caption{Fig.\,\ref{fig:kinematics_apend} continued.}
\end{figure*}

\begin{figure*}
\centering
    \includegraphics[width=0.95\hsize]{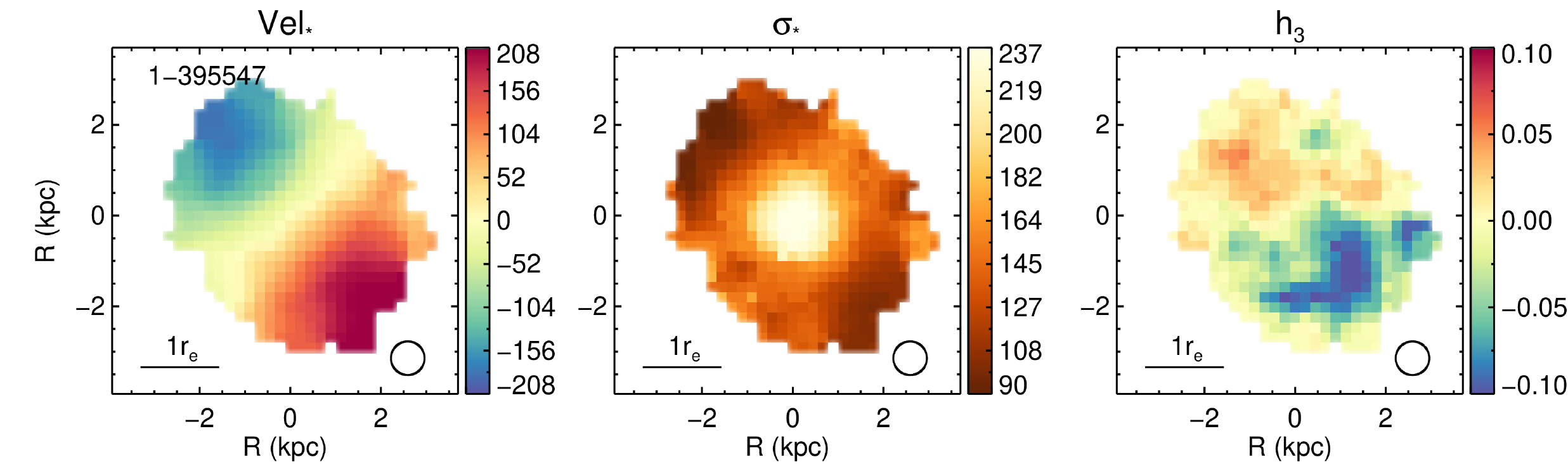}
    \includegraphics[width=0.95\hsize]{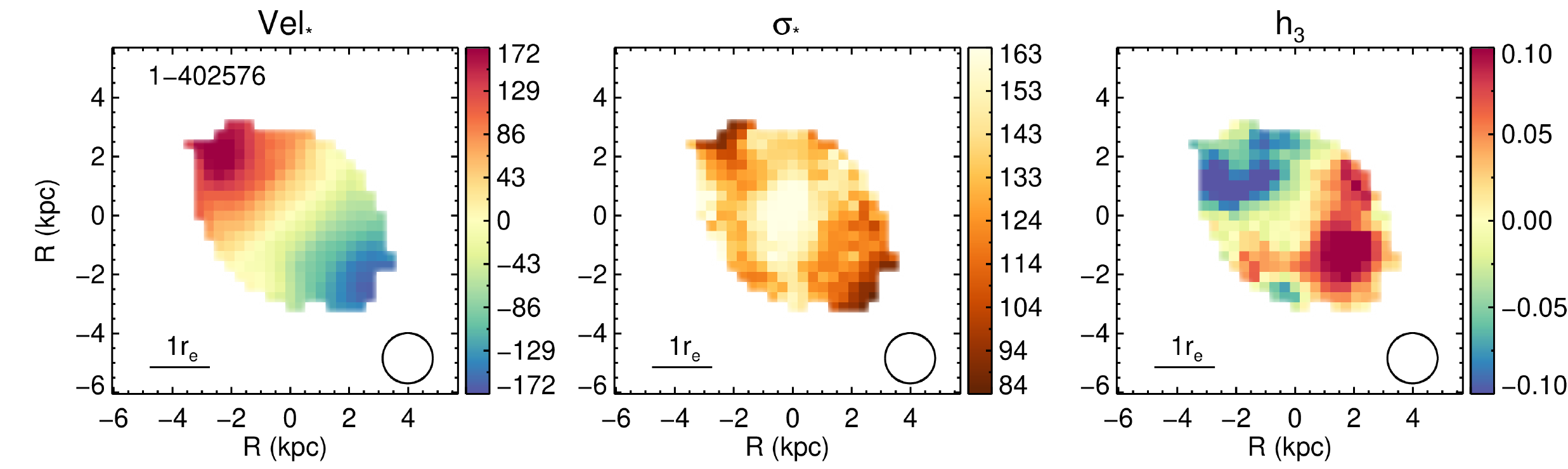}
    \includegraphics[width=0.95\hsize]{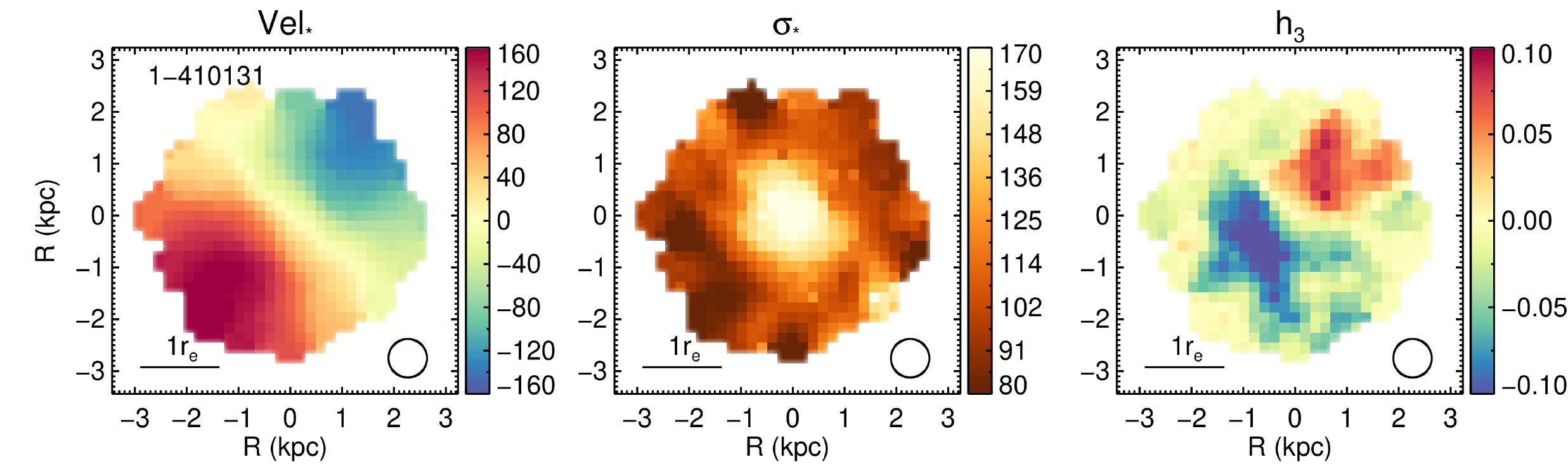}
    \includegraphics[width=0.95\hsize]{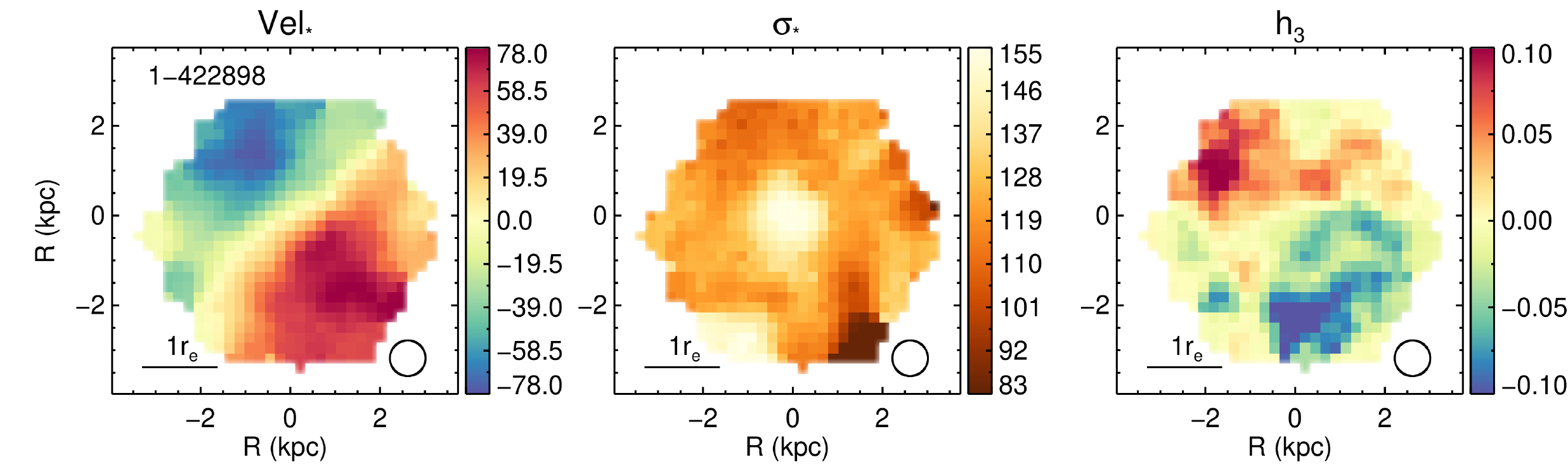}
    \caption{Fig.\,\ref{fig:kinematics_apend} continued.}
\end{figure*}

\begin{figure*}
\centering
    \includegraphics[width=0.95\hsize]{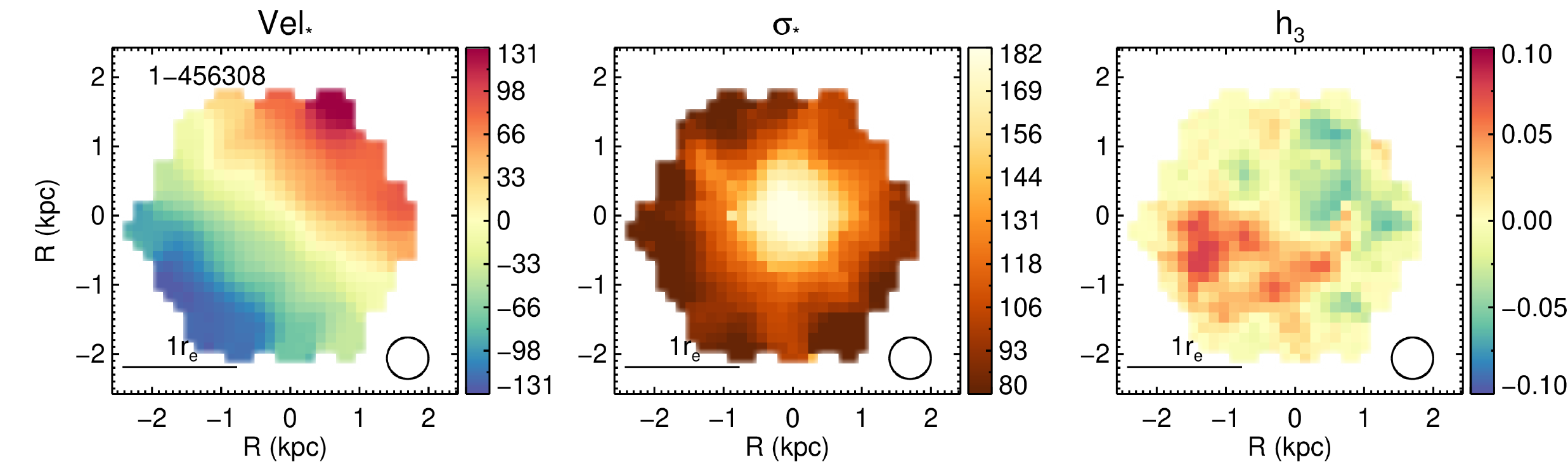}
    \includegraphics[width=0.95\hsize]{maps/1-456768.pdf}
    \includegraphics[width=0.95\hsize]{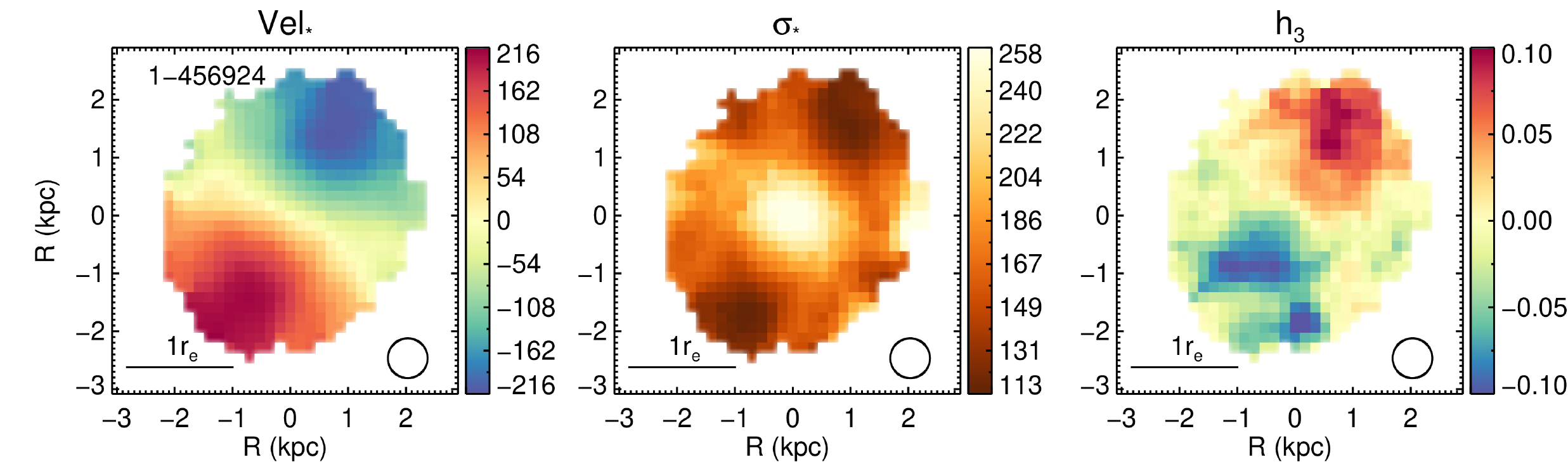}
    \includegraphics[width=0.95\hsize]{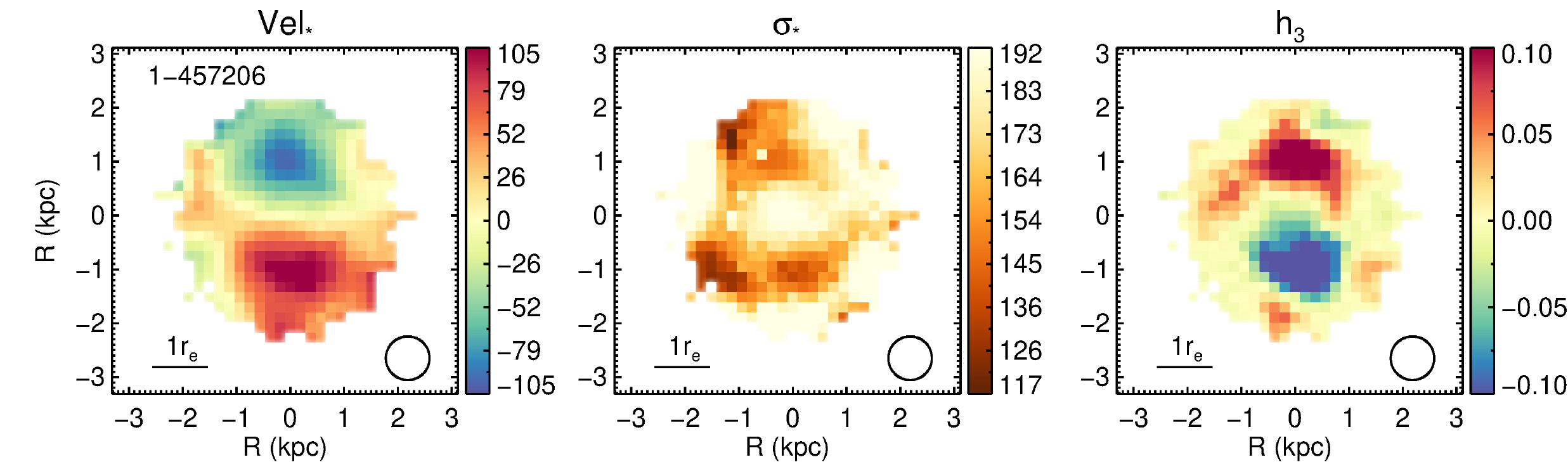}
    \caption{Fig.\,\ref{fig:kinematics_apend} continued.}
\end{figure*}

\begin{figure*}
\centering
    \includegraphics[width=0.95\hsize]{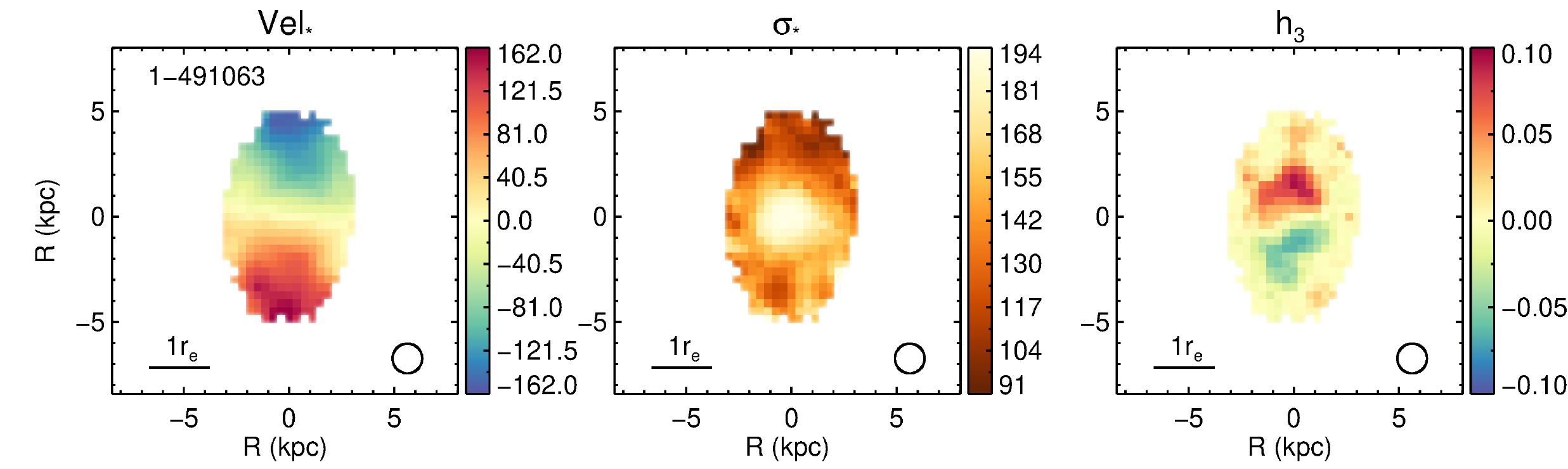}
    \includegraphics[width=0.95\hsize]{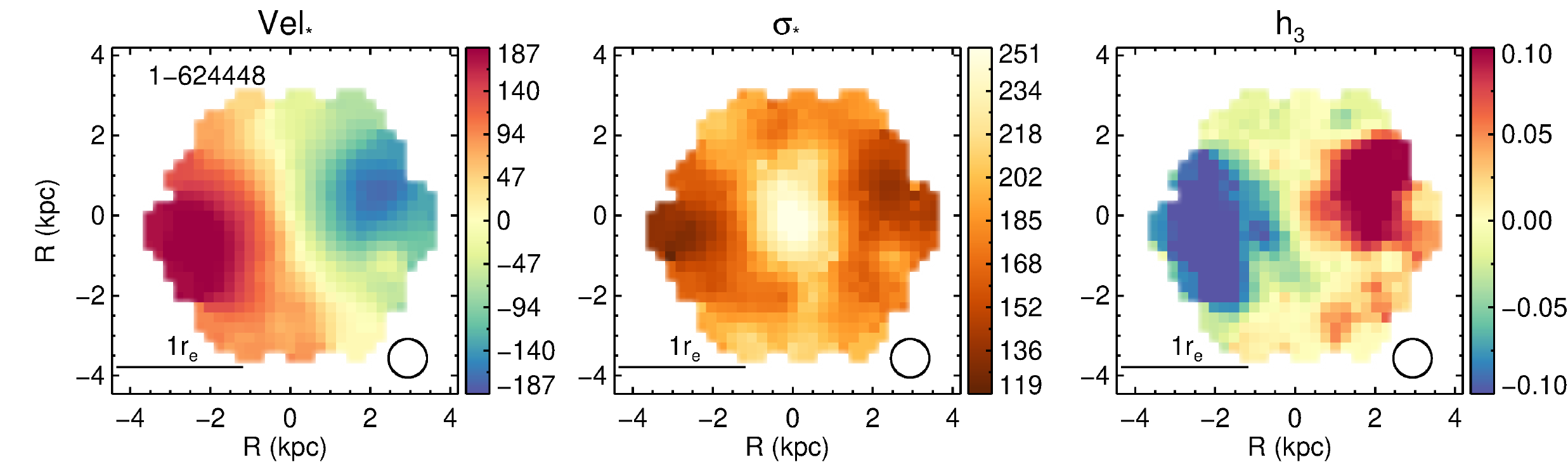}
    \caption{Fig.\,\ref{fig:kinematics_apend} continued.}
\end{figure*}


\end{document}